\documentclass[%
 twocolumn,prl,
 superscriptaddress,
 amsmath,amssymb,
 aps,
floatfix,
]{revtex4-2}

\usepackage{graphicx}
\usepackage{dcolumn}
\usepackage{bm}
\usepackage{grffile} 
\usepackage{subfigure}
\usepackage[marginal]{footmisc}
\usepackage{epstopdf}
\usepackage{amssymb}
\usepackage{mathrsfs}
\usepackage{amsmath,mathtools,amsfonts,bm,graphicx}
\usepackage{upgreek}
\usepackage{color}
\usepackage{etoolbox}
\usepackage[colorlinks=true,linkcolor=blue,urlcolor=blue,citecolor=blue]{hyperref} 
\usepackage{multirow}
\usepackage[normalem]{ulem}

\begin{document}

\preprint{APS/123-QED}

\title{Fate of thermalization of ultracold fermions with two-body dissipation}

\author{Xin-Yuan Gao}
\affiliation{%
Department of Physics, The Chinese University of Hong Kong, Shatin, New Territories, Hong Kong, China
}%

\author{Yangqian Yan}%
 \email{yqyan@cuhk.edu.hk}
\affiliation{%
Department of Physics, The Chinese University of Hong Kong, Shatin, New Territories, Hong Kong, China
}
\affiliation{
The Chinese University of Hong Kong Shenzhen Research Institute, 518057 Shenzhen, China
}%

\date{\today}

\begin{abstract}
    Two-body dissipation due to chemical reactions occurs in both ultracold fermionic and bosonic molecular gases. Despite recent advances in achieving quantum degeneracy, the loss dynamics are typically described phenomenologically using rate equations, often assuming thermalization during chemical reactions. From the first principles, we analyze particle loss, temperature evolution, and momentum distributions in single-component Fermi gases using the inelastic quantum Boltzmann equation. Our results prove that the conventional two-body loss model is valid for trapped systems, though it fails to describe the dynamics in homogeneous systems accurately. Interestingly, we find that systems prepared near or above quantum degeneracy can thermalize spontaneously, even in the absence of elastic collisions, while systems initialized deep in degeneracy display non-equilibrium behavior. Our calculations are in good agreement with recent experimental data from trapped systems and could be further tested in atomic systems with induced two-body loss in box potentials.
\end{abstract}

\maketitle

\textit{Introduction.---}
In classical hydrodynamic theory, dissipation, which originates from the inelastic collisions between particles and leads to heating, has been extensively studied in granular fluids~\cite{sela1998hydrodynamic,garzo1999dense,ben-naim1999shocklike,bobylev2000properties,ben-naim2002scaling}. Correspondingly, similar dissipative phenomena have also been observed in the quantum regime. A vitial feature of ultracold molecular gases---one of the candidates for quantum simulation~\cite{micheli2006toolbox,carr2009cold,osterloh2007strongly,buchler2007strongly,gorshkov2011tunable,gorshkov2011quantum,baranov2012condensed,cornish2024quantum} and quantum computation~\cite{zadoyan2001manipulation,demille2002quantum,andre2006coherent,rabl2006hybrid}--- is their two-body losses due to either chemical reactions or light-assisted chemical reaction, which are intrinsic to both fermions~\cite{ospelkaus2010quantumstate,park2015ultracold,demarco2019degenerate,hu2019directa,schindewolf2022evaporation,duda2023longlived} and bosons~\cite{molony2014creation,takekoshi2014ultracold,ye2018collisionsa,gregory2019stickya,gersema2021probing,bigagli2024observation}. 

After the preparation of the dissipative ultracold molecular systems, it has been observed that the system can be fitted with thermal profiles, with a gradually increasing temperature. This process has been dubbed anti-evaporation~\cite{ni2010dipolar,zhu2013evaporative}. While the system appears thermal and this heating phenomenon has been captured using phenomenological rate equations~\cite{soding1998giant,weber2003threebodya,ye2018collisionsa,guo2018dipolar,gregory2019stickya,horvath2024boseeinstein}, it is not yet fully clear whether the system stays in a genuine thermal state throughout the dynamics. For single-component bosonic systems with inelastic $s$-wave interactions, where the potential has no momentum dependence in Fourier space, particles with different momenta are removed at the same rate, leaving the shape of momentum distribution unchanged. Conversely, for single-component fermionic systems with inelastic $p$-wave interactions, the intrinsic momentum dependence of the interaction leads to particles with different momenta being removed at different rates. Thus, two-body $p$-wave loss could push the system out of thermal equilibrium and lead to exotic physics.

We now focus on single-component fermions with $p$-wave interactions. Although local thermal equilibrium could be assumed for classicial fluids. This may not be the case for ultracold molecules. 
It has been verified that a molecular gas could be thermalized with atom-molecule collision~\cite{tobias2020thermalization}, and theories assuming thermalized density matrix remain valid in a short time window~\cite{braaten2013universal,he2020universal,pan2020nonhermitian,gao2023temperaturedependent}. However, thermal equilibrium may not last a long time for a pure fermionic molecular quantum gas because the thermalization/relaxation time (e.g. several minutes~\cite{SM}) could be much longer than the characteristic two-body decay time (seconds). This seemingly forbids thermal equilibrium in long-term dynamics. There have been numerical studies of long-term dynamics using Lindblad equations~\cite{braaten2017lindblad,he2020exploring}, yet agreement with experiments has not been achieved under typical experimental parameters.

This letter attempts to address the fate of thermalization for single-component Fermi gases with two-body $p-$wave inelastic loss at finite temperatures by analyzing the \textit{long-time dynamics} from first principles. (i) We solve the inelastic quantum Boltzmann equation in the zero and infinitely fast elastic collision limit for systems in harmonic and box potentials. (ii) For harmonically trapped systems, our theory reproduces the long-time dynamics in recent experiments without any fitting parameter. We find that the solution almost agrees with the conventional two-body decay equation for harmonically trapped systems at high initial temperatures. (iii) In box traps, we obtain analytical (numerical) decay dynamics for arbitrary initial temperatures in the zero (strong) elastic collision limit and predict an unconventional particle-number decay equation that could be tested in ultracold atomic experiments.

\textit{Model.---} To describe a weakly interacting and reactive single-component Fermi gas, we begin with the single-channel $p$-wave non-Hermitian interaction 
\begin{equation}
\hat{U}=\frac{3 g}{2 V}\sum_{\mathbf{P},\mathbf{q},\mathbf{q}'}\mathbf{q}\cdot \mathbf{q}'c^\dagger_{\frac{\mathbf{P}}{2}+\mathbf{q}} c^\dagger_{\frac{\mathbf{P}}{2}-\mathbf{q}}c_{\frac{\mathbf{P}}{2}-\mathbf{q}'} c_{\frac{\mathbf{P}}{2}+\mathbf{q}'},
\label{U}
\end{equation}
with coupling strength $g=4\pi\hbar^2[\mathrm{Re}(v_p)+i\mathrm{Im}(v_p)]/M$ derived from low-energy $p$-wave phase shift: $k^3\cot(\delta_p)=-1/v_p+\mathcal{O}(k^2)$. The model is a direct generalization of $s$-wave complex contact interaction~\cite{wang2022complex}. Here, $v_p$, $M$, and $V$ are the complex $p$-wave scattering volume $v_p$, the particle mass, and the system's volume, respectively; $c_\mathbf{k}$ is the fermionic annihilation operator following the canonical anti-commutation relation. By expanding the dynamics of $\langle N_\mathbf{k}\rangle=\langle c^\dagger_\mathbf{k} c_\mathbf{k}\rangle$ up to the second order of the interaction~\cite{SM}, we obtain the inelastic Boltzmann equation that governs the dynamics of a homogeneous system
\begin{equation}
{d n_\mathbf{k}}/{dt}={\mathcal{I}_{\mathrm{inel}}[n_\mathbf{k} V]}/{V}+{\mathcal{I}_{\mathrm{el}}[n_\mathbf{k} V]}/{V},
\label{homo_boltzmann_eq}
\end{equation}
where $n_\mathbf{k}(t)=\langle N_\mathbf{k}\rangle(t)/V$ is the momentum distribution function; $\mathcal{I}{_\mathrm{inel}}[n_\mathbf{k}V]$ and $\mathcal{I}{_\mathrm{el}}[n_\mathbf{k}V]$ are the inelastic and elastic collision integrals, respectively. Specifically, the inelastic collision integral is given by
\begin{equation}
\mathcal{I}_{\mathrm{inel}}[n_\mathbf{k}V]=\frac{12\pi\hbar\mathrm{Im}(v_p)V^2}{M}\int\frac{d^3q}{(2\pi)^3}(q^2+k^2) n_\mathbf{k} n_\mathbf{q},
\label{Iinel}
\end{equation}
and the form of the elastic collision integral $\mathcal{I}_{\mathrm{el}}[n_\mathbf{k}]$ can be found in the companion paper~\cite{SM}. For application to typical molecular Fermi gas in experiments, $\mathcal{I}_{\mathrm{el}}$ can be safely ignored because the relaxation time is much longer than other time scales in the system~\cite{SM}. Equation~(\ref{Iinel}) has also been derived by Ref.~\cite{he2020exploring} using a two-channel model.

To accurately model ultracold gas experiments, which are predominantly conducted in harmonic traps, we adapt Eq.~(\ref{homo_boltzmann_eq}) using the local-density approximation $n_\mathbf{k} V\rightarrow f(\mathbf{k},\mathbf{r})$, where $f(\mathbf{k},\mathbf{r})$ represents the phase space density. In addition to elastic and inelastic collisions, the inhomogeneity prompts the cloud to flow. The inelastic Boltzmann equation in harmonic traps governing the complete set of dynamics is thus
\begin{equation}
\frac{d f}{dt}=\left[-\frac{\hbar \mathbf{k}}{M}\nabla_\mathbf{r}+\frac{\nabla_{\mathbf{r}}U_{\mathrm{ext}}\cdot\nabla_\mathbf{k}}{\hbar}\right]f+\mathcal{I}_{\mathrm{inel}}[f]+\mathcal{I}_{\mathrm{el}}[f],
\label{trap_boltzmann_eq}
\end{equation}
where $U_\mathrm{ext}(\mathbf{r}) = \sum_{i=x,y,z}M\omega_i^2r_i^2/2$ is the external harmonic trapping potential, with $\omega_i$ denoting the trapping angular frequencies in the three spatial directions.

We are interested in the dynamics under weak while complex interactions. In this regime, there are two limits that are easy to solve: no elastic interaction at all, i.e., $\mathrm{Re}(v_p)=0$, or the elastic collisions are much more frequent than the inelastic collisions such that the system is always in thermal equilibrium. We dub the second case the \textit{thermal ansatz}. For typical experimental realizations with weak but finite real parts, the solutions must fall between the two limits. However, as we argue in the companion paper, when the real and imaginary parts are comparable, the pure imaginary interaction, i.e., the first case should be a better approximation~\cite{SM}.

Conventionally, for a two-body loss, denoting the total number of particles to be $N(t)=\int {d^3 k} n_\mathbf{k}(t)/{(2\pi)^3}$ or $\int {d^3rd^3k}f(\mathbf{r},\mathbf{k},t)/{(2\pi)^3}$, it is most natural to expect the dynamics to be~\cite{mehta2009general}
\begin{equation}
{d N(t)}/{dt}\propto - N(t)^\mathcal{N},
\label{N_body_model}
\end{equation}
where $\mathcal{N}=2$. Though this holds for $s$-wave interactions, we will demonstrate later that in single-component Fermi gases with collisions of $p$-wave nature, the two-body loss dynamics may lead to $\mathcal{N}\neq2$.

\begin{figure}
    \centering
    \includegraphics[width=0.495\textwidth]{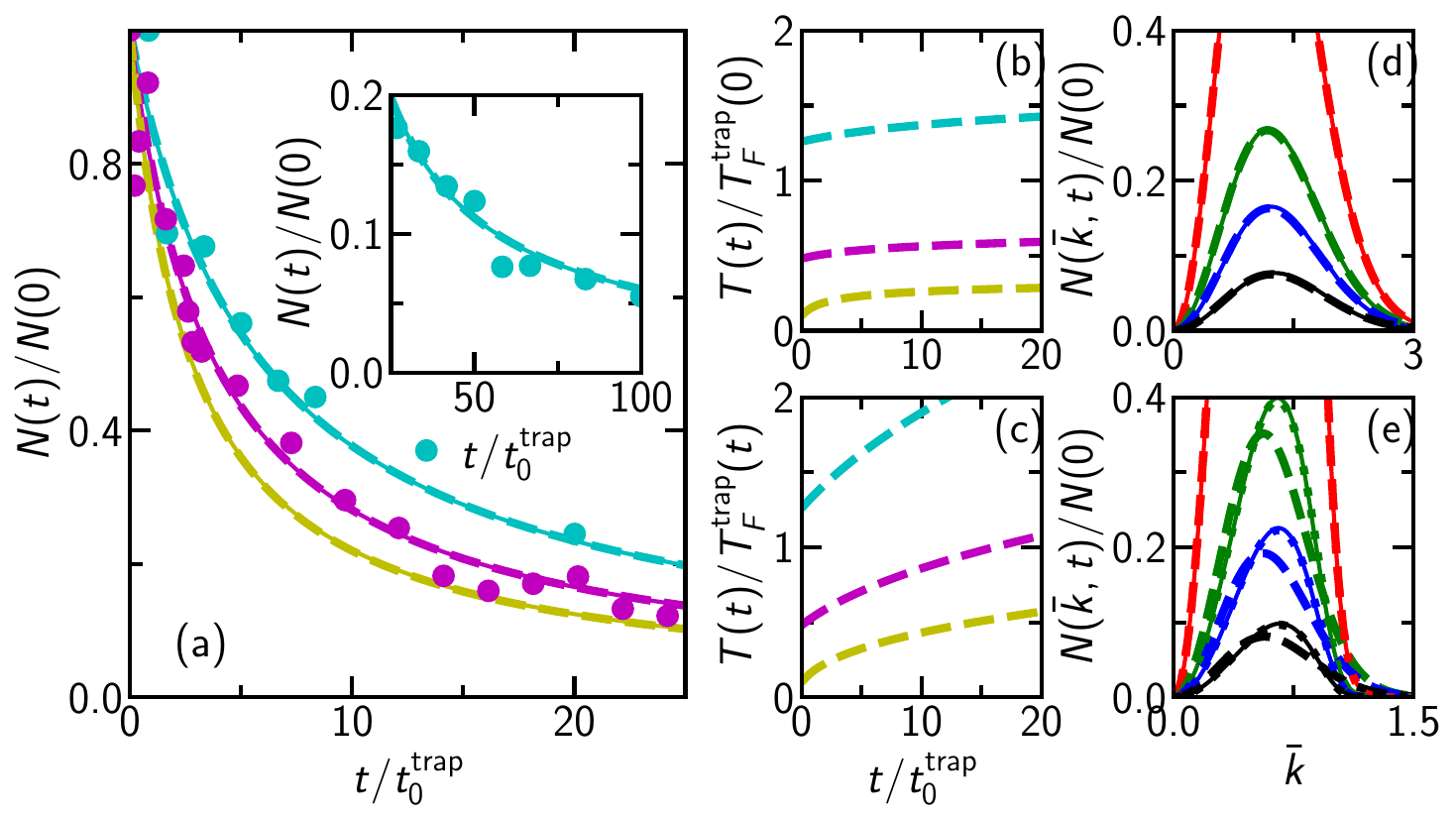}
\caption{
Long-time dynamics of the trapped single-component Fermi gases with inelastic $p$-wave interactions.    
(a) Total number of particles as a function of time. 
Symbols, solid, and dashed lines are experimental data from Ref.~\cite{demarco2019replication}, numerical results solved from the inelastic Boltzmann equation [Eq.~(\ref{trap_boltzmann_eq})] under the fast-flow approximation, and results obtained from thermal ansatz, respectively. From top to bottom, cyan, magenta, and
yellow denote initial temperatures $T=1.26, 0.48$ and $0.1 T_F^\mathrm{trap}(0)$. Inset: Longer time dynamics for $T(0)/T_F^\mathrm{trap}(0)=1.26$. (b) and (c): Dynamics of physical and reduced temperatures of thermal ansatz. Colors have the same meaning as those in (a). (d) and (e): the evolution of radial momentum distribution for $T(0)/T_F^\mathrm{trap}(0)=1.26$ and $0.1$, respectively. From top to bottom at $\Bar{k}=0.7$, red, green, blue, and black colors denote $t/t_0^\mathrm{trap}=0,10,20,50$. Solid lines and dotted lines represent numerical results under fast-flow approximation and its quasi-thermal profile fit, respectively; the dashed line is for the thermal ansatz.
}
    \label{fig_trap}
\end{figure}

\textit{Harmonically Trapped Systems.---}We first consider systems loaded into a harmonic trap, where the dynamics are described by Eq.~(\ref{trap_boltzmann_eq}). As a general feature of the Boltzmann equation, Eq.~(\ref{trap_boltzmann_eq}) has an irreducible six-dimensional spatial complexity, making it hard to solve. Nevertheless, there exists a separation of time scales, which simplifies the problem. The harmonic trap tends to average the phase space distribution $f(\mathbf{r},\mathbf{k},t)$ such that it has a spherical symmetry in the whole phase space, i.e., defining $R(t)=\sqrt{\sum_{i=x,y,z}[({k_i}/{k_F^\mathrm{trap}(t)})^2+({\omega_i r_i}/{\omega r_F(t)})^2]}$, $f(\mathbf{r},\mathbf{k},t)\equiv f(R,t)$, where $k_F^\mathrm{trap}(t)=[48N(t)]^{1/6}\sqrt{{M\omega}/{\hbar}}$ and $r_F(t)=[48N(t)]^{1/6}\sqrt{{\hbar}/{M\omega}}$ are Fermi momentum and Thomas-Fermi radius of harmonically trapped system with $\omega=(\omega_x\omega_y\omega_z)^{1/3}$ the geometric mean of angular frequencies of the harmonic trap. Indeed, this is the steady-state solution without the inelastic term.
Despite the fact that the inelastic collisions drive the phase space density into asymmetric forms, under typical experimental parameters, two-body dissipation has a much longer time scale (usually of the order of $\mathrm{s}$) compared to that of cloud-flowing determined by trap frequencies (typically of the order of $\mathrm{ms}$). 
Thus, we propose a fast-flowing approximation, which assumes the phase space distribution is always in the spherical form. This could be understood as the following: though the inelastic collision tries to bring asymmetry to the distribution, the trap term restores the symmetry at a much faster rate. 
We have validated this approximation through a quasi-1D analogy of Eq.~(\ref{trap_boltzmann_eq}) in the companion paper~\cite{SM}. 
It is noted that the fast-flowing approximation is closely related to the basic assumption of Chapman-Enskog expansion in classic hydrodynamics, where the collision, rather than flowing, is assumed to have the fastest time-scale enforcing system obeying the steady state of the collision integral. 

Under the fast-flowing approximation, we numerically solve Eq.~(\ref{trap_boltzmann_eq}) and demonstrate results starting with systems above, near and in the deep degeneracy ($T(0)/T_F^\mathrm{trap}=1.26,0.48$ and $0.1$, respectively) in Fig.~\ref{fig_trap}(a).  For convenience, we set the unit $t_0^\mathrm{trap}={\pi M}/{(\hbar [k_F^\text{trap}(0)]^5|\mathrm{Im}(v_p)|)}$. We have compared our calculation with experimental data provided in Ref.~\cite{demarco2019replication} for the first two cases and shown that they agree well for long times (both situations correspond to approximately $6$ seconds in the experiment~\cite{demarco2019degenerate}).

To investigate the system's thermalization during evolution, we also solve the inelastic quantum Boltzmann equation under thermal ansatz and compare the result with our numerical results. The fast-flowing approximation is exact for systems in the equilibrium limit since the thermal ansatz naturally respects spherical symmetry in the phase space. We assume that initially, the system starts from an equilibrated state at a temperature $T$. The thermal ansatz then means that  $f^\mathrm{th}(\mathbf{k},\mathbf{r})(t)$ always has the shape of a Fermi-Dirac distribution, but with varying temperature $T(t)$ and Fermi temperature $T_F^\mathrm{trap}(t)={\hbar^2[k_F^\mathrm{trap}(t)]^2}/{2Mk_B}$, i.e.,
$
f^\mathrm{th}(\mathbf{k},\mathbf{r})\equiv f^\mathrm{th}(R(t),t)=\left\{\exp\left[{R(t)^2}/{[T(t)/T_F^\mathrm{trap}(t)]}\right](z^{\mathrm{th},\mathrm{trap}})^{-1}+1\right\}^{-1}
$, where $z^{\mathrm{th},\mathrm{trap}}=-\mathrm{Li}_{3}(-z^{\mathrm{th},\mathrm{trap}})=[T(t)/T_F^\mathrm{trap}(t)]^{-3}/6$ with $\mathrm{Li}_s$ denoting the polylogarithm function. 
Fig.~\ref{fig_trap}(a) shows that the thermal ansatz closely matches the fast-flowing approximation predictions for particle-number dynamics. This observation provides insight into why a phenomenological two-body decay equation describes the system well in this case: In the high-initial-temperature limit $T(0)/T_F^\mathrm{trap}(0)\rightarrow\infty$, we derive an analytical solution based on the thermal ansatz~\cite{SM}:
\begin{equation}
N(t)={N(0)}/{\left[1+{C(t/t_0^\mathrm{trap})}\right]^{0.960}},
\label{trap_thermo_highT}
\end{equation}
where $C=0.198418/{\sqrt{T(0)/T_F^\mathrm{trap}(0)}}$. Applying Eq.~(\ref{N_body_model}) to Eq.~(\ref{trap_thermo_highT}) yields $\mathcal{N}=2.04167$, which is nearly the conventional two-body behavior ($\mathcal{N}\approx2$). This behavior is independent of the trap frequencies as long as they are larger than the characteristic two-body loss frequency.

Through the temperature dynamics of thermal ansatz, we can also reproduce the anti-evaporation. We present the results in two complementary ways. First, Fig.~\ref{fig_trap}(b) illustrates the dynamics of the ``physical temperature," using $T_F^\mathrm{trap}(0)$ as a fixed unit to characterize the average particle energy. Second, Fig.~\ref{fig_trap}(c) displays the ``reduced temperature," where $T_F^\mathrm{trap}(t)$ evolves with the total particle number, indicating whether the system is in a ``high-temperature" or ``low-temperature" regime. Analysis of both physical and reduced temperature evolution [Fig.~\ref{fig_trap}(b) and (c)] shows that both temperatures increase regardless of initial conditions, which is consistent with the anti-evaporation phenomena observed in experiments. 

Finally, we examine the momentum distribution evolution [$N(\Bar{k},t)=4\pi \Bar{k}^2 \int d^3r f(\mathbf{r},\mathbf{k},t)/(2\pi)^3$, where $\Bar{k}=k/k_F^\mathrm{trap}(0)$].
Systems starting above degeneracy are well-described by the corresponding thermal ansatz [Fig.~\ref{fig_trap}(d)], while those beginning in deep degeneracy exhibit non-equilibrium behavior [Fig.~\ref{fig_trap}(e)]. It is worthwhile to note that though thermal ansatz does not agree well with the low-temperature evolution, a ``quasi-thermal profile,'' i.e., fit time-of-flight images to a general Fermi-Dirac distribution with three parameters: prefactor, $T$, and fugacity $z$., can still be made and agrees with the momentum distribution well. In fact, this could be a typical practice in experiments as the prefactor originates from absorption imaging. Fig.~\ref{fig_trap}(d) and (e) demonstrate that this quasi-thermal profile fits the fast-flowing approximation results well, even for low initial temperatures and long-time evolution. Consequently, experimental measurements may conclude that the system is ``thermalized'' based on this criterion. 

\textit{Box potential.---} Though our theory explains existing experiments without fitting parameters, they could still be approximated well by thermal ansatz or even a naive rate equation. This ``disappointing'' result comes from the trap averaging: the harmonic trap provides a force in the momentum space such that it induces a flow in phase space seemingly towards a thermal state. Fortunately, nowadays, experiments are also performed in box traps~\cite{navon2021quantuma}, where this unwanted flow does not exist. In this case, the problem is even simpler as we only need to evolve the three-dimensional momentum distribution instead of the six-dimensional phase space distribution.

Again, we assume the system starts from an equilibrated state at temperature $T$. Thus, $n_{\mathbf{k}}(0)$ simply follows the Fermi-Dirac distribution $n_{\mathbf{k}}(0)=[\exp({\hbar^2k^2}/{2M k_B T})z(T)^{-1}+1]^{-1}$. The fugacity $z(T)$ is implicitly determined by $-\mathrm{Li}_{3/2}[-z(T)]=4\Bar{T}^{-3/2}/3\sqrt{\pi}$, where $\Bar{T}=T/T_F(0)$ and $T_F(0)=\hbar^2(6\pi^2N(0)/V)^{2/3}/2M$.  A systematic analysis of Eq.~(\ref{homo_boltzmann_eq}) without the elastic collision can be achieved by Mellin transforming the momentum distribution, which decomposes the integro-differential equation into infinitely many coupled ordinary differential equations~\cite{SM}. Utilizing this method, we find an accurate approximation of $N(t)$ starting from an equilibrium state at $T$ to be
\begin{equation}
N(t)=N(0)\left[{1+\left(F_1+{F_2}/{F_1}\right)({t}/{t_0})}\right]^{\frac{-2F_1^2}{F_1^2+F_2}},
\label{homo_noneq_sol}
\end{equation}
where 
$
F_j=-\frac{3}{2}\Bar{T}^{\frac{3}{2}+j}\Gamma\left({3}/{2}+j\right)\mathrm{Li}_{\frac{3}{2}+j}[-z(T)];
\label{Fj0}
$
$t_0={M}/({12\pi{\hbar k_F^5(0)}|\mathrm{Im}(v_p)|})$ is the time unit and the homogeneous Fermi momentum is $k_F(t)=(6\pi^2N(t)/V)^{1/3}$.
\begin{figure}[t]
\centering
\includegraphics[width=0.495\textwidth]{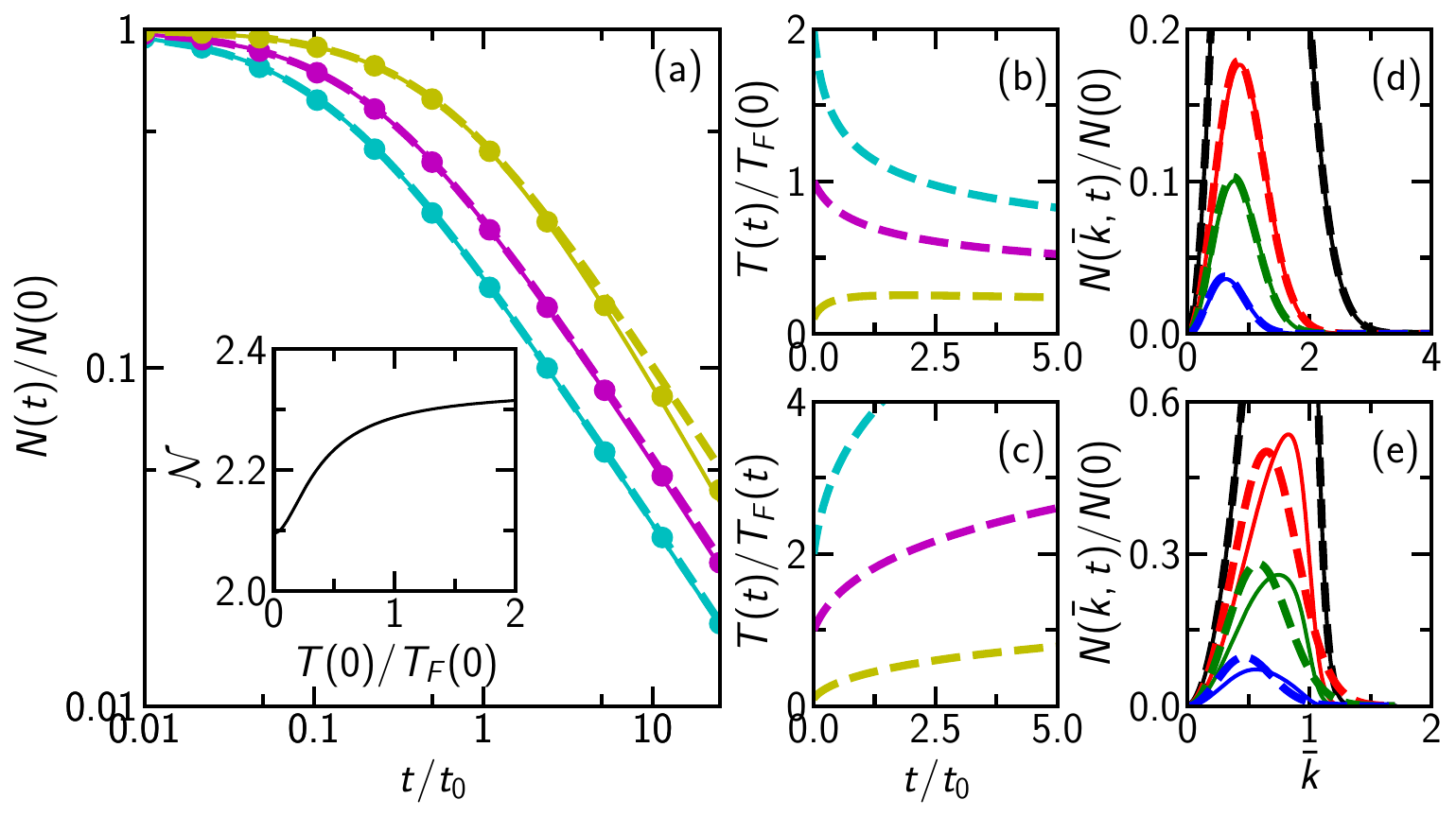}
\caption{
Long-time dynamics of single-component Fermi gases with inelastic $p$-wave collisions in box potential. (a) Total number of particles as a function of time $t$. Circles, solid and dashed lines represent the numerical solution of the inelastic Boltzmann equation [Eq.~(\ref{homo_boltzmann_eq}) ignoring elastic collision integral], the analytical approximation [Eq.~(\ref{homo_noneq_sol})], the solution based on thermal ansatz, respectively. From top to bottom, yellow, magenta, and
cyan denote the initial temperatures $T(0)=0.1, 1$ and $2T_F(0)$, respectively. Inset: The $N$-body indicator [Eq.~(\ref{homo_scN})] as a function of the initial temperature $T(0)$.
(b) and (c) show the physical temperature in units of $T_F(0)$ and the reduced temperature as a function of time, respectively; from top to bottom, the yellow, magenta, and cyan lines correspond to initial temperatures $T(0) = 0.1, 1$, and $2T_F(0)$, respectively.
(d) and (e) show the dynamics of the radial momentum distribution of the system with initial temperature $T(0)=2T_F(0)$ and $T(0)=0.1T_F(0)$, respectively. Solid and dashed lines denote brute-force numerical solution of the inelastic Boltzmann equation and results based on thermal ansatz. Black, red, green, and blue colors are for $t/t_0=0,2,5,25$, respectively.
}
\label{fig_homo}
\end{figure}

Solid lines and symbols in Fig.~\ref{fig_homo}(a) demonstrate the comparison between Eq.~(\ref{homo_noneq_sol}) and brute-force numerical solution, showing that the approximation is satisfactory within a reasonably long time window. It is noted that Eq.~(\ref{homo_noneq_sol}) is the solution of Eq.~(\ref{N_body_model}) with
\begin{equation}
\mathcal{N}=3/2 + {F_2}/({2F_1^2}),
\label{homo_scN}
\end{equation}
thus, one may regard the dissipation dynamics as an $\mathcal{N}$-body process in the conventional sense. The inset of Fig.~\ref{fig_homo}(a) shows $\mathcal{N}$ against $\Bar{T}$, where we observe that $\mathcal{N}$ monotonically decreases with $\Bar{T}$. It is straightforward to obtain the limits $\mathcal{N}(\Bar{T}\rightarrow0)=44/21$ and $\mathcal{N}(\Bar{T}\rightarrow\infty)=7/3$. It is worth mentioning that Eq.~(\ref{homo_noneq_sol}) is exact in the limit $\Bar{T}\rightarrow\infty$, where the full dynamics of $n_\mathbf{k}(t)$ is~\cite{SM}
\begin{equation}
n_\mathbf{k}(t)=\dfrac{4\exp\left[-{\Bar{k}^2}\left(1+{4\Bar{T}t}\right)^{\frac{1}{4}}/{\Bar{T}t_0}\right]}{3\sqrt{\pi}N(0)\Bar{T}^{\frac{3}{2}}\left(1+{4\Bar{T}t/t_0}\right)^{\frac{8}{3}}},
\label{homo_noneq_nk}
\end{equation}
where $\Bar{k}=k/k_F(0)$.

To rigorously check whether the system thermalizes after a long time evolution, we solve Eq.~(\ref{homo_boltzmann_eq}) under the thermal ansatz with varying $T(t)$ and $T_F(t)={\hbar^2[k_F(t)]^2}/{2Mk_B}$, i.e.,
$
    n_\mathbf{k}^\mathrm{th}(t)=\left\{\exp\left[{({k}/{k}_F(t))^2}/{[{T}(t)/{T}_F(t)]}\right]z^\mathrm{th}(t)^{-1}+1\right\}^{-1},
$
where $-\mathrm{Li}_{3/2}[-z^\mathrm{th}(t)]=4[T(t)/T_F(t)]^{-3/2}/3\sqrt{\pi}$. In the high-initial-temperature limit $T(0)/T_F(0)\rightarrow\infty$, there are simple solutions~\cite{SM}:
$
N(t)={N(0)}/{\left[1+{4\Bar{T}(0)t/t_0}\right]^{\frac{3}{4}}}
$ and $
T(t)={T(0)}/{\left[1+{4\Bar{T}(0)t/t_0}\right]^{\frac{1}{4}}}
$.
Substituting back to the thermal ansatz and taking the high-temperature limit, one recovers Eq.~(\ref{homo_noneq_nk}), confirming that the system continually evolves under thermalized profiles even strictly without elastic collisions.

For systems with arbitrary initial temperatures, we numerically evaluate the thermal ansatz to calculate $N(t)/N(0)$ and compare it with Eq.~(\ref{homo_noneq_sol}). This comparison allows us to determine whether systems with arbitrary initial temperatures thermalize similarly to those in the high-initial-temperature limit. We first analyze the temperature dynamics of the thermal ansatz to gain deeper insights into its general behavior. Similar to harmonically trapped systems, we plot the results in terms of ``physical temperatures'' [Fig.~\ref{fig_homo}(b)] and ``reduced temperatures'' [Fig.~\ref{fig_homo}(c)]. The physical temperature dynamics strongly depend on the initial temperature: The physical temperature decreases for systems initially above or near quantum degeneracy. This behavior arises from the centrifugal barrier in the $p$-wave inter-molecular potential. Particles with higher momenta are more likely to tunnel through this barrier and undergo collisions, as reflected in the momentum dependence of Eq.~(\ref{Iinel}). Consequently, $p$-wave collisions preferentially remove higher-momentum, higher-energy particles, reducing the average energy and, thus, the physical temperature. In deeply degenerate systems, any perturbation to the Fermi sea structure, including dissipation, tends to heat the system, increasing the physical temperature. In contrast, reduced temperatures consistently increase, indicating that $p$-wave inelastic collisions invariably drive the system away from quantum degeneracy and ultimately into the high-temperature regime.

Returning to particle-number dynamics, dashed lines in Fig.~\ref{fig_homo}(a) show the thermal-ansatz predictions. For systems near or above quantum degeneracy ($\Bar{T}=1$ and $2$), $\mathcal{N}$ is approximately the same value at the high-temperature limit $7/3$, and we also observe that the solution from Eq.~(\ref{homo_noneq_sol}) agrees well with the thermal ansatz. However, in deeply degenerate systems ($\Bar{T}=0.1$), $\mathcal{N}$ is approximately 2.1, and a notable discrepancy in the number of particles emerges in the long-time tail. 
Correspondingly, we present the evolution of the radial momentum distribution $N(\Bar{k})=4\pi\Bar{k}^2n_\mathbf{k}V/(2\pi)^3$ in Fig.~\ref{fig_homo}(d) and (e) for this two distinct cases: $\Bar{T}=1$ and $\Bar{T}=0.1$. At $\Bar{T}=1$, the two are virtually indistinguishable, underscoring the close approximation to thermalization in near-degenerate conditions. Conversely, a discernible difference emerges between the two profiles at the deeply degenerate temperature $\Bar{T}=0.1$. 

One may wonder whether the ``quasi-thermal profile'' strategy in harmonically trapped systems can also be applied to this case. Interestingly, in homogeneous systems, particularly those starting from low initial temperatures, this approach fails to provide an excellent fit. Our analysis shows that the long-time asymptote of the radial momentum distribution in homogeneous systems follows $N(\Bar{k},t\rightarrow\infty)\propto \exp(-\alpha \Bar{k}^2)\Bar{k}^{19/2}$ in the zero initial temperature limit (see our companion paper~\cite{SM}), where $\alpha$ is a constant. In contrast, the quasi-thermal profile can only fit $\exp(-\alpha \Bar{k}^2)\Bar{k}^{2}$, further highlighting the distinct behavior between homogeneous and harmonically trapped systems. 

\textit{Conclusion.---} To summarize, we derived the inelastic quantum Boltzmann equation for single-component fermi gases with two-body loss and theoretically studied the long-time dynamics, including the particle number, temperature, and momentum distribution evolution, and its thermalization. We demonstrate that $p$-wave inelastic collisions tend to evolve the system under thermal profiles at high temperatures. In contrast, at low initial temperatures, ``quasi-thermal'' fits capture the non-equilibrium behavior for trapped systems but not homogeneous ones. Our first-principle calculations explain experiments in traps without fitting parameters and provide physical insights into the ``anti-evaporation'' phenomena. Our prediction that the low-temperature lossy system could be non-thermal could be further verified using atoms such as Li and K loaded in optical boxes, where the two-body loss could be induced using optical Feshbach resonance or dipolar relaxation~\cite{regal2003tuning,zhang2004pwave,schunck2005feshbach,gunter2005pwave,inada2008collisional,fuchs2008binding,maier2010radiofrequency,gerken2019observationa,kiefer2023ultracold,venu2023unitary}. Besides, our work serves as valuable benchmarks for calibrating relevant Direct Simulation Monte Carlo simulations~\cite{wang2024simulations} and contact measurements using two-body loss in $p$-wave BCS-BEC crossover studies in the future. 

\begin{acknowledgments}
We would like to acknowledge financial support from the National Natural Science Foundation of China under Grant No. 12204395, Hong Kong RGC Early Career Scheme (Grant No. 24308323) and Collaborative Research Fund (Grant No. C4050-23GF), the Space Application System of China Manned Space Program, and CUHK Direct Grant No. 4053676. We thank Zhigang Wu, Doerte Blume, and Kaiyuen Lee for their helpful discussions.
\end{acknowledgments}


\begin{thebibliography}{58}%
    \makeatletter
    \providecommand \@ifxundefined [1]{%
     \@ifx{#1\undefined}
    }%
    \providecommand \@ifnum [1]{%
     \ifnum #1\expandafter \@firstoftwo
     \else \expandafter \@secondoftwo
     \fi
    }%
    \providecommand \@ifx [1]{%
     \ifx #1\expandafter \@firstoftwo
     \else \expandafter \@secondoftwo
     \fi
    }%
    \providecommand \natexlab [1]{#1}%
    \providecommand \enquote  [1]{#1}%
    \providecommand \bibnamefont  [1]{#1}%
    \providecommand \bibfnamefont [1]{#1}%
    \providecommand \citenamefont [1]{#1}%
    \providecommand \href@noop [0]{\@secondoftwo}%
    \providecommand \href [0]{\begingroup \@sanitize@url \@href}%
    \providecommand \@href[1]{\@@startlink{#1}\@@href}%
    \providecommand \@@href[1]{\endgroup#1\@@endlink}%
    \providecommand \@sanitize@url [0]{\catcode `\\12\catcode `\$12\catcode
      `\&12\catcode `\#12\catcode `\^12\catcode `\_12\catcode `\%12\relax}%
    \providecommand \@@startlink[1]{}%
    \providecommand \@@endlink[0]{}%
    \providecommand \url  [0]{\begingroup\@sanitize@url \@url }%
    \providecommand \@url [1]{\endgroup\@href {#1}{\urlprefix }}%
    \providecommand \urlprefix  [0]{URL }%
    \providecommand \Eprint [0]{\href }%
    \providecommand \doibase [0]{https://dx.doi.org}%
    \providecommand \selectlanguage [0]{\@gobble}%
    \providecommand \bibinfo  [0]{\@secondoftwo}%
    \providecommand \bibfield  [0]{\@secondoftwo}%
    \providecommand \translation [1]{[#1]}%
    \providecommand \BibitemOpen [0]{}%
    \providecommand \bibitemStop [0]{}%
    \providecommand \bibitemNoStop [0]{.\EOS\space}%
    \providecommand \EOS [0]{\spacefactor3000\relax}%
    \providecommand \BibitemShut  [1]{\csname bibitem#1\endcsname}%
    \let\auto@bib@innerbib\@empty
    \bibitem [{\citenamefont {Sela}\ and\ \citenamefont
      {Goldhirsch}(1998)}]{sela1998hydrodynamic}%
      \BibitemOpen
      \bibfield  {author} {\bibinfo {author} {\bibfnamefont {N.}~\bibnamefont
      {Sela}}\ and\ \bibinfo {author} {\bibfnamefont {I.}~\bibnamefont
      {Goldhirsch}},\ }\bibfield  {title} {\bibinfo {title} {{Hydrodynamic
      Equations for Rapid Flows of Smooth Inelastic Spheres, to Burnett Order}},\
      }\href {\doibase10.1017/S0022112098008660} {\bibfield  {journal} {\bibinfo
      {journal} {J. Fluid Mech.}\ }\textbf {\bibinfo {volume} {361}},\ \bibinfo
      {pages} {41} (\bibinfo {year} {1998})}\BibitemShut {NoStop}%
    \bibitem [{\citenamefont {Garz{\'o}}\ and\ \citenamefont
      {Dufty}(1999)}]{garzo1999dense}%
      \BibitemOpen
      \bibfield  {author} {\bibinfo {author} {\bibfnamefont {V.}~\bibnamefont
      {Garz{\'o}}}\ and\ \bibinfo {author} {\bibfnamefont {J.~W.}\ \bibnamefont
      {Dufty}},\ }\bibfield  {title} {\bibinfo {title} {Dense fluid transport for
      inelastic hard spheres},\ }\href {\doibase10.1103/PhysRevE.59.5895}
      {\bibfield  {journal} {\bibinfo  {journal} {Phys. Rev. E}\ }\textbf {\bibinfo
      {volume} {59}},\ \bibinfo {pages} {5895} (\bibinfo {year}
      {1999})}\BibitemShut {NoStop}%
    \bibitem [{\citenamefont {{Ben-Naim}}\ \emph {et~al.}(1999)\citenamefont
      {{Ben-Naim}}, \citenamefont {Chen}, \citenamefont {Doolen},\ and\
      \citenamefont {Redner}}]{ben-naim1999shocklike}%
      \BibitemOpen
      \bibfield  {author} {\bibinfo {author} {\bibfnamefont {E.}~\bibnamefont
      {{Ben-Naim}}}, \bibinfo {author} {\bibfnamefont {S.~Y.}\ \bibnamefont
      {Chen}}, \bibinfo {author} {\bibfnamefont {G.~D.}\ \bibnamefont {Doolen}}, \
      and\ \bibinfo {author} {\bibfnamefont {S.}~\bibnamefont {Redner}},\
      }\bibfield  {title} {\bibinfo {title} {Shocklike {{Dynamics}} of {{Inelastic
      Gases}}},\ }\href {\doibase10.1103/PhysRevLett.83.4069} {\bibfield  {journal}
      {\bibinfo  {journal} {Phys. Rev. Lett.}\ }\textbf {\bibinfo {volume} {83}},\
      \bibinfo {pages} {4069} (\bibinfo {year} {1999})}\BibitemShut {NoStop}%
    \bibitem [{\citenamefont {Bobylev}\ \emph {et~al.}(2000)\citenamefont
      {Bobylev}, \citenamefont {Carrillo},\ and\ \citenamefont
      {Gamba}}]{bobylev2000properties}%
      \BibitemOpen
      \bibfield  {author} {\bibinfo {author} {\bibfnamefont {A.~V.}\ \bibnamefont
      {Bobylev}}, \bibinfo {author} {\bibfnamefont {J.~A.}\ \bibnamefont
      {Carrillo}}, \ and\ \bibinfo {author} {\bibfnamefont {I.~M.}\ \bibnamefont
      {Gamba}},\ }\bibfield  {title} {\bibinfo {title} {On {{Some Properties}} of
      {{Kinetic}} and {{Hydrodynamic Equations}} for {{Inelastic Interactions}}},\
      }\href {\doibase10.1023/A:1018627625800} {\bibfield  {journal} {\bibinfo
      {journal} {J. Stat. Phys.}\ }\textbf {\bibinfo {volume} {98}},\ \bibinfo
      {pages} {743} (\bibinfo {year} {2000})}\BibitemShut {NoStop}%
    \bibitem [{\citenamefont {{Ben-Naim}}\ and\ \citenamefont
      {Krapivsky}(2002)}]{ben-naim2002scaling}%
      \BibitemOpen
      \bibfield  {author} {\bibinfo {author} {\bibfnamefont {E.}~\bibnamefont
      {{Ben-Naim}}}\ and\ \bibinfo {author} {\bibfnamefont {P.~L.}\ \bibnamefont
      {Krapivsky}},\ }\bibfield  {title} {\bibinfo {title} {Scaling, multiscaling,
      and nontrivial exponents in inelastic collision processes},\ }\href
      {\doibase10.1103/PhysRevE.66.011309} {\bibfield  {journal} {\bibinfo
      {journal} {Phys. Rev. E}\ }\textbf {\bibinfo {volume} {66}},\ \bibinfo
      {pages} {011309} (\bibinfo {year} {2002})}\BibitemShut {NoStop}%
    \bibitem [{\citenamefont {Micheli}\ \emph {et~al.}(2006)\citenamefont
      {Micheli}, \citenamefont {Brennen},\ and\ \citenamefont
      {Zoller}}]{micheli2006toolbox}%
      \BibitemOpen
      \bibfield  {author} {\bibinfo {author} {\bibfnamefont {A.}~\bibnamefont
      {Micheli}}, \bibinfo {author} {\bibfnamefont {G.~K.}\ \bibnamefont
      {Brennen}}, \ and\ \bibinfo {author} {\bibfnamefont {P.}~\bibnamefont
      {Zoller}},\ }\bibfield  {title} {\bibinfo {title} {A toolbox for lattice-spin
      models with polar molecules},\ }\href {\doibase10.1038/nphys287} {\bibfield
      {journal} {\bibinfo  {journal} {Nat. Phys.}\ }\textbf {\bibinfo {volume}
      {2}},\ \bibinfo {pages} {341} (\bibinfo {year} {2006})}\BibitemShut {NoStop}%
    \bibitem [{\citenamefont {Carr}\ \emph {et~al.}(2009)\citenamefont {Carr},
      \citenamefont {DeMille}, \citenamefont {Krems},\ and\ \citenamefont
      {Ye}}]{carr2009cold}%
      \BibitemOpen
      \bibfield  {author} {\bibinfo {author} {\bibfnamefont {L.~D.}\ \bibnamefont
      {Carr}}, \bibinfo {author} {\bibfnamefont {D.}~\bibnamefont {DeMille}},
      \bibinfo {author} {\bibfnamefont {R.~V.}\ \bibnamefont {Krems}}, \ and\
      \bibinfo {author} {\bibfnamefont {J.}~\bibnamefont {Ye}},\ }\bibfield
      {title} {\bibinfo {title} {Cold and ultracold molecules: Science, technology
      and applications},\ }\href {\doibase10.1088/1367-2630/11/5/055049} {\bibfield
       {journal} {\bibinfo  {journal} {New J. Phys.}\ }\textbf {\bibinfo {volume}
      {11}},\ \bibinfo {pages} {055049} (\bibinfo {year} {2009})}\BibitemShut
      {NoStop}%
    \bibitem [{\citenamefont {Osterloh}\ \emph {et~al.}(2007)\citenamefont
      {Osterloh}, \citenamefont {Barberán},\ and\ \citenamefont
      {Lewenstein}}]{osterloh2007strongly}%
      \BibitemOpen
      \bibfield  {author} {\bibinfo {author} {\bibfnamefont {K.}~\bibnamefont
      {Osterloh}}, \bibinfo {author} {\bibfnamefont {N.}~\bibnamefont {Barberán}},
      \ and\ \bibinfo {author} {\bibfnamefont {M.}~\bibnamefont {Lewenstein}},\
      }\bibfield  {title} {\bibinfo {title} {Strongly {{Correlated States}} of
      {{Ultracold Rotating Dipolar Fermi Gases}}},\ }\href
      {\doibase10.1103/PhysRevLett.99.160403} {\bibfield  {journal} {\bibinfo
      {journal} {Phys. Rev. Lett.}\ }\textbf {\bibinfo {volume} {99}},\ \bibinfo
      {pages} {160403} (\bibinfo {year} {2007})}\BibitemShut {NoStop}%
    \bibitem [{\citenamefont {Büchler}\ \emph {et~al.}(2007)\citenamefont
      {Büchler}, \citenamefont {Demler}, \citenamefont {Lukin}, \citenamefont
      {Micheli}, \citenamefont {Prokof’ev}, \citenamefont {Pupillo},\ and\
      \citenamefont {Zoller}}]{buchler2007strongly}%
      \BibitemOpen
      \bibfield  {author} {\bibinfo {author} {\bibfnamefont {H.~P.}\ \bibnamefont
      {Büchler}}, \bibinfo {author} {\bibfnamefont {E.}~\bibnamefont {Demler}},
      \bibinfo {author} {\bibfnamefont {M.}~\bibnamefont {Lukin}}, \bibinfo
      {author} {\bibfnamefont {A.}~\bibnamefont {Micheli}}, \bibinfo {author}
      {\bibfnamefont {N.}~\bibnamefont {Prokof’ev}}, \bibinfo {author}
      {\bibfnamefont {G.}~\bibnamefont {Pupillo}}, \ and\ \bibinfo {author}
      {\bibfnamefont {P.}~\bibnamefont {Zoller}},\ }\bibfield  {title} {\bibinfo
      {title} {Strongly {{Correlated 2D Quantum Phases}} with {{Cold Polar
      Molecules}}: {{Controlling}} the {{Shape}} of the {{Interaction
      Potential}}},\ }\href {\doibase10.1103/PhysRevLett.98.060404} {\bibfield
      {journal} {\bibinfo  {journal} {Phys. Rev. Lett.}\ }\textbf {\bibinfo
      {volume} {98}},\ \bibinfo {pages} {060404} (\bibinfo {year}
      {2007})}\BibitemShut {NoStop}%
    \bibitem [{\citenamefont {Gorshkov}\ \emph
      {et~al.}(2011{\natexlab{a}})\citenamefont {Gorshkov}, \citenamefont
      {Manmana}, \citenamefont {Chen}, \citenamefont {Ye}, \citenamefont {Demler},
      \citenamefont {Lukin},\ and\ \citenamefont {Rey}}]{gorshkov2011tunable}%
      \BibitemOpen
      \bibfield  {author} {\bibinfo {author} {\bibfnamefont {A.~V.}\ \bibnamefont
      {Gorshkov}}, \bibinfo {author} {\bibfnamefont {S.~R.}\ \bibnamefont
      {Manmana}}, \bibinfo {author} {\bibfnamefont {G.}~\bibnamefont {Chen}},
      \bibinfo {author} {\bibfnamefont {J.}~\bibnamefont {Ye}}, \bibinfo {author}
      {\bibfnamefont {E.}~\bibnamefont {Demler}}, \bibinfo {author} {\bibfnamefont
      {M.~D.}\ \bibnamefont {Lukin}}, \ and\ \bibinfo {author} {\bibfnamefont
      {A.~M.}\ \bibnamefont {Rey}},\ }\bibfield  {title} {\bibinfo {title} {Tunable
      {{Superfluidity}} and {{Quantum Magnetism}} with {{Ultracold Polar
      Molecules}}},\ }\href {\doibase10.1103/PhysRevLett.107.115301} {\bibfield
      {journal} {\bibinfo  {journal} {Phys. Rev. Lett.}\ }\textbf {\bibinfo
      {volume} {107}},\ \bibinfo {pages} {115301} (\bibinfo {year}
      {2011}{\natexlab{a}})}\BibitemShut {NoStop}%
    \bibitem [{\citenamefont {Gorshkov}\ \emph
      {et~al.}(2011{\natexlab{b}})\citenamefont {Gorshkov}, \citenamefont
      {Manmana}, \citenamefont {Chen}, \citenamefont {Demler}, \citenamefont
      {Lukin},\ and\ \citenamefont {Rey}}]{gorshkov2011quantum}%
      \BibitemOpen
      \bibfield  {author} {\bibinfo {author} {\bibfnamefont {A.~V.}\ \bibnamefont
      {Gorshkov}}, \bibinfo {author} {\bibfnamefont {S.~R.}\ \bibnamefont
      {Manmana}}, \bibinfo {author} {\bibfnamefont {G.}~\bibnamefont {Chen}},
      \bibinfo {author} {\bibfnamefont {E.}~\bibnamefont {Demler}}, \bibinfo
      {author} {\bibfnamefont {M.~D.}\ \bibnamefont {Lukin}}, \ and\ \bibinfo
      {author} {\bibfnamefont {A.~M.}\ \bibnamefont {Rey}},\ }\bibfield  {title}
      {\bibinfo {title} {Quantum magnetism with polar alkali-metal dimers},\ }\href
      {\doibase10.1103/PhysRevA.84.033619} {\bibfield  {journal} {\bibinfo
      {journal} {Phys. Rev. A}\ }\textbf {\bibinfo {volume} {84}},\ \bibinfo
      {pages} {033619} (\bibinfo {year} {2011}{\natexlab{b}})}\BibitemShut
      {NoStop}%
    \bibitem [{\citenamefont {Baranov}\ \emph {et~al.}(2012)\citenamefont
      {Baranov}, \citenamefont {Dalmonte}, \citenamefont {Pupillo},\ and\
      \citenamefont {Zoller}}]{baranov2012condensed}%
      \BibitemOpen
      \bibfield  {author} {\bibinfo {author} {\bibfnamefont {M.~A.}\ \bibnamefont
      {Baranov}}, \bibinfo {author} {\bibfnamefont {M.}~\bibnamefont {Dalmonte}},
      \bibinfo {author} {\bibfnamefont {G.}~\bibnamefont {Pupillo}}, \ and\
      \bibinfo {author} {\bibfnamefont {P.}~\bibnamefont {Zoller}},\ }\bibfield
      {title} {\bibinfo {title} {Condensed {{Matter Theory}} of {{Dipolar Quantum
      Gases}}},\ }\href {\doibase10.1021/cr2003568} {\bibfield  {journal} {\bibinfo
       {journal} {Chem. Rev.}\ }\textbf {\bibinfo {volume} {112}},\ \bibinfo
      {pages} {5012} (\bibinfo {year} {2012})}\BibitemShut {NoStop}%
    \bibitem [{\citenamefont {Cornish}\ \emph {et~al.}(2024)\citenamefont
      {Cornish}, \citenamefont {Tarbutt},\ and\ \citenamefont
      {Hazzard}}]{cornish2024quantum}%
      \BibitemOpen
      \bibfield  {author} {\bibinfo {author} {\bibfnamefont {S.~L.}\ \bibnamefont
      {Cornish}}, \bibinfo {author} {\bibfnamefont {M.~R.}\ \bibnamefont
      {Tarbutt}}, \ and\ \bibinfo {author} {\bibfnamefont {K.~R.~A.}\ \bibnamefont
      {Hazzard}},\ }\bibfield  {title} {\bibinfo {title} {Quantum computation and
      quantum simulation with ultracold molecules},\ }\href
      {\doibase10.1038/s41567-024-02453-9} {\bibfield  {journal} {\bibinfo
      {journal} {Nat. Phys.}\ }\textbf {\bibinfo {volume} {20}},\ \bibinfo {pages}
      {730} (\bibinfo {year} {2024})}\BibitemShut {NoStop}%
    \bibitem [{\citenamefont {Zadoyan}\ \emph {et~al.}(2001)\citenamefont
      {Zadoyan}, \citenamefont {Kohen}, \citenamefont {Lidar},\ and\ \citenamefont
      {Apkarian}}]{zadoyan2001manipulation}%
      \BibitemOpen
      \bibfield  {author} {\bibinfo {author} {\bibfnamefont {R.}~\bibnamefont
      {Zadoyan}}, \bibinfo {author} {\bibfnamefont {D.}~\bibnamefont {Kohen}},
      \bibinfo {author} {\bibfnamefont {D.~A.}\ \bibnamefont {Lidar}}, \ and\
      \bibinfo {author} {\bibfnamefont {V.~A.}\ \bibnamefont {Apkarian}},\
      }\bibfield  {title} {\bibinfo {title} {The manipulation of massive
      ro-vibronic superpositions using time–frequency-resolved coherent
      anti-{{Stokes Raman}} scattering ({{TFRCARS}}): From quantum control to
      quantum computing},\ }\href {\doibase10.1016/S0301-0104(01)00270-1}
      {\bibfield  {journal} {\bibinfo  {journal} {Chem. Phys.}\ }\textbf {\bibinfo
      {volume} {266}},\ \bibinfo {pages} {323} (\bibinfo {year}
      {2001})}\BibitemShut {NoStop}%
    \bibitem [{\citenamefont {DeMille}(2002)}]{demille2002quantum}%
      \BibitemOpen
      \bibfield  {author} {\bibinfo {author} {\bibfnamefont {D.}~\bibnamefont
      {DeMille}},\ }\bibfield  {title} {\bibinfo {title} {Quantum {{Computation}}
      with {{Trapped Polar Molecules}}},\ }\href
      {\doibase10.1103/PhysRevLett.88.067901} {\bibfield  {journal} {\bibinfo
      {journal} {Phys. Rev. Lett.}\ }\textbf {\bibinfo {volume} {88}},\ \bibinfo
      {pages} {067901} (\bibinfo {year} {2002})}\BibitemShut {NoStop}%
    \bibitem [{\citenamefont {André}\ \emph {et~al.}(2006)\citenamefont {André},
      \citenamefont {DeMille}, \citenamefont {Doyle}, \citenamefont {Lukin},
      \citenamefont {Maxwell}, \citenamefont {Rabl}, \citenamefont {Schoelkopf},\
      and\ \citenamefont {Zoller}}]{andre2006coherent}%
      \BibitemOpen
      \bibfield  {author} {\bibinfo {author} {\bibfnamefont {A.}~\bibnamefont
      {André}}, \bibinfo {author} {\bibfnamefont {D.}~\bibnamefont {DeMille}},
      \bibinfo {author} {\bibfnamefont {J.~M.}\ \bibnamefont {Doyle}}, \bibinfo
      {author} {\bibfnamefont {M.~D.}\ \bibnamefont {Lukin}}, \bibinfo {author}
      {\bibfnamefont {S.~E.}\ \bibnamefont {Maxwell}}, \bibinfo {author}
      {\bibfnamefont {P.}~\bibnamefont {Rabl}}, \bibinfo {author} {\bibfnamefont
      {R.~J.}\ \bibnamefont {Schoelkopf}}, \ and\ \bibinfo {author} {\bibfnamefont
      {P.}~\bibnamefont {Zoller}},\ }\bibfield  {title} {\bibinfo {title} {A
      coherent all-electrical interface between polar molecules and mesoscopic
      superconducting resonators},\ }\href {\doibase10.1038/nphys386} {\bibfield
      {journal} {\bibinfo  {journal} {Nat. Phys.}\ }\textbf {\bibinfo {volume}
      {2}},\ \bibinfo {pages} {636} (\bibinfo {year} {2006})}\BibitemShut {NoStop}%
    \bibitem [{\citenamefont {Rabl}\ \emph {et~al.}(2006)\citenamefont {Rabl},
      \citenamefont {DeMille}, \citenamefont {Doyle}, \citenamefont {Lukin},
      \citenamefont {Schoelkopf},\ and\ \citenamefont {Zoller}}]{rabl2006hybrid}%
      \BibitemOpen
      \bibfield  {author} {\bibinfo {author} {\bibfnamefont {P.}~\bibnamefont
      {Rabl}}, \bibinfo {author} {\bibfnamefont {D.}~\bibnamefont {DeMille}},
      \bibinfo {author} {\bibfnamefont {J.~M.}\ \bibnamefont {Doyle}}, \bibinfo
      {author} {\bibfnamefont {M.~D.}\ \bibnamefont {Lukin}}, \bibinfo {author}
      {\bibfnamefont {R.~J.}\ \bibnamefont {Schoelkopf}}, \ and\ \bibinfo {author}
      {\bibfnamefont {P.}~\bibnamefont {Zoller}},\ }\bibfield  {title} {\bibinfo
      {title} {Hybrid {{Quantum Processors}}: {{Molecular Ensembles}} as {{Quantum
      Memory}} for {{Solid State Circuits}}},\ }\href
      {\doibase10.1103/PhysRevLett.97.033003} {\bibfield  {journal} {\bibinfo
      {journal} {Phys. Rev. Lett.}\ }\textbf {\bibinfo {volume} {97}},\ \bibinfo
      {pages} {033003} (\bibinfo {year} {2006})}\BibitemShut {NoStop}%
    \bibitem [{\citenamefont {Ospelkaus}\ \emph {et~al.}(2010)\citenamefont
      {Ospelkaus}, \citenamefont {Ni}, \citenamefont {Wang}, \citenamefont
      {De~Miranda}, \citenamefont {Neyenhuis}, \citenamefont {Quéméner},
      \citenamefont {Julienne}, \citenamefont {Bohn}, \citenamefont {Jin},\ and\
      \citenamefont {Ye}}]{ospelkaus2010quantumstate}%
      \BibitemOpen
      \bibfield  {author} {\bibinfo {author} {\bibfnamefont {S.}~\bibnamefont
      {Ospelkaus}}, \bibinfo {author} {\bibfnamefont {K.-K.}\ \bibnamefont {Ni}},
      \bibinfo {author} {\bibfnamefont {D.}~\bibnamefont {Wang}}, \bibinfo {author}
      {\bibfnamefont {M.~H.~G.}\ \bibnamefont {De~Miranda}}, \bibinfo {author}
      {\bibfnamefont {B.}~\bibnamefont {Neyenhuis}}, \bibinfo {author}
      {\bibfnamefont {.~G.}\ \bibnamefont {Quéméner}}, \bibinfo {author}
      {\bibfnamefont {P.~S.}\ \bibnamefont {Julienne}}, \bibinfo {author}
      {\bibfnamefont {J.~L.}\ \bibnamefont {Bohn}}, \bibinfo {author}
      {\bibfnamefont {D.~S.}\ \bibnamefont {Jin}}, \ and\ \bibinfo {author}
      {\bibfnamefont {J.}~\bibnamefont {Ye}},\ }\bibfield  {title} {\bibinfo
      {title} {Quantum-state controlled chemical reactions of ultracold
      potassium-rubidium molecules},\ }\href {\doibase10.1126/science.1184121}
      {\bibfield  {journal} {\bibinfo  {journal} {Science}\ }\textbf {\bibinfo
      {volume} {327}},\ \bibinfo {pages} {853} (\bibinfo {year}
      {2010})}\BibitemShut {NoStop}%
    \bibitem [{\citenamefont {Park}\ \emph {et~al.}(2015)\citenamefont {Park},
      \citenamefont {Will},\ and\ \citenamefont {Zwierlein}}]{park2015ultracold}%
      \BibitemOpen
      \bibfield  {author} {\bibinfo {author} {\bibfnamefont {J.~W.}\ \bibnamefont
      {Park}}, \bibinfo {author} {\bibfnamefont {S.~A.}\ \bibnamefont {Will}}, \
      and\ \bibinfo {author} {\bibfnamefont {M.~W.}\ \bibnamefont {Zwierlein}},\
      }\bibfield  {title} {\bibinfo {title} {Ultracold {{Dipolar Gas}} of
      {{Fermionic}} ${}^{23}\mathrm{Na}{}^{40}\mathrm{K}$ {{Molecules}} in {{Their
      Absolute Ground State}}},\ }\href {\doibase10.1103/PhysRevLett.114.205302}
      {\bibfield  {journal} {\bibinfo  {journal} {Phys. Rev. Lett.}\ }\textbf
      {\bibinfo {volume} {114}},\ \bibinfo {pages} {205302} (\bibinfo {year}
      {2015})}\BibitemShut {NoStop}%
    \bibitem [{\citenamefont {De~Marco}\ \emph
      {et~al.}(2019{\natexlab{a}})\citenamefont {De~Marco}, \citenamefont
      {Valtolina}, \citenamefont {Matsuda}, \citenamefont {Tobias}, \citenamefont
      {Covey},\ and\ \citenamefont {Ye}}]{demarco2019degenerate}%
      \BibitemOpen
      \bibfield  {author} {\bibinfo {author} {\bibfnamefont {L.}~\bibnamefont
      {De~Marco}}, \bibinfo {author} {\bibfnamefont {G.}~\bibnamefont {Valtolina}},
      \bibinfo {author} {\bibfnamefont {K.}~\bibnamefont {Matsuda}}, \bibinfo
      {author} {\bibfnamefont {W.~G.}\ \bibnamefont {Tobias}}, \bibinfo {author}
      {\bibfnamefont {J.~P.}\ \bibnamefont {Covey}}, \ and\ \bibinfo {author}
      {\bibfnamefont {J.}~\bibnamefont {Ye}},\ }\bibfield  {title} {\bibinfo
      {title} {A degenerate {{Fermi}} gas of polar molecules},\ }\href
      {\doibase10.1126/science.aau7230} {\bibfield  {journal} {\bibinfo  {journal}
      {Science}\ }\textbf {\bibinfo {volume} {363}},\ \bibinfo {pages} {853}
      (\bibinfo {year} {2019}{\natexlab{a}})}\BibitemShut {NoStop}%
    \bibitem [{\citenamefont {Hu}\ \emph {et~al.}(2019)\citenamefont {Hu},
      \citenamefont {Liu}, \citenamefont {Grimes}, \citenamefont {Lin},
      \citenamefont {Gheorghe}, \citenamefont {Vexiau}, \citenamefont
      {{Bouloufa-Maafa}}, \citenamefont {Dulieu}, \citenamefont {Rosenband},\ and\
      \citenamefont {Ni}}]{hu2019directa}%
      \BibitemOpen
      \bibfield  {author} {\bibinfo {author} {\bibfnamefont {M.-G.}\ \bibnamefont
      {Hu}}, \bibinfo {author} {\bibfnamefont {Y.}~\bibnamefont {Liu}}, \bibinfo
      {author} {\bibfnamefont {D.~D.}\ \bibnamefont {Grimes}}, \bibinfo {author}
      {\bibfnamefont {Y.-W.}\ \bibnamefont {Lin}}, \bibinfo {author} {\bibfnamefont
      {A.~H.}\ \bibnamefont {Gheorghe}}, \bibinfo {author} {\bibfnamefont
      {R.}~\bibnamefont {Vexiau}}, \bibinfo {author} {\bibfnamefont
      {N.}~\bibnamefont {{Bouloufa-Maafa}}}, \bibinfo {author} {\bibfnamefont
      {O.}~\bibnamefont {Dulieu}}, \bibinfo {author} {\bibfnamefont
      {T.}~\bibnamefont {Rosenband}}, \ and\ \bibinfo {author} {\bibfnamefont
      {K.-K.}\ \bibnamefont {Ni}},\ }\bibfield  {title} {\bibinfo {title} {Direct
      observation of bimolecular reactions of ultracold {{KRb}} molecules},\ }\href
      {\doibase10.1126/science.aay9531} {\bibfield  {journal} {\bibinfo  {journal}
      {Science}\ }\textbf {\bibinfo {volume} {366}},\ \bibinfo {pages} {1111}
      (\bibinfo {year} {2019})}\BibitemShut {NoStop}%
    \bibitem [{\citenamefont {Schindewolf}\ \emph {et~al.}(2022)\citenamefont
      {Schindewolf}, \citenamefont {Bause}, \citenamefont {Chen}, \citenamefont
      {Duda}, \citenamefont {Karman}, \citenamefont {Bloch},\ and\ \citenamefont
      {Luo}}]{schindewolf2022evaporation}%
      \BibitemOpen
      \bibfield  {author} {\bibinfo {author} {\bibfnamefont {A.}~\bibnamefont
      {Schindewolf}}, \bibinfo {author} {\bibfnamefont {R.}~\bibnamefont {Bause}},
      \bibinfo {author} {\bibfnamefont {X.-Y.}\ \bibnamefont {Chen}}, \bibinfo
      {author} {\bibfnamefont {M.}~\bibnamefont {Duda}}, \bibinfo {author}
      {\bibfnamefont {T.}~\bibnamefont {Karman}}, \bibinfo {author} {\bibfnamefont
      {I.}~\bibnamefont {Bloch}}, \ and\ \bibinfo {author} {\bibfnamefont {X.-Y.}\
      \bibnamefont {Luo}},\ }\bibfield  {title} {\bibinfo {title} {Evaporation of
      microwave-shielded polar molecules to quantum degeneracy},\ }\href
      {\doibase10.1038/s41586-022-04900-0} {\bibfield  {journal} {\bibinfo
      {journal} {Nature}\ }\textbf {\bibinfo {volume} {607}},\ \bibinfo {pages}
      {677} (\bibinfo {year} {2022})}\BibitemShut {NoStop}%
    \bibitem [{\citenamefont {Duda}\ \emph {et~al.}(2023)\citenamefont {Duda},
      \citenamefont {Chen}, \citenamefont {Bause}, \citenamefont {Schindewolf},
      \citenamefont {Bloch},\ and\ \citenamefont {Luo}}]{duda2023longlived}%
      \BibitemOpen
      \bibfield  {author} {\bibinfo {author} {\bibfnamefont {M.}~\bibnamefont
      {Duda}}, \bibinfo {author} {\bibfnamefont {X.-Y.}\ \bibnamefont {Chen}},
      \bibinfo {author} {\bibfnamefont {R.}~\bibnamefont {Bause}}, \bibinfo
      {author} {\bibfnamefont {A.}~\bibnamefont {Schindewolf}}, \bibinfo {author}
      {\bibfnamefont {I.}~\bibnamefont {Bloch}}, \ and\ \bibinfo {author}
      {\bibfnamefont {X.-Y.}\ \bibnamefont {Luo}},\ }\bibfield  {title} {\bibinfo
      {title} {{Long-Lived Fermionic Feshbach Molecules with Tunable $p$-Wave
      Interactions}},\ }\href {\doibase10.1103/PhysRevA.107.053322} {\bibfield
      {journal} {\bibinfo  {journal} {Phys. Rev. A}\ }\textbf {\bibinfo {volume}
      {107}},\ \bibinfo {pages} {053322} (\bibinfo {year} {2023})}\BibitemShut
      {NoStop}%
    \bibitem [{\citenamefont {Molony}\ \emph {et~al.}(2014)\citenamefont {Molony},
      \citenamefont {Gregory}, \citenamefont {Ji}, \citenamefont {Lu},
      \citenamefont {K{\"o}ppinger}, \citenamefont {Le~Sueur}, \citenamefont
      {Blackley}, \citenamefont {Hutson},\ and\ \citenamefont
      {Cornish}}]{molony2014creation}%
      \BibitemOpen
      \bibfield  {author} {\bibinfo {author} {\bibfnamefont {P.~K.}\ \bibnamefont
      {Molony}}, \bibinfo {author} {\bibfnamefont {P.~D.}\ \bibnamefont {Gregory}},
      \bibinfo {author} {\bibfnamefont {Z.}~\bibnamefont {Ji}}, \bibinfo {author}
      {\bibfnamefont {B.}~\bibnamefont {Lu}}, \bibinfo {author} {\bibfnamefont
      {M.~P.}\ \bibnamefont {K{\"o}ppinger}}, \bibinfo {author} {\bibfnamefont
      {C.~R.}\ \bibnamefont {Le~Sueur}}, \bibinfo {author} {\bibfnamefont {C.~L.}\
      \bibnamefont {Blackley}}, \bibinfo {author} {\bibfnamefont {J.~M.}\
      \bibnamefont {Hutson}}, \ and\ \bibinfo {author} {\bibfnamefont {S.~L.}\
      \bibnamefont {Cornish}},\ }\bibfield  {title} {\bibinfo {title} {Creation of
      {{Ultracold}} ${}^{87}\mathrm{Rb}{}^{133}\mathrm{Cs}$ {{Molecules}} in the
      {{Rovibrational Ground State}}},\ }\href
      {\doibase10.1103/PhysRevLett.113.255301} {\bibfield  {journal} {\bibinfo
      {journal} {Phys. Rev. Lett.}\ }\textbf {\bibinfo {volume} {113}},\ \bibinfo
      {pages} {255301} (\bibinfo {year} {2014})}\BibitemShut {NoStop}%
    \bibitem [{\citenamefont {Takekoshi}\ \emph {et~al.}(2014)\citenamefont
      {Takekoshi}, \citenamefont {Reichs{\"o}llner}, \citenamefont {Schindewolf},
      \citenamefont {Hutson}, \citenamefont {Le~Sueur}, \citenamefont {Dulieu},
      \citenamefont {Ferlaino}, \citenamefont {Grimm},\ and\ \citenamefont
      {N{\"a}gerl}}]{takekoshi2014ultracold}%
      \BibitemOpen
      \bibfield  {author} {\bibinfo {author} {\bibfnamefont {T.}~\bibnamefont
      {Takekoshi}}, \bibinfo {author} {\bibfnamefont {L.}~\bibnamefont
      {Reichs{\"o}llner}}, \bibinfo {author} {\bibfnamefont {A.}~\bibnamefont
      {Schindewolf}}, \bibinfo {author} {\bibfnamefont {J.~M.}\ \bibnamefont
      {Hutson}}, \bibinfo {author} {\bibfnamefont {C.~R.}\ \bibnamefont
      {Le~Sueur}}, \bibinfo {author} {\bibfnamefont {O.}~\bibnamefont {Dulieu}},
      \bibinfo {author} {\bibfnamefont {F.}~\bibnamefont {Ferlaino}}, \bibinfo
      {author} {\bibfnamefont {R.}~\bibnamefont {Grimm}}, \ and\ \bibinfo {author}
      {\bibfnamefont {H.-C.}\ \bibnamefont {N{\"a}gerl}},\ }\bibfield  {title}
      {\bibinfo {title} {Ultracold {{Dense Samples}} of {{Dipolar RbCs Molecules}}
      in the {{Rovibrational}} and {{Hyperfine Ground State}}},\ }\href
      {\doibase10.1103/PhysRevLett.113.205301} {\bibfield  {journal} {\bibinfo
      {journal} {Phys. Rev. Lett.}\ }\textbf {\bibinfo {volume} {113}},\ \bibinfo
      {pages} {205301} (\bibinfo {year} {2014})}\BibitemShut {NoStop}%
    \bibitem [{\citenamefont {Ye}\ \emph {et~al.}(2018)\citenamefont {Ye},
      \citenamefont {Guo}, \citenamefont {{González-Martínez}}, \citenamefont
      {Quéméner},\ and\ \citenamefont {Wang}}]{ye2018collisionsa}%
      \BibitemOpen
      \bibfield  {author} {\bibinfo {author} {\bibfnamefont {X.}~\bibnamefont
      {Ye}}, \bibinfo {author} {\bibfnamefont {M.}~\bibnamefont {Guo}}, \bibinfo
      {author} {\bibfnamefont {M.~L.}\ \bibnamefont {{González-Martínez}}},
      \bibinfo {author} {\bibfnamefont {G.}~\bibnamefont {Quéméner}}, \ and\
      \bibinfo {author} {\bibfnamefont {D.}~\bibnamefont {Wang}},\ }\bibfield
      {title} {\bibinfo {title} {Collisions of ultracold
      $^{23}\mathrm{Na}^{87}\mathrm{Rb}$ molecules with controlled chemical
      reactivities},\ }\href {\doibase10.1126/sciadv.aaq0083} {\bibfield  {journal}
      {\bibinfo  {journal} {Sci. Adv.}\ }\textbf {\bibinfo {volume} {4}},\ \bibinfo
      {pages} {eaaq0083} (\bibinfo {year} {2018})}\BibitemShut {NoStop}%
    \bibitem [{\citenamefont {Gregory}\ \emph {et~al.}(2019)\citenamefont
      {Gregory}, \citenamefont {Frye}, \citenamefont {Blackmore}, \citenamefont
      {Bridge}, \citenamefont {Sawant}, \citenamefont {Hutson},\ and\ \citenamefont
      {Cornish}}]{gregory2019stickya}%
      \BibitemOpen
      \bibfield  {author} {\bibinfo {author} {\bibfnamefont {P.~D.}\ \bibnamefont
      {Gregory}}, \bibinfo {author} {\bibfnamefont {M.~D.}\ \bibnamefont {Frye}},
      \bibinfo {author} {\bibfnamefont {J.~A.}\ \bibnamefont {Blackmore}}, \bibinfo
      {author} {\bibfnamefont {E.~M.}\ \bibnamefont {Bridge}}, \bibinfo {author}
      {\bibfnamefont {R.}~\bibnamefont {Sawant}}, \bibinfo {author} {\bibfnamefont
      {J.~M.}\ \bibnamefont {Hutson}}, \ and\ \bibinfo {author} {\bibfnamefont
      {S.~L.}\ \bibnamefont {Cornish}},\ }\bibfield  {title} {\bibinfo {title}
      {Sticky collisions of ultracold {{RbCs}} molecules},\ }\href
      {\doibase10.1038/s41467-019-11033-y} {\bibfield  {journal} {\bibinfo
      {journal} {Nat. Commun.}\ }\textbf {\bibinfo {volume} {10}},\ \bibinfo
      {pages} {3104} (\bibinfo {year} {2019})}\BibitemShut {NoStop}%
    \bibitem [{\citenamefont {Gersema}\ \emph {et~al.}(2021)\citenamefont
      {Gersema}, \citenamefont {Voges}, \citenamefont {{Meyer zum Alten Borgloh}},
      \citenamefont {Koch}, \citenamefont {Hartmann}, \citenamefont {Zenesini},
      \citenamefont {Ospelkaus}, \citenamefont {Lin}, \citenamefont {He},\ and\
      \citenamefont {Wang}}]{gersema2021probing}%
      \BibitemOpen
      \bibfield  {author} {\bibinfo {author} {\bibfnamefont {P.}~\bibnamefont
      {Gersema}}, \bibinfo {author} {\bibfnamefont {K.~K.}\ \bibnamefont {Voges}},
      \bibinfo {author} {\bibfnamefont {M.}~\bibnamefont {{Meyer zum Alten
      Borgloh}}}, \bibinfo {author} {\bibfnamefont {L.}~\bibnamefont {Koch}},
      \bibinfo {author} {\bibfnamefont {T.}~\bibnamefont {Hartmann}}, \bibinfo
      {author} {\bibfnamefont {A.}~\bibnamefont {Zenesini}}, \bibinfo {author}
      {\bibfnamefont {S.}~\bibnamefont {Ospelkaus}}, \bibinfo {author}
      {\bibfnamefont {J.}~\bibnamefont {Lin}}, \bibinfo {author} {\bibfnamefont
      {J.}~\bibnamefont {He}}, \ and\ \bibinfo {author} {\bibfnamefont
      {D.}~\bibnamefont {Wang}},\ }\bibfield  {title} {\bibinfo {title} {Probing
      photoinduced two-body loss of ultracold nonreactive bosonic
      $^{23}\mathrm{Na}^{87}\mathrm{Rb}$ and $^{23}\mathrm{Na}^{39}\mathrm{K}$
      molecules},\ }\href {\doibase10.1103/PhysRevLett.127.163401} {\bibfield
      {journal} {\bibinfo  {journal} {Phys. Rev. Lett.}\ }\textbf {\bibinfo
      {volume} {127}},\ \bibinfo {pages} {163401} (\bibinfo {year}
      {2021})}\BibitemShut {NoStop}%
    \bibitem [{\citenamefont {Bigagli}\ \emph {et~al.}(2024)\citenamefont
      {Bigagli}, \citenamefont {Yuan}, \citenamefont {Zhang}, \citenamefont
      {Bulatovic}, \citenamefont {Karman}, \citenamefont {Stevenson},\ and\
      \citenamefont {Will}}]{bigagli2024observation}%
      \BibitemOpen
      \bibfield  {author} {\bibinfo {author} {\bibfnamefont {N.}~\bibnamefont
      {Bigagli}}, \bibinfo {author} {\bibfnamefont {W.}~\bibnamefont {Yuan}},
      \bibinfo {author} {\bibfnamefont {S.}~\bibnamefont {Zhang}}, \bibinfo
      {author} {\bibfnamefont {B.}~\bibnamefont {Bulatovic}}, \bibinfo {author}
      {\bibfnamefont {T.}~\bibnamefont {Karman}}, \bibinfo {author} {\bibfnamefont
      {I.}~\bibnamefont {Stevenson}}, \ and\ \bibinfo {author} {\bibfnamefont
      {S.}~\bibnamefont {Will}},\ }\bibfield  {title} {\bibinfo {title}
      {{Observation of {{Bose}}–{{Einstein}} Condensation of Dipolar
      Molecules}},\ }\href {\doibase10.1038/s41586-024-07492-z} {\bibfield
      {journal} {\bibinfo  {journal} {Nature (London)}\ ,\ \bibinfo {pages} {1}}
      (\bibinfo {year} {2024})}\BibitemShut {NoStop}%
    \bibitem [{\citenamefont {Ni}\ \emph {et~al.}(2010)\citenamefont {Ni},
      \citenamefont {Ospelkaus}, \citenamefont {Wang}, \citenamefont
      {Qu{\'e}m{\'e}ner}, \citenamefont {Neyenhuis}, \citenamefont {{de Miranda}},
      \citenamefont {Bohn}, \citenamefont {Ye},\ and\ \citenamefont
      {Jin}}]{ni2010dipolar}%
      \BibitemOpen
      \bibfield  {author} {\bibinfo {author} {\bibfnamefont {K.-K.}\ \bibnamefont
      {Ni}}, \bibinfo {author} {\bibfnamefont {S.}~\bibnamefont {Ospelkaus}},
      \bibinfo {author} {\bibfnamefont {D.}~\bibnamefont {Wang}}, \bibinfo {author}
      {\bibfnamefont {G.}~\bibnamefont {Qu{\'e}m{\'e}ner}}, \bibinfo {author}
      {\bibfnamefont {B.}~\bibnamefont {Neyenhuis}}, \bibinfo {author}
      {\bibfnamefont {M.~H.~G.}\ \bibnamefont {{de Miranda}}}, \bibinfo {author}
      {\bibfnamefont {J.~L.}\ \bibnamefont {Bohn}}, \bibinfo {author}
      {\bibfnamefont {J.}~\bibnamefont {Ye}}, \ and\ \bibinfo {author}
      {\bibfnamefont {D.~S.}\ \bibnamefont {Jin}},\ }\bibfield  {title} {\bibinfo
      {title} {Dipolar collisions of polar molecules in the quantum regime},\
      }\href {\doibase10.1038/nature08953} {\bibfield  {journal} {\bibinfo
      {journal} {Nature (London)}\ }\textbf {\bibinfo {volume} {464}},\ \bibinfo
      {pages} {1324} (\bibinfo {year} {2010})}\BibitemShut {NoStop}%
    \bibitem [{\citenamefont {Zhu}\ \emph {et~al.}(2013)\citenamefont {Zhu},
      \citenamefont {Qu{\'e}m{\'e}ner}, \citenamefont {Rey},\ and\ \citenamefont
      {Holland}}]{zhu2013evaporative}%
      \BibitemOpen
      \bibfield  {author} {\bibinfo {author} {\bibfnamefont {B.}~\bibnamefont
      {Zhu}}, \bibinfo {author} {\bibfnamefont {G.}~\bibnamefont
      {Qu{\'e}m{\'e}ner}}, \bibinfo {author} {\bibfnamefont {A.~M.}\ \bibnamefont
      {Rey}}, \ and\ \bibinfo {author} {\bibfnamefont {M.~J.}\ \bibnamefont
      {Holland}},\ }\bibfield  {title} {\bibinfo {title} {Evaporative cooling of
      reactive polar molecules confined in a two-dimensional geometry},\ }\href
      {\doibase10.1103/PhysRevA.88.063405} {\bibfield  {journal} {\bibinfo
      {journal} {Phys. Rev. A}\ }\textbf {\bibinfo {volume} {88}},\ \bibinfo
      {pages} {063405} (\bibinfo {year} {2013})}\BibitemShut {NoStop}%
    \bibitem [{\citenamefont {S{\"o}ding}\ \emph {et~al.}(1998)\citenamefont
      {S{\"o}ding}, \citenamefont {{Gu{\'e}ry-Odelin}}, \citenamefont {Desbiolles},
      \citenamefont {Ferrari},\ and\ \citenamefont {Dalibard}}]{soding1998giant}%
      \BibitemOpen
      \bibfield  {author} {\bibinfo {author} {\bibfnamefont {J.}~\bibnamefont
      {S{\"o}ding}}, \bibinfo {author} {\bibfnamefont {D.}~\bibnamefont
      {{Gu{\'e}ry-Odelin}}}, \bibinfo {author} {\bibfnamefont {P.}~\bibnamefont
      {Desbiolles}}, \bibinfo {author} {\bibfnamefont {G.}~\bibnamefont {Ferrari}},
      \ and\ \bibinfo {author} {\bibfnamefont {J.}~\bibnamefont {Dalibard}},\
      }\bibfield  {title} {\bibinfo {title} {Giant {{Spin Relaxation}} of an
      {{Ultracold Cesium Gas}}},\ }\href {\doibase10.1103/PhysRevLett.80.1869}
      {\bibfield  {journal} {\bibinfo  {journal} {Phys. Rev. Lett.}\ }\textbf
      {\bibinfo {volume} {80}},\ \bibinfo {pages} {1869} (\bibinfo {year}
      {1998})}\BibitemShut {NoStop}%
    \bibitem [{\citenamefont {Weber}\ \emph {et~al.}(2003)\citenamefont {Weber},
      \citenamefont {Herbig}, \citenamefont {Mark}, \citenamefont {Nägerl},\ and\
      \citenamefont {Grimm}}]{weber2003threebodya}%
      \BibitemOpen
      \bibfield  {author} {\bibinfo {author} {\bibfnamefont {T.}~\bibnamefont
      {Weber}}, \bibinfo {author} {\bibfnamefont {J.}~\bibnamefont {Herbig}},
      \bibinfo {author} {\bibfnamefont {M.}~\bibnamefont {Mark}}, \bibinfo {author}
      {\bibfnamefont {H.-C.}\ \bibnamefont {Nägerl}}, \ and\ \bibinfo {author}
      {\bibfnamefont {R.}~\bibnamefont {Grimm}},\ }\bibfield  {title} {\bibinfo
      {title} {Three-{{Body Recombination}} at {{Large Scattering Lengths}} in an
      {{Ultracold Atomic Gas}}},\ }\href {\doibase10.1103/PhysRevLett.91.123201}
      {\bibfield  {journal} {\bibinfo  {journal} {Phys. Rev. Lett.}\ }\textbf
      {\bibinfo {volume} {91}},\ \bibinfo {pages} {123201} (\bibinfo {year}
      {2003})}\BibitemShut {NoStop}%
    \bibitem [{\citenamefont {Guo}\ \emph {et~al.}(2018)\citenamefont {Guo},
      \citenamefont {Ye}, \citenamefont {He}, \citenamefont
      {{González-Martínez}}, \citenamefont {Vexiau}, \citenamefont {Quéméner},\
      and\ \citenamefont {Wang}}]{guo2018dipolar}%
      \BibitemOpen
      \bibfield  {author} {\bibinfo {author} {\bibfnamefont {M.}~\bibnamefont
      {Guo}}, \bibinfo {author} {\bibfnamefont {X.}~\bibnamefont {Ye}}, \bibinfo
      {author} {\bibfnamefont {J.}~\bibnamefont {He}}, \bibinfo {author}
      {\bibfnamefont {M.~L.}\ \bibnamefont {{González-Martínez}}}, \bibinfo
      {author} {\bibfnamefont {R.}~\bibnamefont {Vexiau}}, \bibinfo {author}
      {\bibfnamefont {G.}~\bibnamefont {Quéméner}}, \ and\ \bibinfo {author}
      {\bibfnamefont {D.}~\bibnamefont {Wang}},\ }\bibfield  {title} {\bibinfo
      {title} {Dipolar {{Collisions}} of {{Ultracold Ground-State Bosonic
      Molecules}}},\ }\href {\doibase10.1103/PhysRevX.8.041044} {\bibfield
      {journal} {\bibinfo  {journal} {Phys. Rev. X}\ }\textbf {\bibinfo {volume}
      {8}},\ \bibinfo {pages} {041044} (\bibinfo {year} {2018})}\BibitemShut
      {NoStop}%
    \bibitem [{\citenamefont {Horvath}\ \emph {et~al.}(2024)\citenamefont
      {Horvath}, \citenamefont {Dhar}, \citenamefont {Das}, \citenamefont {Frye},
      \citenamefont {Guo}, \citenamefont {Hutson}, \citenamefont {Landini},\ and\
      \citenamefont {Nägerl}}]{horvath2024boseeinstein}%
      \BibitemOpen
      \bibfield  {author} {\bibinfo {author} {\bibfnamefont {M.}~\bibnamefont
      {Horvath}}, \bibinfo {author} {\bibfnamefont {S.}~\bibnamefont {Dhar}},
      \bibinfo {author} {\bibfnamefont {A.}~\bibnamefont {Das}}, \bibinfo {author}
      {\bibfnamefont {M.~D.}\ \bibnamefont {Frye}}, \bibinfo {author}
      {\bibfnamefont {Y.}~\bibnamefont {Guo}}, \bibinfo {author} {\bibfnamefont
      {J.~M.}\ \bibnamefont {Hutson}}, \bibinfo {author} {\bibfnamefont
      {M.}~\bibnamefont {Landini}}, \ and\ \bibinfo {author} {\bibfnamefont
      {H.-C.}\ \bibnamefont {Nägerl}},\ }\bibfield  {title} {\bibinfo {title}
      {Bose-{{Einstein}} condensation of non-ground-state caesium atoms},\ }\href
      {\doibase10.1038/s41467-024-47760-0} {\bibfield  {journal} {\bibinfo
      {journal} {Nat Commun}\ }\textbf {\bibinfo {volume} {15}},\ \bibinfo {pages}
      {3739} (\bibinfo {year} {2024})}\BibitemShut {NoStop}%
    \bibitem [{\citenamefont {Tobias}\ \emph {et~al.}(2020)\citenamefont {Tobias},
      \citenamefont {Matsuda}, \citenamefont {Valtolina}, \citenamefont {De~Marco},
      \citenamefont {Li},\ and\ \citenamefont {Ye}}]{tobias2020thermalization}%
      \BibitemOpen
      \bibfield  {author} {\bibinfo {author} {\bibfnamefont {W.~G.}\ \bibnamefont
      {Tobias}}, \bibinfo {author} {\bibfnamefont {K.}~\bibnamefont {Matsuda}},
      \bibinfo {author} {\bibfnamefont {G.}~\bibnamefont {Valtolina}}, \bibinfo
      {author} {\bibfnamefont {L.}~\bibnamefont {De~Marco}}, \bibinfo {author}
      {\bibfnamefont {J.-R.}\ \bibnamefont {Li}}, \ and\ \bibinfo {author}
      {\bibfnamefont {J.}~\bibnamefont {Ye}},\ }\bibfield  {title} {\bibinfo
      {title} {Thermalization and sub-{{Poissonian}} density fluctuations in a
      degenerate molecular {{Fermi}} gas},\ }\href
      {\doibase10.1103/PhysRevLett.124.033401} {\bibfield  {journal} {\bibinfo
      {journal} {Phys. Rev. Lett.}\ }\textbf {\bibinfo {volume} {124}},\ \bibinfo
      {pages} {033401} (\bibinfo {year} {2020})}\BibitemShut {NoStop}%
    \bibitem [{\citenamefont {Braaten}\ and\ \citenamefont
      {Hammer}(2013)}]{braaten2013universal}%
      \BibitemOpen
      \bibfield  {author} {\bibinfo {author} {\bibfnamefont {E.}~\bibnamefont
      {Braaten}}\ and\ \bibinfo {author} {\bibfnamefont {H.~W.}\ \bibnamefont
      {Hammer}},\ }\bibfield  {title} {\bibinfo {title} {Universal relation for the
      inelastic two-body loss rate},\ }\href
      {\doibase10.1088/0953-4075/46/21/215203} {\bibfield  {journal} {\bibinfo
      {journal} {J. Phys. B: At. Mol. Opt. Phys.}\ }\textbf {\bibinfo {volume}
      {46}},\ \bibinfo {pages} {215203} (\bibinfo {year} {2013})}\BibitemShut
      {NoStop}%
    \bibitem [{\citenamefont {He}\ \emph {et~al.}(2020{\natexlab{a}})\citenamefont
      {He}, \citenamefont {Lv}, \citenamefont {Lin},\ and\ \citenamefont
      {Zhou}}]{he2020universal}%
      \BibitemOpen
      \bibfield  {author} {\bibinfo {author} {\bibfnamefont {M.}~\bibnamefont
      {He}}, \bibinfo {author} {\bibfnamefont {C.}~\bibnamefont {Lv}}, \bibinfo
      {author} {\bibfnamefont {H.-Q.}\ \bibnamefont {Lin}}, \ and\ \bibinfo
      {author} {\bibfnamefont {Q.}~\bibnamefont {Zhou}},\ }\bibfield  {title}
      {\bibinfo {title} {Universal relations for ultracold reactive molecules},\
      }\href {\doibase10.1126/sciadv.abd4699} {\bibfield  {journal} {\bibinfo
      {journal} {Sci. Adv.}\ }\textbf {\bibinfo {volume} {6}},\ \bibinfo {pages}
      {eabd4699} (\bibinfo {year} {2020}{\natexlab{a}})}\BibitemShut {NoStop}%
    \bibitem [{\citenamefont {Pan}\ \emph {et~al.}(2020)\citenamefont {Pan},
      \citenamefont {Chen}, \citenamefont {Chen},\ and\ \citenamefont
      {Zhai}}]{pan2020nonhermitian}%
      \BibitemOpen
      \bibfield  {author} {\bibinfo {author} {\bibfnamefont {L.}~\bibnamefont
      {Pan}}, \bibinfo {author} {\bibfnamefont {X.}~\bibnamefont {Chen}}, \bibinfo
      {author} {\bibfnamefont {Y.}~\bibnamefont {Chen}}, \ and\ \bibinfo {author}
      {\bibfnamefont {H.}~\bibnamefont {Zhai}},\ }\bibfield  {title} {\bibinfo
      {title} {Non-{{Hermitian}} linear response theory},\ }\href
      {\doibase10.1038/s41567-020-0889-6} {\bibfield  {journal} {\bibinfo
      {journal} {Nat. Phys.}\ }\textbf {\bibinfo {volume} {16}},\ \bibinfo {pages}
      {767} (\bibinfo {year} {2020})}\BibitemShut {NoStop}%
    \bibitem [{\citenamefont {Gao}\ \emph {et~al.}(2023)\citenamefont {Gao},
      \citenamefont {Blume},\ and\ \citenamefont
      {Yan}}]{gao2023temperaturedependent}%
      \BibitemOpen
      \bibfield  {author} {\bibinfo {author} {\bibfnamefont {X.-Y.}\ \bibnamefont
      {Gao}}, \bibinfo {author} {\bibfnamefont {D.}~\bibnamefont {Blume}}, \ and\
      \bibinfo {author} {\bibfnamefont {Y.}~\bibnamefont {Yan}},\ }\bibfield
      {title} {\bibinfo {title} {Temperature-{{Dependent Contact}} of {{Weakly
      Interacting Single-Component Fermi Gases}} and {{Loss Rate}} of {{Degenerate
      Polar Molecules}}},\ }\href {\doibase10.1103/PhysRevLett.131.043401}
      {\bibfield  {journal} {\bibinfo  {journal} {Phys. Rev. Lett.}\ }\textbf
      {\bibinfo {volume} {131}},\ \bibinfo {pages} {043401} (\bibinfo {year}
      {2023})}\BibitemShut {NoStop}%
    \bibitem [{SM()}]{SM}%
      \BibitemOpen
      \bibinfo {note} {{The companion paper at to-be-inserted-by-the-editor.
      }}\BibitemShut {NoStop}%
    \bibitem [{\citenamefont {Braaten}\ \emph {et~al.}(2017)\citenamefont
      {Braaten}, \citenamefont {Hammer},\ and\ \citenamefont
      {Lepage}}]{braaten2017lindblad}%
      \BibitemOpen
      \bibfield  {author} {\bibinfo {author} {\bibfnamefont {E.}~\bibnamefont
      {Braaten}}, \bibinfo {author} {\bibfnamefont {H.-W.}\ \bibnamefont {Hammer}},
      \ and\ \bibinfo {author} {\bibfnamefont {G.~P.}\ \bibnamefont {Lepage}},\
      }\bibfield  {title} {\bibinfo {title} {Lindblad equation for the inelastic
      loss of ultracold atoms},\ }\href {\doibase10.1103/PhysRevA.95.012708}
      {\bibfield  {journal} {\bibinfo  {journal} {Phys. Rev. A}\ }\textbf {\bibinfo
      {volume} {95}},\ \bibinfo {pages} {012708} (\bibinfo {year}
      {2017})}\BibitemShut {NoStop}%
    \bibitem [{\citenamefont {He}\ \emph {et~al.}(2020{\natexlab{b}})\citenamefont
      {He}, \citenamefont {Bilitewski}, \citenamefont {Greene},\ and\ \citenamefont
      {Rey}}]{he2020exploring}%
      \BibitemOpen
      \bibfield  {author} {\bibinfo {author} {\bibfnamefont {P.}~\bibnamefont
      {He}}, \bibinfo {author} {\bibfnamefont {T.}~\bibnamefont {Bilitewski}},
      \bibinfo {author} {\bibfnamefont {C.~H.}\ \bibnamefont {Greene}}, \ and\
      \bibinfo {author} {\bibfnamefont {A.~M.}\ \bibnamefont {Rey}},\ }\bibfield
      {title} {\bibinfo {title} {Exploring chemical reactions in a quantum
      degenerate gas of polar molecules via complex formation},\ }\href
      {\doibase10.1103/PhysRevA.102.063322} {\bibfield  {journal} {\bibinfo
      {journal} {Phys. Rev. A}\ }\textbf {\bibinfo {volume} {102}},\ \bibinfo
      {pages} {063322} (\bibinfo {year} {2020}{\natexlab{b}})}\BibitemShut
      {NoStop}%
    \bibitem [{\citenamefont {Wang}\ \emph {et~al.}(2022)\citenamefont {Wang},
      \citenamefont {Liu},\ and\ \citenamefont {Shi}}]{wang2022complex}%
      \BibitemOpen
      \bibfield  {author} {\bibinfo {author} {\bibfnamefont {C.}~\bibnamefont
      {Wang}}, \bibinfo {author} {\bibfnamefont {C.}~\bibnamefont {Liu}}, \ and\
      \bibinfo {author} {\bibfnamefont {Z.-Y.}\ \bibnamefont {Shi}},\ }\bibfield
      {title} {\bibinfo {title} {Complex {{Contact Interaction}} for {{Systems}}
      with {{Short-Range Two-Body Losses}}},\ }\href
      {\doibase10.1103/PhysRevLett.129.203401} {\bibfield  {journal} {\bibinfo
      {journal} {Phys. Rev. Lett.}\ }\textbf {\bibinfo {volume} {129}},\ \bibinfo
      {pages} {203401} (\bibinfo {year} {2022})}\BibitemShut {NoStop}%
    \bibitem [{\citenamefont {Mehta}\ \emph {et~al.}(2009)\citenamefont {Mehta},
      \citenamefont {Rittenhouse}, \citenamefont {D’Incao}, \citenamefont {{von
      Stecher}},\ and\ \citenamefont {Greene}}]{mehta2009general}%
      \BibitemOpen
      \bibfield  {author} {\bibinfo {author} {\bibfnamefont {N.~P.}\ \bibnamefont
      {Mehta}}, \bibinfo {author} {\bibfnamefont {S.~T.}\ \bibnamefont
      {Rittenhouse}}, \bibinfo {author} {\bibfnamefont {J.~P.}\ \bibnamefont
      {D’Incao}}, \bibinfo {author} {\bibfnamefont {J.}~\bibnamefont {{von
      Stecher}}}, \ and\ \bibinfo {author} {\bibfnamefont {C.~H.}\ \bibnamefont
      {Greene}},\ }\bibfield  {title} {\bibinfo {title} {General {{Theoretical
      Description}} of ${N}$-{{Body Recombination}}},\ }\href
      {\doibase10.1103/PhysRevLett.103.153201} {\bibfield  {journal} {\bibinfo
      {journal} {Phys. Rev. Lett.}\ }\textbf {\bibinfo {volume} {103}},\ \bibinfo
      {pages} {153201} (\bibinfo {year} {2009})}\BibitemShut {NoStop}%
    \bibitem [{\citenamefont {De~Marco}\ \emph
      {et~al.}(2019{\natexlab{b}})\citenamefont {De~Marco}, \citenamefont
      {Valtolina}, \citenamefont {Matsuda}, \citenamefont {Tobias}, \citenamefont
      {Covey},\ and\ \citenamefont {Ye}}]{demarco2019replication}%
      \BibitemOpen
      \bibfield  {author} {\bibinfo {author} {\bibfnamefont {L.}~\bibnamefont
      {De~Marco}}, \bibinfo {author} {\bibfnamefont {G.}~\bibnamefont {Valtolina}},
      \bibinfo {author} {\bibfnamefont {K.}~\bibnamefont {Matsuda}}, \bibinfo
      {author} {\bibfnamefont {W.}~\bibnamefont {Tobias}}, \bibinfo {author}
      {\bibfnamefont {J.}~\bibnamefont {Covey}}, \ and\ \bibinfo {author}
      {\bibfnamefont {J.}~\bibnamefont {Ye}},\ }\href {\doibase10.7910/DVN/RLOBHV}
      {\bibinfo {title} {{Replication Data for: A Degenerate Fermi Gas of Polar
      Molecules}},\ } (\bibinfo {year} {2019}{\natexlab{b}})\BibitemShut {NoStop}%
    \bibitem [{\citenamefont {Navon}\ \emph {et~al.}(2021)\citenamefont {Navon},
      \citenamefont {Smith},\ and\ \citenamefont {Hadzibabic}}]{navon2021quantuma}%
      \BibitemOpen
      \bibfield  {author} {\bibinfo {author} {\bibfnamefont {N.}~\bibnamefont
      {Navon}}, \bibinfo {author} {\bibfnamefont {R.~P.}\ \bibnamefont {Smith}}, \
      and\ \bibinfo {author} {\bibfnamefont {Z.}~\bibnamefont {Hadzibabic}},\
      }\bibfield  {title} {\bibinfo {title} {Quantum gases in optical boxes},\
      }\href {\doibase10.1038/s41567-021-01403-z} {\bibfield  {journal} {\bibinfo
      {journal} {Nat. Phys.}\ }\textbf {\bibinfo {volume} {17}},\ \bibinfo {pages}
      {1334} (\bibinfo {year} {2021})}\BibitemShut {NoStop}%
    \bibitem [{\citenamefont {Regal}\ \emph {et~al.}(2003)\citenamefont {Regal},
      \citenamefont {Ticknor}, \citenamefont {Bohn},\ and\ \citenamefont
      {Jin}}]{regal2003tuning}%
      \BibitemOpen
      \bibfield  {author} {\bibinfo {author} {\bibfnamefont {C.~A.}\ \bibnamefont
      {Regal}}, \bibinfo {author} {\bibfnamefont {C.}~\bibnamefont {Ticknor}},
      \bibinfo {author} {\bibfnamefont {J.~L.}\ \bibnamefont {Bohn}}, \ and\
      \bibinfo {author} {\bibfnamefont {D.~S.}\ \bibnamefont {Jin}},\ }\bibfield
      {title} {\bibinfo {title} {Tuning $p$-{{Wave Interactions}} in an {{Ultracold
      Fermi Gas}} of {{Atoms}}},\ }\href {\doibase10.1103/PhysRevLett.90.053201}
      {\bibfield  {journal} {\bibinfo  {journal} {Phys. Rev. Lett.}\ }\textbf
      {\bibinfo {volume} {90}},\ \bibinfo {pages} {053201} (\bibinfo {year}
      {2003})}\BibitemShut {NoStop}%
    \bibitem [{\citenamefont {Zhang}\ \emph {et~al.}(2004)\citenamefont {Zhang},
      \citenamefont {{van Kempen}}, \citenamefont {Bourdel}, \citenamefont
      {Khaykovich}, \citenamefont {Cubizolles}, \citenamefont {Chevy},
      \citenamefont {Teichmann}, \citenamefont {Tarruell}, \citenamefont
      {Kokkelmans},\ and\ \citenamefont {Salomon}}]{zhang2004pwave}%
      \BibitemOpen
      \bibfield  {author} {\bibinfo {author} {\bibfnamefont {J.}~\bibnamefont
      {Zhang}}, \bibinfo {author} {\bibfnamefont {E.~G.~M.}\ \bibnamefont {{van
      Kempen}}}, \bibinfo {author} {\bibfnamefont {T.}~\bibnamefont {Bourdel}},
      \bibinfo {author} {\bibfnamefont {L.}~\bibnamefont {Khaykovich}}, \bibinfo
      {author} {\bibfnamefont {J.}~\bibnamefont {Cubizolles}}, \bibinfo {author}
      {\bibfnamefont {F.}~\bibnamefont {Chevy}}, \bibinfo {author} {\bibfnamefont
      {M.}~\bibnamefont {Teichmann}}, \bibinfo {author} {\bibfnamefont
      {L.}~\bibnamefont {Tarruell}}, \bibinfo {author} {\bibfnamefont {S.~J. J.
      M.~F.}\ \bibnamefont {Kokkelmans}}, \ and\ \bibinfo {author} {\bibfnamefont
      {C.}~\bibnamefont {Salomon}},\ }\bibfield  {title} {\bibinfo {title}
      {${{P}}$-wave {{Feshbach}} resonances of ultracold ${}^{6}\mathrm{Li}$},\
      }\href {\doibase10.1103/PhysRevA.70.030702} {\bibfield  {journal} {\bibinfo
      {journal} {Phys. Rev. A}\ }\textbf {\bibinfo {volume} {70}},\ \bibinfo
      {pages} {030702} (\bibinfo {year} {2004})}\BibitemShut {NoStop}%
    \bibitem [{\citenamefont {Schunck}\ \emph {et~al.}(2005)\citenamefont
      {Schunck}, \citenamefont {Zwierlein}, \citenamefont {Stan}, \citenamefont
      {Raupach}, \citenamefont {Ketterle}, \citenamefont {Simoni}, \citenamefont
      {Tiesinga}, \citenamefont {Williams},\ and\ \citenamefont
      {Julienne}}]{schunck2005feshbach}%
      \BibitemOpen
      \bibfield  {author} {\bibinfo {author} {\bibfnamefont {C.~H.}\ \bibnamefont
      {Schunck}}, \bibinfo {author} {\bibfnamefont {M.~W.}\ \bibnamefont
      {Zwierlein}}, \bibinfo {author} {\bibfnamefont {C.~A.}\ \bibnamefont {Stan}},
      \bibinfo {author} {\bibfnamefont {S.~M.~F.}\ \bibnamefont {Raupach}},
      \bibinfo {author} {\bibfnamefont {W.}~\bibnamefont {Ketterle}}, \bibinfo
      {author} {\bibfnamefont {A.}~\bibnamefont {Simoni}}, \bibinfo {author}
      {\bibfnamefont {E.}~\bibnamefont {Tiesinga}}, \bibinfo {author}
      {\bibfnamefont {C.~J.}\ \bibnamefont {Williams}}, \ and\ \bibinfo {author}
      {\bibfnamefont {P.~S.}\ \bibnamefont {Julienne}},\ }\bibfield  {title}
      {\bibinfo {title} {Feshbach resonances in fermionic ${}^{6}\mathrm{Li}$},\
      }\href {\doibase10.1103/PhysRevA.71.045601} {\bibfield  {journal} {\bibinfo
      {journal} {Phys. Rev. A}\ }\textbf {\bibinfo {volume} {71}},\ \bibinfo
      {pages} {045601} (\bibinfo {year} {2005})}\BibitemShut {NoStop}%
    \bibitem [{\citenamefont {Günter}\ \emph {et~al.}(2005)\citenamefont
      {Günter}, \citenamefont {Stöferle}, \citenamefont {Moritz}, \citenamefont
      {Köhl},\ and\ \citenamefont {Esslinger}}]{gunter2005pwave}%
      \BibitemOpen
      \bibfield  {author} {\bibinfo {author} {\bibfnamefont {K.}~\bibnamefont
      {Günter}}, \bibinfo {author} {\bibfnamefont {T.}~\bibnamefont {Stöferle}},
      \bibinfo {author} {\bibfnamefont {H.}~\bibnamefont {Moritz}}, \bibinfo
      {author} {\bibfnamefont {M.}~\bibnamefont {Köhl}}, \ and\ \bibinfo {author}
      {\bibfnamefont {T.}~\bibnamefont {Esslinger}},\ }\bibfield  {title} {\bibinfo
      {title} {$p$-{{Wave Interactions}} in {{Low-Dimensional Fermionic Gases}}},\
      }\href {\doibase10.1103/PhysRevLett.95.230401} {\bibfield  {journal}
      {\bibinfo  {journal} {Phys. Rev. Lett.}\ }\textbf {\bibinfo {volume} {95}},\
      \bibinfo {pages} {230401} (\bibinfo {year} {2005})}\BibitemShut {NoStop}%
    \bibitem [{\citenamefont {Inada}\ \emph {et~al.}(2008)\citenamefont {Inada},
      \citenamefont {Horikoshi}, \citenamefont {Nakajima}, \citenamefont
      {{Kuwata-Gonokami}}, \citenamefont {Ueda},\ and\ \citenamefont
      {Mukaiyama}}]{inada2008collisional}%
      \BibitemOpen
      \bibfield  {author} {\bibinfo {author} {\bibfnamefont {Y.}~\bibnamefont
      {Inada}}, \bibinfo {author} {\bibfnamefont {M.}~\bibnamefont {Horikoshi}},
      \bibinfo {author} {\bibfnamefont {S.}~\bibnamefont {Nakajima}}, \bibinfo
      {author} {\bibfnamefont {M.}~\bibnamefont {{Kuwata-Gonokami}}}, \bibinfo
      {author} {\bibfnamefont {M.}~\bibnamefont {Ueda}}, \ and\ \bibinfo {author}
      {\bibfnamefont {T.}~\bibnamefont {Mukaiyama}},\ }\bibfield  {title} {\bibinfo
      {title} {Collisional {{Properties}} of $p$-{{Wave Feshbach Molecules}}},\
      }\href {\doibase10.1103/PhysRevLett.101.100401} {\bibfield  {journal}
      {\bibinfo  {journal} {Phys. Rev. Lett.}\ }\textbf {\bibinfo {volume} {101}},\
      \bibinfo {pages} {100401} (\bibinfo {year} {2008})}\BibitemShut {NoStop}%
    \bibitem [{\citenamefont {Fuchs}\ \emph {et~al.}(2008)\citenamefont {Fuchs},
      \citenamefont {Ticknor}, \citenamefont {Dyke}, \citenamefont {Veeravalli},
      \citenamefont {Kuhnle}, \citenamefont {Rowlands}, \citenamefont {Hannaford},\
      and\ \citenamefont {Vale}}]{fuchs2008binding}%
      \BibitemOpen
      \bibfield  {author} {\bibinfo {author} {\bibfnamefont {J.}~\bibnamefont
      {Fuchs}}, \bibinfo {author} {\bibfnamefont {C.}~\bibnamefont {Ticknor}},
      \bibinfo {author} {\bibfnamefont {P.}~\bibnamefont {Dyke}}, \bibinfo {author}
      {\bibfnamefont {G.}~\bibnamefont {Veeravalli}}, \bibinfo {author}
      {\bibfnamefont {E.}~\bibnamefont {Kuhnle}}, \bibinfo {author} {\bibfnamefont
      {W.}~\bibnamefont {Rowlands}}, \bibinfo {author} {\bibfnamefont
      {P.}~\bibnamefont {Hannaford}}, \ and\ \bibinfo {author} {\bibfnamefont
      {C.~J.}\ \bibnamefont {Vale}},\ }\bibfield  {title} {\bibinfo {title}
      {Binding energies of ${}^{6}\mathrm{Li}$ $p$-wave {{Feshbach}} molecules},\
      }\href {\doibase10.1103/PhysRevA.77.053616} {\bibfield  {journal} {\bibinfo
      {journal} {Phys. Rev. A}\ }\textbf {\bibinfo {volume} {77}},\ \bibinfo
      {pages} {053616} (\bibinfo {year} {2008})}\BibitemShut {NoStop}%
    \bibitem [{\citenamefont {Maier}\ \emph {et~al.}(2010)\citenamefont {Maier},
      \citenamefont {Marzok}, \citenamefont {Zimmermann},\ and\ \citenamefont
      {Courteille}}]{maier2010radiofrequency}%
      \BibitemOpen
      \bibfield  {author} {\bibinfo {author} {\bibfnamefont {R.~A.~W.}\
      \bibnamefont {Maier}}, \bibinfo {author} {\bibfnamefont {C.}~\bibnamefont
      {Marzok}}, \bibinfo {author} {\bibfnamefont {C.}~\bibnamefont {Zimmermann}},
      \ and\ \bibinfo {author} {\bibfnamefont {{\relax Ph}.~W.}\ \bibnamefont
      {Courteille}},\ }\bibfield  {title} {\bibinfo {title} {Radio-frequency
      spectroscopy of ${}^{6}\mathrm{Li}$ $p$-wave molecules: {{Towards}}
      photoemission spectroscopy of a $p$-wave superfluid},\ }\href
      {\doibase10.1103/PhysRevA.81.064701} {\bibfield  {journal} {\bibinfo
      {journal} {Phys. Rev. A}\ }\textbf {\bibinfo {volume} {81}},\ \bibinfo
      {pages} {064701} (\bibinfo {year} {2010})}\BibitemShut {NoStop}%
    \bibitem [{\citenamefont {Gerken}\ \emph {et~al.}(2019)\citenamefont {Gerken},
      \citenamefont {Tran}, \citenamefont {Häfner}, \citenamefont {Tiemann},
      \citenamefont {Zhu},\ and\ \citenamefont
      {Weidemüller}}]{gerken2019observationa}%
      \BibitemOpen
      \bibfield  {author} {\bibinfo {author} {\bibfnamefont {M.}~\bibnamefont
      {Gerken}}, \bibinfo {author} {\bibfnamefont {B.}~\bibnamefont {Tran}},
      \bibinfo {author} {\bibfnamefont {S.}~\bibnamefont {Häfner}}, \bibinfo
      {author} {\bibfnamefont {E.}~\bibnamefont {Tiemann}}, \bibinfo {author}
      {\bibfnamefont {B.}~\bibnamefont {Zhu}}, \ and\ \bibinfo {author}
      {\bibfnamefont {M.}~\bibnamefont {Weidemüller}},\ }\bibfield  {title}
      {\bibinfo {title} {Observation of dipolar splittings in high-resolution
      atom-loss spectroscopy of ${}^{6}\mathrm{Li}$ $p$-wave {{Feshbach}}
      resonances},\ }\href {\doibase10.1103/PhysRevA.100.050701} {\bibfield
      {journal} {\bibinfo  {journal} {Phys. Rev. A}\ }\textbf {\bibinfo {volume}
      {100}},\ \bibinfo {pages} {050701} (\bibinfo {year} {2019})}\BibitemShut
      {NoStop}%
    \bibitem [{\citenamefont {Kiefer}\ \emph {et~al.}(2023)\citenamefont {Kiefer},
      \citenamefont {Hachmann},\ and\ \citenamefont
      {Hemmerich}}]{kiefer2023ultracold}%
      \BibitemOpen
      \bibfield  {author} {\bibinfo {author} {\bibfnamefont {Y.}~\bibnamefont
      {Kiefer}}, \bibinfo {author} {\bibfnamefont {M.}~\bibnamefont {Hachmann}}, \
      and\ \bibinfo {author} {\bibfnamefont {A.}~\bibnamefont {Hemmerich}},\
      }\bibfield  {title} {\bibinfo {title} {Ultracold {{Feshbach}} molecules in an
      orbital optical lattice},\ }\href {\doibase10.1038/s41567-023-01994-9}
      {\bibfield  {journal} {\bibinfo  {journal} {Nat. Phys.}\ }\textbf {\bibinfo
      {volume} {19}},\ \bibinfo {pages} {794} (\bibinfo {year} {2023})}\BibitemShut
      {NoStop}%
    \bibitem [{\citenamefont {Venu}\ \emph {et~al.}(2023)\citenamefont {Venu},
      \citenamefont {Xu}, \citenamefont {Mamaev}, \citenamefont {Corapi},
      \citenamefont {Bilitewski}, \citenamefont {D’Incao}, \citenamefont
      {Fujiwara}, \citenamefont {Rey},\ and\ \citenamefont
      {Thywissen}}]{venu2023unitary}%
      \BibitemOpen
      \bibfield  {author} {\bibinfo {author} {\bibfnamefont {V.}~\bibnamefont
      {Venu}}, \bibinfo {author} {\bibfnamefont {P.}~\bibnamefont {Xu}}, \bibinfo
      {author} {\bibfnamefont {M.}~\bibnamefont {Mamaev}}, \bibinfo {author}
      {\bibfnamefont {F.}~\bibnamefont {Corapi}}, \bibinfo {author} {\bibfnamefont
      {T.}~\bibnamefont {Bilitewski}}, \bibinfo {author} {\bibfnamefont {J.~P.}\
      \bibnamefont {D’Incao}}, \bibinfo {author} {\bibfnamefont {C.~J.}\
      \bibnamefont {Fujiwara}}, \bibinfo {author} {\bibfnamefont {A.~M.}\
      \bibnamefont {Rey}}, \ and\ \bibinfo {author} {\bibfnamefont {J.~H.}\
      \bibnamefont {Thywissen}},\ }\bibfield  {title} {\bibinfo {title} {Unitary
      p-wave interactions between fermions in an optical lattice},\ }\href
      {\doibase10.1038/s41586-022-05405-6} {\bibfield  {journal} {\bibinfo
      {journal} {Nature (London)}\ }\textbf {\bibinfo {volume} {613}},\ \bibinfo
      {pages} {262} (\bibinfo {year} {2023})}\BibitemShut {NoStop}%
    \bibitem [{\citenamefont {Wang}\ \emph {et~al.}(2024)\citenamefont {Wang},
      \citenamefont {Biswas}, \citenamefont {Eppelt}, \citenamefont {Deng},
      \citenamefont {Luo},\ and\ \citenamefont {Bohn}}]{wang2024simulations}%
      \BibitemOpen
      \bibfield  {author} {\bibinfo {author} {\bibfnamefont {R.~R.~W.}\
      \bibnamefont {Wang}}, \bibinfo {author} {\bibfnamefont {S.}~\bibnamefont
      {Biswas}}, \bibinfo {author} {\bibfnamefont {S.}~\bibnamefont {Eppelt}},
      \bibinfo {author} {\bibfnamefont {F.}~\bibnamefont {Deng}}, \bibinfo {author}
      {\bibfnamefont {X.-Y.}\ \bibnamefont {Luo}}, \ and\ \bibinfo {author}
      {\bibfnamefont {J.~L.}\ \bibnamefont {Bohn}},\ }\href@noop {} {\bibinfo
      {title} {Simulations of evaporation to deep {{Fermi}} degeneracy in
      microwave-shielded molecules},\ } (\bibinfo {year} {2024}),\ \Eprint
      {https://arxiv.org/abs/2407.14466} {arXiv:2407.14466 [cond-mat,
      physics:physics, physics:quant-ph]} \BibitemShut {NoStop}%
    \end{thebibliography}

%

\end{document}


\preprint{APS/123-QED}

\title{Analysis on inelastic quantum Boltzmann equation of single-component Fermi gases}

\author{Xin-Yuan Gao}
\affiliation{%
Department of Physics, The Chinese University of Hong Kong, Shatin, New Territories, Hong Kong, China
}%
\author{Yangqian Yan}%
 \email{yqyan@cuhk.edu.hk}
\affiliation{%
Department of Physics, The Chinese University of Hong Kong, Shatin, New Territories, Hong Kong, China
}
\affiliation{
The Chinese University of Hong Kong Shenzhen Research Institute, 518057 Shenzhen, China
}%

\date{\today}

\begin{abstract}
This work is a companion paper to \{FILLED-BY-THE-EDITOR\}, where we discuss the non-equilibrium two-body loss dynamics of a single-component ultracold Fermi gas and its possible thermalization. The present paper provides detailed information on the derivation and analysis of the inelastic quantum Boltzmann equation (IQBE) used to describe the system. We demonstrate that the Mellin transform is a powerful tool for solving and approximating the IQBE for free-space systems. In this case, the particle number dynamics are beyond the description of the widely used phenomenological two-body equation. For harmonically trapped systems, we propose a fast-flowing approximation (FFA) to simplify the numerical evaluation of the IQBE. We verify the approximation in an analogous quasi-1D system and apply it to 3D calculations, obtaining satisfactory agreement with recent experimental results. Furthermore, we compare the non-equilibrium results with those obtained using a thermal ansatz in both situations, providing a systematic understanding of the anti-evaporation phenomena observed in such systems.
\end{abstract}

\maketitle

\section{Introduction}
\label{intro}

As the key concept in the kinetic theory, the Boltzmann equation describes the evolution of a classic many-body system towards thermal equilibrium~\cite{huang2008statistical}. An analog can be obtained for normal-phase quantum gases or even Bose-Einstein condensates~\cite{zaremba1999dynamics} and fermionic superfluid~\cite{urban2006dynamics,urban2007coupling}, known as the quantum Boltzmann equation. It is notoriously hard to solve such equations directly due to their six-dimensional spatial complexity and the sophisticated form of the collision integral. Under the collision integral's relaxation approximation (or BGK approximation~\cite{puppo2019kinetic}), the Boltzmann equation has been widely adopted to study hydrodynamic expansion~\cite{you1996ballistica,guery-odelin2002meanfield}, collective modes\cite{guery-odelin1999collective,pedri2003dynamics}, and spin waves~\cite{nikuni2002linear} in harmonically trapped quantum gases. Various numerical methods also tackle collision integral without approximation, mainly based on the Direct Simulation Monte Carlo method~\cite{bird1994molecular, pareschi2001introduction} and its variants~\cite{wu1996direct,jackson2002modeling,lepers2010numerical,barletta2010direct,wade2011directa}. In all usual applications above, the particles are assumed to be featureless and only possess translational motions.
Consequently, the collision integral only captures the system's elastic collisions. However, inelastic collisions can also play an essential role in certain situations, especially in the study of ultracold molecular gases, where the inelastic collision due to intrinsic chemical reactions affects the efficiency of evaporative cooling and the lifetime of the gas~\cite{ospelkaus2010quantumstate}. In the Boltzmann equations describing classic granular fluids, the inelastic collision is implemented by breaking momentum and energy conservation in the collision integral~\cite{sela1998hydrodynamic,garzo1999dense,ben-naim1999shocklike,bobylev2000properties,ben-naim2002scaling}. Nevertheless, such generalization cannot be directly applied to quantum Boltzmann equations since the inelastic collision changes the particles' internal states, which indicates not only the dissipation of energy but also the loss of the total number of particles (if we only concern the internal states where particles start from).  Note that non-Hermitian linear response theory predicts (scratched) exponential loss for a one-body loss and broadening of momentum distribution for two-body atom-number-conserving dissipation in a short-time scale~\cite{pan2020nonhermitian}. In comparison, this work considers the long-time dynamics as well as the effect of the two-body lossy dissipator.

In this article, we focus on single-component ultracold molecular Fermi gas, a simple but highly non-trivial quantum gas system with inelastic collision, to demonstrate how to model the system's dynamics with a proper inelastic quantum Boltzmann equation (IQBE). Experimentally, this is related to two molecular platforms, including ${}^{40}\mathrm{K}{}^{87}\mathrm{Rb}$ (e.g. See Ref.~\cite{demarco2019degenerate}) and ${}^{23}\mathrm{Na}{}^{40}\mathrm{K}$ (e.g. See Ref.~\cite{duda2023longlived}) Fermi gases with two-body loss induced by chemical reactions and two atomic platforms: Li and K near with two-body loss induced by dipolar relaxation. Here are some general properties of those systems: (I) Although the detailed mechanism may differ, the outcomes are the same. The outgoing particles of inelastic scattering will gain enough kinetic energy to escape from the trap. As a result, two particles will be lost per inelastic collision event. (II) The systems are prepared with initial temperatures typically below $1~\mu \text{K}$. In this ultracold regime, the two-body collision follows threshold behavior, i.e., only the lowest possible partial wave channel dominates the phase shift. This is true for both elastic and inelastic collisions. (III) Unlike the usual Fermi gas with balanced spin-up and spin-down components, all particles in such systems are typically prepared in the same rovibrational ground state. Because of the Pauli exclusion principle, the natural lowest partial wave channel, $s$-wave collision, is prohibited. Combined with the abovementioned, the $p$-wave channel will dominate, providing the interparticle potential with a centrifugal barrier. As detailed in this work, the relative momentum selection caused by the centrifugal barrier is the origin of the intriguing dynamics of such systems. (IV) Without Feshbach resonance, the bare scattering volume of elastic collision is extremely small. Furthermore, when the system is away from the equilibrium state, $p$-wave collision cannot equilibrate the system in a typical experimental time scale. The argument will also be elaborated in the following sections.

This work will be mainly divided into three parts. In the first part, we introduce the model, derive the IQBE from scratch, and comment on the system's relaxation time. In the second part, we focus on homogeneous systems with spatial translational symmetry. In the third part, we work on harmonically trapped systems closely related to recent experiments.

\section{Non-Hermitian Hamiltonian}

As explained in Sec.~\ref{intro}, we must construct a model with $p$-wave interaction and include inelastic collision. The simplest option is to use one-channel non-hermitian Hamiltonian
\begin{equation}
\begin{split}
    &\hat{H}=\hat{T}+\hat{U},\\
    &\hat{T}=\sum_\mathbf{k} \frac{\hbar^2k^2}{2M}c_\mathbf{k}^\dagger c_\mathbf{k},\\
    &\hat{U}=\frac{3 g}{2 V}\sum_{\mathbf{P},\mathbf{q},\mathbf{q}'}\mathbf{q}\cdot \mathbf{q}'c^\dagger_{\frac{\mathbf{P}}{2}+\mathbf{q}} c^\dagger_{\frac{\mathbf{P}}{2}-\mathbf{q}}c_{\frac{\mathbf{P}}{2}-\mathbf{q}'} c_{\frac{\mathbf{P}}{2}+\mathbf{q}'},
\end{split}
\label{Hamiltonian}
\end{equation}
where $M$ is the mass of particles, $V$ the volume of the system, and $c_\mathbf{k}$ the annihilation operator of a particle. The momentum dependence in the interaction reflects the $p$-wave symmetry. Importantly, the coupling constant $g$ is a complex number, accounting for both elastic and inelastic collisions. The non-hermitian Hamiltonian here is an effective field theory from a hermitian model, integrating the product channel of two-body chemical reaction~\cite{braaten2017lindblad}. For low-temperature threshold collision, the phase shift of the $p$-wave channel $\delta_p$ can be expanded as $k^3\cot(\delta_p)=-1/v_p+\mathcal{O}(k^2)$, where $v_p$ is defined as the scattering volume. Because the system has inelastic collisions, both $\delta_p$ and $v_p$ are expected to be complex. Following the standard renormalization procedure of comparing $\mathrm{T}$ matrix calculated from Eq.~(\ref{Hamiltonian}) with scattering amplitude, one has
\begin{equation}
    \frac{1}{g}=\frac{M}{4\pi\hbar^2 v_p}+\frac{M}{2\pi^2\hbar^2}\int_{0}^{\infty}dqq^{2}.
\end{equation}
For our aim in this work, it is enough only to take the normal part of the renormalization condition, which is
\begin{equation}
    g=4\pi\hbar^2[\mathrm{Re}(v_p)+i\mathrm{Im}(v_p)]/M.
\end{equation}

Most literature discussing many-body systems with $p$-wave collision stresses the importance of effective range $R$ besides $v_p$. It is defined through the next order expansion of $p$-wave phase shift $k^3\cot(\delta_p)=-1/v_p-k^2/R+\mathcal{O}(k^4)$. The reason why $R$ should be further considered is that the bound state's binding energy is determined by both $v_p$ and $R$, i.e. $E_b=\hbar^2 R/Mv_p$~\cite{ho2005fermion,yu2015universal}. Nonetheless, the bound state is irrelevant since we do not consider applying Feshbach resonance to the system. As a result, in our model, it is enough to set only one parameter to determine $v_p$ and ignore $R$. \break

\section{Inelastic Quantum Boltzmann Equation}
\label{IQBE}

In this section, we provide a deviation of the inelastic quantum Boltzmann equation we will use to describe the dissipation dynamics in this work. The standard form of a Boltzmann equation is an equation of motion of phase space density $f(\mathbf{k},\mathbf{r})$, composed of Vlasov equation for modeling diffusion and drift flow and collision integral $\mathcal{I}_\mathrm{coll}[f]$ depicting the collision
\begin{equation}
    \frac{d f}{dt}+\left[\frac{\hbar \mathbf{k}}{M}\nabla_\mathbf{r}-\frac{\nabla_{\mathbf{r}}U_{\mathrm{ext}}\cdot\nabla_\mathbf{k}}{\hbar}\right]f=\mathcal{I}_\mathrm{coll}[f].
    \label{boltzmann_eqn_form}
\end{equation}
The task is to obtain the form of $\mathcal{I}_\mathrm{coll}[f]$ from the first principle, e.g. the Schr\"odinger equation. 

We start with homogeneous systems, where we assume $f(\mathbf{k},\mathbf{r})=V n_\mathbf{k}$ with $n_\mathbf{k}$ the momentum distribution of the system. It simplifies Eq.~(\ref{boltzmann_eqn_form}) to 
\begin{equation}
    \frac{d n_\mathbf{k}}{dt}=\frac{1}{V}\mathcal{I}_\mathrm{coll}[V n_\mathbf{k}].
    \label{general_homo_IQBE}
\end{equation}
After obtaining $\mathcal{I}_\mathrm{coll}[V n_\mathbf{k}]/V$ for homogeneous systems, we immediately have the form $\mathcal{I}_\mathrm{coll}$ which also applied to the arbitrary inhomogeneous system under local density approximation, which means one treat each local regimes a separate small homogeneous piece. The approximation is exact in the thermodynamic limit.

We use the following symbols to denote the time dependence of operators, states, and the density matrix in different pictures:
\begin{align*}
    &\text{Schr\"odinger picture: }\hat{A},~|\psi_t\rangle,~\rho_t;\\
    &\text{Heisenburg picture: }\hat{A}_t,~|\psi\rangle,~\rho;\\
    &\text{Interaction picture: }\hat{A}(t),~|\psi(t)\rangle,
\end{align*}
Investigating the evolution of $n_\mathbf{k}$ is equivalent to studying the dynamics of number operator $N_\mathbf{k}=c_\mathbf{k}^\dagger c_\mathbf{k}$. In a small time window $t$, the change of expectation $\langle N_\mathbf{k}\rangle$ is
\begin{equation}
\begin{split}
    &\Delta \langle N_\mathbf{k}\rangle=\mathrm{Tr}(\rho_t N_\mathbf{k})-\mathrm{Tr}(\rho_0 N_\mathbf{k})\\
    &=\sum_n \rho_{nn}\left(\langle n(t)|N_\mathbf{k}|n(t) \rangle-\langle n_0|N_\mathbf{k}|n_0 \rangle\right)\\
    &=\sum_n \rho_{nn}\langle n |e^{\frac{i}{\hbar}\mathcal{T}\int \hat{U}^\dagger(t)dt}\\
    &\times\left(N_\mathbf{k}e^{-\frac{i}{\hbar}\mathcal{T}\int^t \hat{U}(t')dt'}-e^{-\frac{i}{\hbar}\mathcal{T}\int^t \hat{U}^\dagger(t')dt'} N_\mathbf{k}\right)|n\rangle.
\end{split}
\end{equation}
Here, we use the symbol $|n\rangle$ to denote a general Fock state, and $\mathcal{T}$ is the time-ordered operator. Based on the definition of $\mathcal{T}$~\cite{fetter2012quantum}, one can expand $\Delta \langle N_\mathbf{k}\rangle$ up to the second-order in terms of the interaction,
\begin{equation}
    \Delta \langle N_\mathbf{k}\rangle=\Delta \langle N_\mathbf{k}\rangle^{(1)}+\Delta \langle N_\mathbf{k}\rangle^{(2)}_1+\Delta \langle N_\mathbf{k}\rangle^{(2)}_2,
\end{equation}
where the first-order term is
\begin{equation}
    \langle N_\mathbf{k}\rangle^{(1)}=\sum_n \frac{\rho_{nn}}{i\hbar}\langle n| N_\mathbf{k}\int_0^t \hat{U}(t')dt'- \int_0^t \hat{U}^\dagger(t')dt'N_\mathbf{k}|n\rangle.
\end{equation}
And there are two second-order terms, which are
\begin{equation}
\begin{split}
&\Delta \langle N_\mathbf{k}\rangle^{(2)}_1=\sum_n \frac{\rho_{nn}}{\hbar^2}\langle n|\int_0^t \hat{U}^\dagger(t')dt'\\
&\times\left(N_\mathbf{k}\int_0^t \hat{U}(t')dt'- \int_0^t \hat{U}^\dagger(t')dt'N_\mathbf{k}\right)|n\rangle, 
\end{split}
\end{equation}
and
\begin{equation}
\begin{split}
\Delta \langle &N_\mathbf{k}\rangle^{(2)}_2=-\sum_n \frac{\rho_{nn}}{\hbar^2}\langle n|N_\mathbf{k}\int_0^tdt'\int_0^{t'}dt''\hat{U}(t')\hat{U}(t'')\\
&-\int_0^tdt'\int_0^{t'}dt''\hat{U}^\dagger(t')\hat{U}^\dagger(t'')N_\mathbf{k}|n\rangle.
\end{split}
\end{equation}
It is worth emphasizing that since $\hat{U}$ is non-hermitian, $\hat{U}^\dagger$ is distinguishable from $\hat{U}$. The expansion up to the second order is justified by the fact that interaction is weak enough, which will be further confirmed in Sec.~\ref{relax_tau}.

Let us focus on the $\langle N_\mathbf{k}\rangle^{(1)}$ first. Decomposing the term by separating real and imaginary parts of interaction and taking the short-time limit, we have
\begin{equation}
\begin{split}
    &\Delta \langle N_\mathbf{k}\rangle^{(1)}\xrightarrow[]{t\rightarrow0}\frac{t}{i\hbar}\sum_n\rho_{nn}\langle n|[N_\mathbf{k},\mathrm{Re}(\hat{U})]|n\rangle\\
    &+\frac{t}{\hbar}\sum_n\rho_{nn}\langle n|\{N_\mathbf{k},\mathrm{Im}(\hat{U})\}|n\rangle.
\end{split}
\label{first_order_deltank_wrong}
\end{equation}
Before further proceeding, an important observation is that the second term above is not physical, since if we replace $N_\mathbf{k}$ with identity operator $\mathbb{I}$ (in the previous steps, we do not explicitly use the form of $N_\mathbf{k}$, so this is available), 
\begin{equation}
    \Delta\langle\mathbb{I}\rangle^{(1)}=\Delta \mathrm{Tr}(\rho)\stackrel{?}{=}\frac{2t}{\hbar}\mathrm{Im}(\hat{U}).
\end{equation}
Because the trace of the density matrix should always be one, $\Delta \mathrm{Tr}(\rho)=0$, which is inconsistent with the right-hand side of the equation. To fix the problem, we should artificially add a Lindblad jump term to Eq.~(\ref{first_order_deltank_wrong}),
\begin{equation}
\begin{split}
    &\Delta \langle N_\mathbf{k}\rangle^{(1)}=\frac{t}{i\hbar}\sum_n\rho_{nn}\langle n|[N_\mathbf{k},\mathrm{Re}(\hat{U})]|n\rangle\\
    &+\frac{t}{\hbar}\sum_n\rho_{nn}\langle n|\{N_\mathbf{k},\mathrm{Im}(\hat{U})\}|n\rangle\\
    &-\frac{2t}{\hbar}\sum_{n,\mathbf{P},m}\rho_{nn}\langle n|\hat{L}_{\mathbf{P},m}^\dagger N_\mathbf{k}\hat{L}_{\mathbf{P},m}|n\rangle,
\end{split}
\label{first_order_deltank}
\end{equation}
where $\sum_{\mathbf{P},m} L^\dagger_{\mathbf{P},m} L_{\mathbf{P},m}=\mathrm{Im}(\hat{U})$. One can check that if $N_\mathbf{k}$ is replaced by identity operator $\mathbb{I}$ in Eq.~(\ref{first_order_deltank}), the trace of the density matrix is conserved. To find the form of $L_{\mathbf{P},m}$, we use the addition theorem of Legendre polynomial,
\begin{equation}
    \mathbf{q}\cdot \mathbf{q}'=qq'P_1(\hat{\mathbf{q}}\cdot \hat{\mathbf{q}}')=\frac{4\pi}{3}\sum_{m=-1}^l{Y}_{1m}(\hat{\mathbf{q}}){Y}^*_{1m}(\hat{\mathbf{q}'}),
\end{equation}
where ${Y}_{lm}$ are the spherical harmonics. Combining with Eq.~(\ref{Hamiltonian}), we read off that the explicit form of $L_{\mathbf{P},m}$ is
\begin{equation}
    \hat{L}_{\mathbf{P},m}=\sum_{\mathbf{q}}2\sqrt{\pi\mathrm{Im}(g)} q Y_{1m}(\hat{\mathbf{q}})c_{\frac{\mathbf{P}}{2}-\mathbf{q}} c_{\frac{\mathbf{P}}{2}+\mathbf{q}}.
\end{equation}
Then, a straightforward while lengthy calculation gives
\begin{equation}
\begin{split}
    \Delta \langle N_\mathbf{k}\rangle^{(1)}=&-\frac{6t}{\hbar V}\sum_n\rho_{nn}\sum_{\mathbf{P},\mathbf{q}}\left(\mathbf{q}-\frac{\mathbf{P}}{2}\right)\cdot\left(\mathbf{k}-\frac{\mathbf{P}}{2}\right)\\
    &\times\big[\mathrm{Re}(g)\mathrm{Im}\langle n| c^\dagger_{\mathbf{k}} c^\dagger_{\mathbf{P}-\mathbf{k}}c_{\mathbf{q}} c_{\mathbf{P}-\mathbf{q}}|n \rangle\\
    &+\mathrm{Im}(g)\mathrm{Re}\langle n| c^\dagger_{\mathbf{k}} c^\dagger_{\mathbf{P}-\mathbf{k}}c_{\mathbf{q}} c_{\mathbf{P}-\mathbf{q}}|n \rangle\big].
\end{split}
\label{first_order_deltank_2}
\end{equation}
Recalling that we choose $| n \rangle$ to represent Fock states, the expectation can be evaluated following the canonical anti-commutation relation,
\begin{equation}
\begin{split}
    &\langle n| c^\dagger_{\mathbf{k}} c^\dagger_{\mathbf{P}-\mathbf{k}}c_{\mathbf{q}} c_{\mathbf{P}-\mathbf{q}}|n \rangle \\
    &= (1-\delta_{\mathbf{q},\mathbf{P}/2})(\delta_{\mathbf{q}+\mathbf{k},\mathbf{P}}-\delta_{\mathbf{q},\mathbf{k}})N_{\mathbf{k}}(n)N_{\mathbf{P}-\mathbf{k}}(n)\\
    &+\delta_{\mathbf{q},\mathbf{P}/2}\delta_{\mathbf{k},\mathbf{P}/2}N_{\mathbf{k}}(n)(N_{\mathbf{k}}(n)-1),
\end{split}
\label{factorization}
\end{equation}
where $N_\mathbf{k}(n)$ signifies occupation number of particle with momentum $\mathbf{k}$ of state $| n \rangle$. Substituting the above result to Eq.~(\ref{first_order_deltank_2}), we have
\begin{equation}
\begin{split}
    &\frac{\Delta \langle N_\mathbf{k}^{(1)}\rangle}{t}\xrightarrow[]{t\rightarrow0}\frac{d \langle N_\mathbf{k}\rangle^{(1)}}{dt}\\
    &=-\frac{6\mathrm{Im}(g)}{\hbar V}\sum_n \rho_{nn} \sum_{\mathbf{P},\mathbf{q}}\left(\mathbf{q}-\frac{\mathbf{P}}{2}\right)\cdot\left(\mathbf{k}-\frac{\mathbf{P}}{2}\right)\\
    &\times(\delta_{\mathbf{q}+\mathbf{k},\mathbf{P}}-\delta_{\mathbf{q},\mathbf{k}})N_{\mathbf{k}}(n)N_{\mathbf{P}-\mathbf{k}}(n).
\end{split}
\end{equation}
Changing variable $\mathbf{P}\rightarrow\mathbf{q}'+\mathbf{k}$, one observes that two delta function contribute equally,
\begin{equation}
    \frac{d \langle N_\mathbf{k}\rangle^{(1)}}{dt}=\frac{3\mathrm{Im}(g)}{\hbar V}\sum_n \rho_{nn} \sum_{\mathbf{q}}\left({\mathbf{q}-\mathbf{k}}\right)^2N_{\mathbf{k}}(n)N_{\mathbf{q}}(n).
\end{equation}
Because we assume that the interaction of the system is weak, to the leading order, the expectation of $N_\mathbf{k}$ and $N_\mathbf{q}$ can be factorized~\cite{snoke2012basis},
\begin{equation}
\begin{split}
   \frac{d \langle N_\mathbf{k}\rangle^{(1)}}{dt}&= \frac{3\mathrm{Im}(g)}{\hbar V}\sum_{n,n'} \rho_{nn} \rho_{n'n'}\sum_{\mathbf{q}}\left({\mathbf{q}-\mathbf{k}}\right)^2N_{\mathbf{k}}(n)N_{\mathbf{q}}(n')\\
   &=\frac{3\mathrm{Im}(g)}{\hbar}\int \frac{d^3q}{(2\pi)^3}(\mathbf{q}-\mathbf{k})^2 \langle N_\mathbf{k} \rangle \langle N_\mathbf{q}\rangle.
   \label{first_order_result}
\end{split}
\end{equation}
It is noted that the right-hand side of Eq.~(\ref{first_order_result}) has already given a part of $\mathcal{I}_\mathrm{coll}[V n_\mathbf{k}]/V$. Utilizing the spherical symmetry of $\langle N_\mathbf{k} \rangle$ (it has the symmetry because the interaction $\hat{U}$ here is spherical symmetric) and $\langle N_\mathbf{k}\rangle=V n_\mathbf{k}$,
\begin{equation}
\begin{split}
    &\mathcal{I}_\mathrm{coll}[V n_\mathbf{k}]/V=\mathcal{I}_\mathrm{inel}[V n_\mathbf{k}]/V+\cdots\\
    &=\frac{12\pi\hbar\mathrm{Im}(v_p)V}{M}\int\frac{d^3q}{(2\pi)^3}(q^2+k^2) n_\mathbf{k} n_\mathbf{q}+\cdots.
\end{split}
\label{Iinel}
\end{equation}
One observes that this part of the collision integral neither has a constraint on momentum/energy conservation nor a parameter related to $\mathrm{Re}(v_p)$; thus, it is not a conventional collision integral characterizing the strength of elastic collision, but a part describing the momentum dependence of two-body chemical reaction. We name this part of $\mathcal{I}_\mathrm{coll}$ to be $\mathcal{I}_\mathrm{inel}$, the inelastic collision integral. An equivalence of Eq.~(\ref{Iinel}) has first been obtained by Ref.~\cite{he2020exploring} using a two-channel model. Here, we derive Eq.~(\ref{Iinel}) using our one-channel model differently to make symbols self-contained. There are even more different methods for arriving at an inelastic collision integral in a two-body dissipative system. For example, Ref.~\cite{pan2020nonhermitian} and Ref.~\cite{huang2023modeling} reach similar equations using linear response theory and Keldysh field theory, respectively.

We find that up to the first-order expansion, only inelastic collision is considered. One needs to go to the second-order expansion to account for elastic collision. Since we assume interaction and reaction are weak, we ignore all inelastic collision contributions from the second-order process. Consequently, when consider the second-order process, we take $\hat{U}(t)=\hat{U}(t)^\dagger=\mathrm{Re}(\hat{U}(t))$. Ref.~\cite{snoke2012basis} argues that only $\Delta\langle N_\mathbf{k}\rangle^{(2)}_1$ contributes to the collision integral. With the simplification above,
\begin{equation}
\begin{split}
    &\Delta\langle N_\mathbf{k}\rangle^{(2)}=\Delta\langle N_\mathbf{k}\rangle^{(2)}_1=\sum_n \frac{\rho_{nn}}{\hbar^2}\\ &\times\langle n | \int_0^t \int_0^t dt' dt''\mathrm{Re}(\hat{U}(t))[N_\mathbf{k},\mathrm{Re}(\hat{U}(t))]| n \rangle\\
    &=\sum_{n,l} \rho_{nn}\frac{\sin^2((E_n-E_l)t/2\hbar)}{(E_n-E_l)^2}\langle n |\mathrm{Re}(\hat{U})| l \rangle \\
    &\times\langle l |[N_\mathbf{k},\mathrm{Re}(\hat{U})]| n \rangle,
\end{split}
\end{equation}
where $E_n$ and $E_l$ denote energy of state $|n\rangle$ and $|l\rangle$, respectively. Taking the Markovian limit for the above expression~\cite{snoke2012basis}, we have
\begin{equation}
\begin{split}
    \Delta\langle N_\mathbf{k}\rangle^{(2)}&=\frac{2\pi t}{\hbar}\sum_{n,l} \rho_{nn}\delta(E_n-E_l)\\
    &\times\langle n |\mathrm{Re}(\hat{U})| l \rangle \langle l |[N_\mathbf{k},\mathrm{Re}(\hat{U})]| n \rangle.
    \label{markovian_limit}
\end{split}
\end{equation}
Because both $|n\rangle$ and $|l\rangle$ are fock states, it is possible to directly evaluate expectations in Eq.~(\ref{markovian_limit}), which yields
\begin{equation}
\begin{split}
    \sum_{n,l}\rho_{nn}& \langle n |\mathrm{Re}(\hat{U})| l \rangle \langle l |[N_\mathbf{k},\mathrm{Re}(\hat{U})]| n \rangle\\
    =&\frac{9}{2}\frac{\mathrm{Re}(g)^2}{V^2}\sum_{\mathbf{P},\mathbf{Q},\mathbf{p},\mathbf{p}',\mathbf{q}}(\mathbf{p}\cdot\mathbf{p}')\left[(\frac{\mathbf{Q}}{2}-\mathbf{k})\cdot\mathbf{q}\right]\\
    &\times\big[\delta_{\mathbf{p},\mathbf{q}}\delta_{\mathbf{P},\mathbf{Q}}\delta_{\mathbf{P}/2+\mathbf{p}',\mathbf{k}}(1-\langle N_{\mathbf{Q}-\mathbf{k}}\rangle)\\
&\times(1-\langle N_{\mathbf{k}}\rangle)N_{\mathbf{Q}/2-\mathbf{q}}N_{\mathbf{Q}/2+\mathbf{q}}\\
    &-\delta_{\mathbf{p}',\mathbf{q}}\delta_{\mathbf{P},\mathbf{Q}}\delta_{\mathbf{P}/2-\mathbf{p},\mathbf{k}}(1-\langle N_{\mathbf{Q}/2+\mathbf{q}}\rangle)\\
    &\times(1-\langle N_{\mathbf{Q}/2-\mathbf{q}}\rangle)N_{\mathbf{k}}N_{\mathbf{Q}-\mathbf{k}}\big].
\end{split}
\label{second_order_expectation}
\end{equation}
Again, we factorize expectations of the number of particles with different momentum as what is done in Eq.~(\ref{factorization}). Combining Eqs.~(\ref{markovian_limit}) and (\ref{second_order_expectation}), we have
\begin{equation}
\begin{split}
    &\frac{\Delta\langle N_\mathbf{k}\rangle^{(2)}}{t}\xrightarrow[]{t\rightarrow0}\frac{d \langle N_\mathbf{k}\rangle^{(2)}}{dt}\\
    &=\frac{18\pi M}{\hbar^3}\frac{\mathrm{Re}(g)^2}{V^2} \sum_{\mathbf{Q},\mathbf{q}}\left(\frac{\mathbf{Q}}{2}-\mathbf{k}\right)^2\left(\frac{\mathbf{Q}}{2}-\mathbf{q}\right)^2\\
    &\times\delta\left(k^2-q^2+\mathbf{k}\cdot\mathbf{Q}-\mathbf{q}\cdot\mathbf{Q}\right)[(1-\langle N_{\mathbf{Q}-\mathbf{k}}\rangle)(1-\langle N_{\mathbf{k}}\rangle)\\
    &\langle N_{\mathbf{q}}\rangle \langle N_{\mathbf{Q}-\mathbf{q}}\rangle-\langle N_{\mathbf{Q}-\mathbf{k}}\rangle \langle N_{\mathbf{k}}\rangle (1-\langle N_{\mathbf{q}}\rangle)(1-\langle N_{\mathbf{Q}-\mathbf{q}}\rangle)].
\end{split}
\label{second_order_result}
\end{equation}
Equation~(\ref{second_order_result}) contains the terms describing elastic collision, making $\mathcal{I}_\mathrm{coll}[V n_\mathbf{k}]/V=\mathcal{I}_\mathrm{inel}[V n_\mathbf{k}]/V+\mathcal{I}_\mathrm{el}[V n_\mathbf{k}]/V$, where $\mathcal{I}_\mathrm{el}$ is the elastic collision integral, in contrast to the inelastic contribution. Explicitly, the form is given by
\begin{equation}
\begin{split}
    &\mathcal{I}_\mathrm{el}[V n_\mathbf{k}]/V=\frac{288\pi^3\hbar \mathrm{Re}(v_p)^2V}{M}\int\frac{d^3q}{(2\pi)^3}\int\frac{d^3Q}{(2\pi)^3}\\
    &\times\delta\left(k^2-q^2+\mathbf{k}\cdot\mathbf{Q}-\mathbf{q}\cdot\mathbf{Q}\right)\left(\frac{\mathbf{Q}}{2}-\mathbf{k}\right)^2\left(\frac{\mathbf{Q}}{2}-\mathbf{q}\right)^2\\
    &\times\big[(1-n_{\mathbf{Q}-\mathbf{k}}V)(1-n_\mathbf{k}V)n_\mathbf{q}n_{\mathbf{Q}-\mathbf{q}}\\
    &-n_{\mathbf{Q}-\mathbf{k}}n_\mathbf{k}(1-n_\mathbf{q}V)(1-n_{\mathbf{Q}-\mathbf{q}}V)\big].
\end{split}
\end{equation}

\section{Relaxation Time of $p$-wave Elastic Collision}
\label{relax_tau}

In the introduction, we state that recent experimental systems exhibiting only $p$-wave elastic collisions struggle to rethermalize from a non-equilibrium state. This observation is crucial, as it allows us to safely disregard the elastic collision integral when analyzing IQBE in subsequent discussions. Before validating this statement, it is essential to distinguish between two key assumptions in this work: 1) the system's interaction is weak, and 2) the system can hardly rethermalize. The first assumption implies that the interaction effect can be treated as a perturbation, i.e., $\operatorname{Re}(v_p) k_F^3\ll1$, where $k_F$ is the system's Fermi momentum (divided by $\hbar$). The second assumption means that the relaxation time of elastic collisions is significantly longer than other time scales in the system. It is worth noting that these two assumptions are not inherently related. This section will use typical experimental data to support both assumptions.

We focus on two experimental realizations mentioned in the introduction: rovibrational ground state ${}^{40}\mathrm{K}{}^{87}\mathrm{Rb}$ and ${}^{23}\mathrm{Na}{}^{40}\mathrm{K}$ molecular gases. The real parts of their bare scattering volumes $v_p$ have been determined to be $(118a_0)^3$\cite{idziaszek2010universal} and $(88a_0)^3$\cite{duda2023longlived}, respectively. The Fermi momentum of a harmonically trapped system is defined by $k_F=(48N)^{1/6}\sqrt{M\Bar{\omega}/\hbar}$~\cite{butts1997trapped}. For a conservative estimate, we deliberately overestimate the typical number of particles $N$ in experiments to be $10^5$ and assume the geometric mean of the harmonic trap's angular frequency $\Bar{\omega}$ to be approximately $2\pi\times100~\text{Hz}$. Using these values, we calculate $\operatorname{Re}(v_p)k_F^3$ for both systems, yielding approximately $7.5\times10^{-4}$ and $1.1\times10^{-4}$, respectively. These values are significantly less than 1, supporting the weak interaction assumption.

To estimate the relaxation time, we begin with a simple derivation. We define the $p$-wave scattering volume from the phase shift $\delta_p$, related to the scattering matrix $S$ by:
\begin{equation}
S=e^{2i\delta_p}\approx1+2i\delta_p.
\end{equation}
The expansion is valid because we have already shown that the interaction is weak enough: the phase shift should also be small. The elastic cross-section is then given by:
\begin{equation}
\sigma_{\mathrm{el}}=\frac{3\pi}{k^2}|1-S|^2\approx 12\pi[\mathrm{Re}(v_p)]^2(k_r)^4,
\end{equation}
where $k_r$ is the relative momentum between two scatters in a two-body collision. Defining the scattering energy $E=\hbar^2k_r^2/M$, we can express $\sigma_{\mathrm{el}}$ as:
\begin{equation}
\sigma_{\mathrm{el}}(E)=12\pi[\mathrm{Re}(v_p)]^2\frac{M^2E^2}{\hbar^4}.
\end{equation}
Assuming the trapped system is not in deep degeneracy, we can approximate the phase space density of the cloud with a simple Gaussian distribution:
\begin{equation}
f(\mathbf{k},\mathbf{r})=z\exp\left(-\frac{\hbar^2k^2}{2M k_B T}\right)\exp\left(-\frac{M\sum_{i}\omega_i^2r_i^2}{2 k_B T}\right),
\label{PSD}
\end{equation}
where $i$ represents $x$, $y$, and $z$ directions of the harmonic trap. The fugacity $z={T_F^3}/({6T^3})$, with $T_F=\hbar^2k_F^2/2M$, is determined by the normalization condition:
\begin{equation}
N=\int \frac{d^3k}{(2\pi)^3}d^3r f(\mathbf{k},\mathbf{r};z).
\end{equation}
Integrating the real space dependence in $f$, we obtain the momentum distribution:
\begin{equation}
f(\mathbf{k})=z \left(\frac{2 \pi k_B T}{M \omega^2}\right)^{3/2}\exp\left(-\frac{\hbar^2k^2}{2M k_B T}\right).
\end{equation}
For two colliding particles with momenta $\mathbf{k}_1$ and $\mathbf{k}_2$, the distribution is:
\begin{equation}
f(\mathbf{k}_1)f(\mathbf{k}_2)=f(\mathbf{k}_r)f(\mathbf{k}_R),
\end{equation}
where $\mathbf{k}_r=(\mathbf{k}_1-\mathbf{k}_2)/2$ and $\mathbf{k}_R=\mathbf{k}_1+\mathbf{k}_2$ are relative and center-of-mass momenta. Changing the argument of $f(\mathbf{k}_r)$ to $E$:
\begin{equation}
f(\mathbf{k}_r)=f(E)=z \left(\frac{2 \pi k_B T}{M \omega^2}\right)^{3/2}\exp\left(-\frac{E}{k_BT}\right).
\end{equation}
We then calculate:
\begin{equation}
\begin{split}
\langle\sigma_{\mathrm{el}}v_r\rangle&=\frac{1}{N}\int \frac{d^3 k_r}{(2\pi)^3} f(E) \sigma_{\mathrm{el}}(E) v_r(E)\\
&=1152\sqrt{2\pi}M^{3/2}[\mathrm{Re}(v_p)]^2{(k_B T)^{5/2}}/{\hbar^4},
\label{expectation_sigmav}
\end{split}
\end{equation}
where $v_r$ is the relative speed of colliding particles, defined by $Mv_r^2/4=E$. From Eq.~(\ref{PSD}), we obtain the in-situ average density:
\begin{equation}
\begin{split}
f(\mathbf{r})&=\int \frac{d^3k}{(2\pi)^3} f(\mathbf{k},\mathbf{r}),\\
\langle f(\mathbf{r})\rangle&=\dfrac{\int d^3r f(\mathbf{r})^2}{\int d^3r f(\mathbf{r})}=\frac{N}{8\pi^{3/2}}\Bar{\omega}^3\left(\frac{k_B T}{M}\right)^{-3/2}.
\end{split}
\label{expectation_fr}
\end{equation}
Combining Eqs.~(\ref{expectation_sigmav}) and (\ref{expectation_fr}), we derive the time scale for one elastic collision event:
\begin{equation}
t_{\mathrm{el}}=\dfrac{1}{\langle\sigma_{\mathrm{el}}v_r\rangle\langle f(\mathbf{r})\rangle}=\frac{2^{19/6}\pi}{3^{4/3}N^{1/3}[\mathrm{Re}({v}_p)k_F^3]^2(T/T_F)\Bar{\omega}}.
\label{tel}
\end{equation}
The system's relaxation time should be $\alpha t_{\mathrm{el}}$, where $\alpha$ is typically larger than 1, indicating that each particle should collide more than once on average to reach equilibrium~\cite{wu1996direct,monroe1993measurement}. For our purposes, the exact value of $\alpha$ is less critical, as a sufficiently long $t_\mathrm{el}$ implies an even longer relaxation time.

Using the same parameters as in our previous discussion of $\operatorname{Re}(v_p)k_F^3$ and setting $T=T_F$, we calculate $t_\mathrm{el}$ for the two systems under consideration. The results are approximately $6~\text{min}$ for ${}^{40}\mathrm{K}{}^{87}\mathrm{Rb}$ and $5~\text{hours}$ for ${}^{23}\mathrm{Na}{}^{40}\mathrm{K}$ molecular gases. Given that typical experimental durations are on the order of several seconds, these relaxation times for $p$-wave elastic collisions are significantly longer than other relevant time scales in these systems. This substantial difference in time scales supports our earlier assumption that the systems can hardly rethermalize through elastic collisions alone.

\section{Dynamics of Homogeneous Systems}

In this section, we discuss solutions to the IQBE derived in Sec.~\ref{IQBE}, focusing on homogeneous systems without external potentials. As explained in Sec~\ref{relax_tau}, it is reasonable to ignore the elastic collision integral when modeling realistic cases. Therefore, we analyze the IQBE without elastic collisions. Then, we introduce effective temperatures to assess whether systems heat or cool after dissipation. We also compare our analytical approach with solutions obtained from the thermal ansatz, which assumes continuous thermalization during system dynamics.

For clarity, we employ dimensionless quantities throughout this section, denoted by a bar hat. Our chosen unit system is based on initial Fermi momentum $k_F(0)=(6\pi^2N(0)/V)^{1/3}$, temperature $T_F(0)=\hbar^2 k_F(0)^2/2 M k_B$, and energy $E_F(0)=k_B T_F(0)$:
\begin{align*}
&\text{momentum: } k\rightarrow \Bar{k} k_F(0), \\
&\text{temperature: } T\rightarrow\Bar{T} T_F(0),\\
&\text{time: }t\rightarrow\Bar{t}\hbar/E_F(0)
\end{align*}
It is important to note that we consistently use the \textit{initial} particle number $N(0)$ rather than 
instantaneous particle number $N(t)$ as a reference for our units, hence the argument $0$. This distinction becomes crucial when discussing systems using effective temperatures, where the effective Fermi temperature $T_F(t)$ varies with time. To avoid confusion, our unit system remains constant, i.e., generally, $\Bar{T}_F(t)\neq1$.

\subsection{Mellin Space Analysis}
\label{mellin_analysis}

Ignoring the elastic collision integral, the dimensionless form of IQBE Eq.~(\ref{general_homo_IQBE}) is
\begin{equation}
    \frac{d N(\Bar{k},\tau)}{d \tau}=-N(\Bar{k},\tau)\int d\Bar{q}(\Bar{k}^2+\Bar{q}^2)N(\Bar{q},\tau). 
    \label{homo_boltzmann_eq_dimensionless}
\end{equation}
where 
\begin{equation}
    \tau=-24\pi\mathrm{Im}(\bar{v}_p)\bar{t}=-12\pi\frac{\hbar k_F^5(0)}{M}\mathrm{Im}(v_p)t
\end{equation}
is defined for convenience. $N(\bar{k},\tau)$ is defined to be 
\begin{align*}
N(\bar{k},\tau)=\frac{4\pi\Bar{k}^2 V n_{\bar{\mathbf{k}}}(\tau)}{(2\pi)^3},
\end{align*}
which can be regarded as a 1D projection of the momentum distribution. If we set the system to be thermalized initially, then the initial condition of $N(\bar{k},\tau)$ is 
\begin{equation}
    N(\bar{k},0)=\frac{3\Bar{k}^2}{\exp(\frac{\Bar{k}^2}{\Bar{T}})z^{-1}+1},z=-\mathrm{Li}_{\frac{3}{2}}^{-1}\left(\frac{-4}{3\sqrt{\pi}\Bar{T}^{3/2}}\right),
    \label{Nktau_initial_condition}
\end{equation}
where $\mathrm{Li}$ denotes the polylogarithm function and ${}^{-1}$ signifies an inverse function. The Mellin transform of a function $g(x)$ is defined by
\begin{equation}
    \mathcal{M}[g(x)](s)=\int_0^\infty dx x^{s-1} g(x).
\end{equation}
We observe that Eq.~(\ref{homo_boltzmann_eq_dimensionless}) has a simpler form in Mellin space (of dimensionless momentum $\Bar{k}$):
\begin{equation}
\begin{split}
    &\frac{d \mathcal{M}[N(\bar{k},\tau)](s)}{d\tau}=-\mathcal{M}[N(\bar{k},\tau)](s)\mathcal{M}[N(\bar{k},\tau)](3)\\
    &~~~~~~~~-\mathcal{M}[N(\bar{k},\tau)](s+2)\mathcal{M}[N(\bar{k},\tau)](1).
\end{split}
\end{equation}
Let us define $F_j=\mathcal{M}[N(\bar{k},\tau)](2j+1)$, the above equation becomes:
\begin{equation}
    \frac{d F_j(\tau)}{d\tau}=-F_j(\tau) F_1(\tau)-F_{j+1}(\tau)F_0(\tau).
    \label{F_eqns}
\end{equation}
Notably, the above equation's $F_0$ and $F_1$ are particularly important. It is easy to check that they represent the normalized total number of particles $N(\tau)/N(0)$ and total energy $N(\tau)\Bar{E}(\tau)/N(0)$, respectively. Combining the definition of $F_j$ and Eq.~(\ref{Nktau_initial_condition}), at $\tau=0$, the initial conditions of $F_j$ are given by
\begin{equation}
    F_j(0)=-\frac{3}{2}\Bar{T}^{\frac{3}{2}+j}\Gamma\left(\frac{3}{2}+j\right)\mathrm{Li}_{\frac{3}{2}+j}(-z).
    \label{Fj0_initial_condition}
\end{equation}

\subsubsection{High Initial Temperature Exact Solution}
\label{high_T_exact}

In the high-initial-temperature limit $z\rightarrow0$, Eq.~(\ref{F_eqns}) can be analytically solved. First, it is easy to find two crucial properties of high-order derivatives of $F_j(\tau)$ using Eq.~(\ref{F_eqns}): (1) the $n$-th order derivative of $F_j$ is expressible as a sum of products of $n+1$ $F$ terms, and (2) for each term, the sum of the indices of $F$ equals $j+n$. Summarized, the relationship is:
\begin{equation}
\begin{split}
    \frac{d^n F_j}{d\tau^n}=(-1)^n&\sum_{\substack{\{s_1,s_2,\cdots,s_{n+1}\}\\ \sum_i s_i=j+n}} C[j;\{s_1,s_2,\cdots,s_{n+1}\}]\\
    &\times F_{s_1}F_{s_2}\cdots F_{s_{n+1}}.
    \label{dnFdtn_eqn}
\end{split}
\end{equation}
The index $i$ varies from $1$ to $n+1$, and $C[j;\{s_1,s_2,\cdots,s_{n+1}\}]$ denotes the count of terms corresponding to a set of indices $s_i$. Figuring a general form for $C[j;\{s_1,s_2,\cdots,s_{n+1}\}]$ is generally challenging. However, with the high-temperature limit, the difficulty can be avoided. We do further one order of derivative to Eq.~(\ref{dnFdtn_eqn}):
\begin{equation}
    \begin{split}
        \frac{d^{n+1}}{d\tau^{n+1}}{F}_j&=(-1)^n\sum_{\substack{\{s_1,s_2,\cdots,s_{n+1}\}\\ \sum_i s_i=j+n}}C[j;\{s_1,s_2,\cdots,s_{n+1}\}]\\&\times\dfrac{\frac{d}{d\tau}({F}_{s_1}{F}_{s_2}\cdots{F}_{s_{n+1}})}{{F}_{s_1}{F}_{s_2}\cdots{F}_{s_{n+1}}}{F}_{s_1}{F}_{s_2}\cdots{F}_{s_{n+1}}.
    \end{split}
    \label{dnp1Fdtnp1}
\end{equation}
Using the general expression of Eq.~(\ref{F_eqns}),
\begin{equation}
    \dfrac{\frac{d}{d\tau}({F}_{s_1}{F}_{s_2}\cdots{F}_{s_{n+1}})}{{F}_{s_1}{F}_{s_2}\cdots{F}_{s_{n+1}}}=-\sum_{k=1}^{n+1}\sum_{r=\{0,1\}}\dfrac{{F}_r{F}_{s_k+1-r}}{{F}_{s_k}}.
    \label{product_derivative}
\end{equation}
At high-temperature regime $\Bar{T}\gg1$, using the property of polylogarithm functions~\cite{wood1992computation}
\begin{equation}
    -\mathrm{Li}_s(-z)\xrightarrow[]{z\rightarrow0}z,
    \label{polylog_highT}
\end{equation}
the fugacity in Eq.~(\ref{Nktau_initial_condition}) reduces to
\begin{equation}
    z\xrightarrow[]{\Bar{T}\rightarrow\infty}-\frac{4}{3\sqrt{\pi}\Bar{T}^{3/2}}=\frac{1}{\Gamma(5/2)\Bar{T}^{3/2}}.
\end{equation}
Then, the initial condition Eq.~(\ref{Fj0_initial_condition}) has the asymptote
\begin{equation}
    {F}_j(0)\xrightarrow[]{\Bar{T}\rightarrow\infty}\frac{3 \bar{T}^j}{2}\frac{\Gamma(\frac{3}{2}+j)}{\Gamma(\frac{5}{2})}.
    \label{Fj0highT}
\end{equation}
Substituting Eq.~(\ref{Fj0highT}) into Eq.~(\ref{product_derivative}) and taking $\tau=0$,
\begin{equation}
    \left.\dfrac{\frac{d}{d\tau}({F}_{s_1}{F}_{s_2}\cdots{F}_{s_{n+1}})}{{F}_{s_1}{F}_{s_2}\cdots{F}_{s_{n+1}}}\right|_{\tau=0}=-\bar{T}\sum_{k=1}^{n+1}(3+s_k).
    \label{dFsFs}
\end{equation}
Substituting back to Eq.~(\ref{dnp1Fdtnp1}), 
\begin{equation}
    \begin{split}
        &\left.\frac{d^{n+1}}{d\tau^{n+1}}{F}_j\right|_{\tau=0}=(-1)^n\sum_{\substack{\{s_1,s_2,\cdots,s_{n+1}\}\\ \sum_i s_i=j+n}}C[j;\{s_1,s_2,\cdots,s_{n+1}\}]\\
        &~~~~~~~~\times({F}_{s_1}{F}_{s_2}\cdots{F}_{s_{n+1}})\left[-\bar{T}\left(3(n+1)+\sum_{i=1}^{n+1}s_i\right)\right].
    \end{split}
\end{equation}
Because of the constraint $\sum_i s_i=j+n$ on the outermost summation, 
\begin{equation}
\begin{split}
    &\left.\frac{d^{n+1}}{d\tau^{n+1}}{F}_j\right|_{\tau=0}=(-1)^n\sum_{\substack{\{s_1,s_2,\cdots,s_{n+1}\}\\ \sum_i s_i=j+n}}C[j;\{s_1,s_2,\cdots,s_{n+1}\}]\\
    &~~~~~~~~\times({F}_{s_1}{F}_{s_2}\cdots{F}_{s_{n+1}})\left[-\bar{T}(3(n+1)+(n+j))\right]\\
    &~~~~~~~~=-\bar{T}(3(n+1)+(n+j))\left.\frac{d^{n}}{d\tau^{n}}{F}_j\right|_{\tau=0}.
\end{split}
\end{equation}
The recursion relation can be explicitly solved as
\begin{equation}
\begin{split}
    \left.\frac{d^{n}}{d\tau^{n}}{F}_j\right|_{\tau=0}&=F_j(0)(-\Bar{T})^n\prod_{n'=1}^{n}(4n'-1+j)\\
    &=F_j(0)4^{n}(-\Bar{T})^n\left(\frac{3+j}{4}\right)_{n},
\end{split}
\end{equation}
where $(a)_x=\Gamma(a+x)/\Gamma(a)$ is the Pochhammer symbol. Then, we find that $F_j(\tau)$ can be identified as the generalized hypergeometric function ${}_1F_{0}$,
\begin{equation}
\begin{split}
    F_j(\tau)&=\sum_{n=0}^{\infty}\frac{\tau^n}{n!}\left.\frac{d^{n}}{d\tau^{n}}{F}_j\right|_{\tau=0}\\
    &=F_j(0)\sum_{n=0}^\infty\frac{(-4\Bar{T}\tau)^n}{n!}\left(\frac{3+j}{4}\right)_{n}\\
    &=F_j(0){}_1F_0\left(\frac{3+j}{4};;-4\Bar{T}\tau\right).
\end{split}
\end{equation}
In fact, ${}_1F_{0}$ is nothing but a simple function:
\begin{equation}
    F_j(\tau)=\frac{F_j(0)}{(1+4\Bar{T}\tau)^{{3}/{4}+{j}/{4}}}.
    \label{homo_noneq_high_temperature_Fj}
\end{equation}
Especially, for particle number dynamics $F_0(\tau)=N(\tau)/N(0)$,
\begin{equation}
    \frac{N(\tau)}{N(0)}=\frac{1}{(1+4\Bar{T}\tau)^{3/4}}.
    \label{homo_noneq_high_temperature_NoN0}
\end{equation}

Equation~(\ref{homo_noneq_high_temperature_NoN0}) reveals an unexpected characteristic of the system: contrary to conventional $\mathcal{N}$-body recombination theory~\cite{mehta2009general}, the dissipation dynamics here do not follow a typical ``two-body" loss pattern. The governing equation for an $\mathcal{N}$-body loss is generally expressed as:
\begin{equation}
\frac{d}{d\tau}\left(\frac{N(\tau)}{N(0)}\right)=-K_\mathcal{N}\left(\frac{N(\tau)}{N(0)}\right)^\mathcal{N},
\label{N_body_decay_ansatz}
\end{equation}
where $K_\mathcal{N}$ represents the constant $\mathcal{N}$-body recombination rate coefficient. The general solutions to Eq.~(\ref{N_body_decay_ansatz}) are straightforward:
\begin{equation}
\frac{N(\tau)}{N(0)}=\begin{cases}
\exp (-K_1 \tau) & \text{for } \mathcal{N}=1\\
[1+(\mathcal{N}-1)K_\mathcal{N}\tau]^{-1/(\mathcal{N}-1)} & \text{for } \mathcal{N}> 1
\end{cases}.
\label{N_body_decay_ansatz_sol}
\end{equation}
By comparing Eqs.~(\ref{homo_noneq_high_temperature_NoN0}) and (\ref{N_body_decay_ansatz_sol}), we deduce that 
\begin{equation}
\mathcal{N}=\frac{7}{3},
\label{homo_scN_highT}
\end{equation}
indicating a ``fractional-body" loss process. 

In the high-initial-temperature limit, it is possible to perform the inverse Mellin transform to find the full dynamics of $N(\Bar{k},\tau)$. Equation~(\ref{homo_noneq_high_temperature_Fj}) with Mellin frequency $s$ as the argument is expressed as
\begin{equation}
    \mathcal{M}[N(\Bar{k},\tau)](s)=\frac{2}{\sqrt{\pi}}\Bar{T}^{(s-1)/2}(1+4\Bar{T}\tau)^{-(5+s)/8}\Gamma(s/2+1).
    \label{highT_Mspace_sol}
\end{equation}
We first note that
\begin{equation}
\begin{split}
    &\mathcal{M}^{-1}[(1+4\Bar{T}\tau)^{-(5+s)/8}](\Bar{k})=(1+4\Bar{T}\tau)^{-3/4}\\
    &\qquad\qquad\times\delta[\Bar{k}-(1+4\Bar{T}\tau)^{-1/8}].
\end{split}
    \label{Minv1}
\end{equation}
And from Eq.~(\ref{Nktau_initial_condition}), we know that
\begin{equation}
    \begin{split}
        \left.N(\Bar{k},0)\right|_{z\rightarrow0}=&\mathcal{M}^{-1}[\frac{2}{\sqrt{\pi}}\Bar{T}^{(s-1)/2}\Gamma(s/2+1)](\Bar{k})\\
        =&\frac{4\Bar{k}^2}{\sqrt{\pi}\Bar{T}^{3/2}}\exp\left(\frac{\Bar{k}^2}{\Bar{T}}\right).
    \end{split}
    \label{Minv2}
\end{equation}
Inverse Mellin transform has a property to convert multiplication to convolution~\cite{bateman1954tables}:
\begin{equation}
    \mathcal{M}^{-1}[f(s)g(s)]=\int_0^\infty \frac{dy}{y}\mathcal{M}^{-1}[f]\left(\frac{x}{y}\right)\mathcal{M}^{-1}[g]\left(y\right).
    \label{Minv_property}
\end{equation}
Combining Eqs.~(\ref{Minv1}), (\ref{Minv2}) and (\ref{Minv_property}), we can perform the inverse transform of Eq.~(\ref{highT_Mspace_sol}), which gives
\begin{equation}
    N(\Bar{k},\tau)=\frac{4\Bar{k}^2}{\sqrt{\pi}\Bar{T}^{3/2}(1+4\Bar{T}\tau)^{3/8}}\exp\left(-(1+4\Bar{T}\tau)^{1/4}\frac{\Bar{k}^2}{\Bar{T}}\right).
    \label{homo_noneq_Nktau_dynamics}
\end{equation}
An important observation on Eq.~(\ref{homo_noneq_Nktau_dynamics}) is that the momentum distribution is kept to be a Gaussian (the additional $\bar{k}^2$ comes from measurement $d^3\mathbf{k}$) for long-time dynamics, which motivates us to interpret it as an equilibrated profile. Indeed, we will see in the next subsection, Sec.~(\ref{Teff_and_thermal_ansatz}), Eq.~(\ref{homo_noneq_Nktau_dynamics}) has the same expression as its corresponding thermal ansatz. In other words, the system can automatically be kept in equilibrium even without any elastic collision in the high-initial-temperature limit. 

\subsubsection{Pad\'e Approximant Method for Arbitrary Initial Temperatures}

\begin{figure}[t!]
    \centering
    \includegraphics[width=0.4\textwidth]{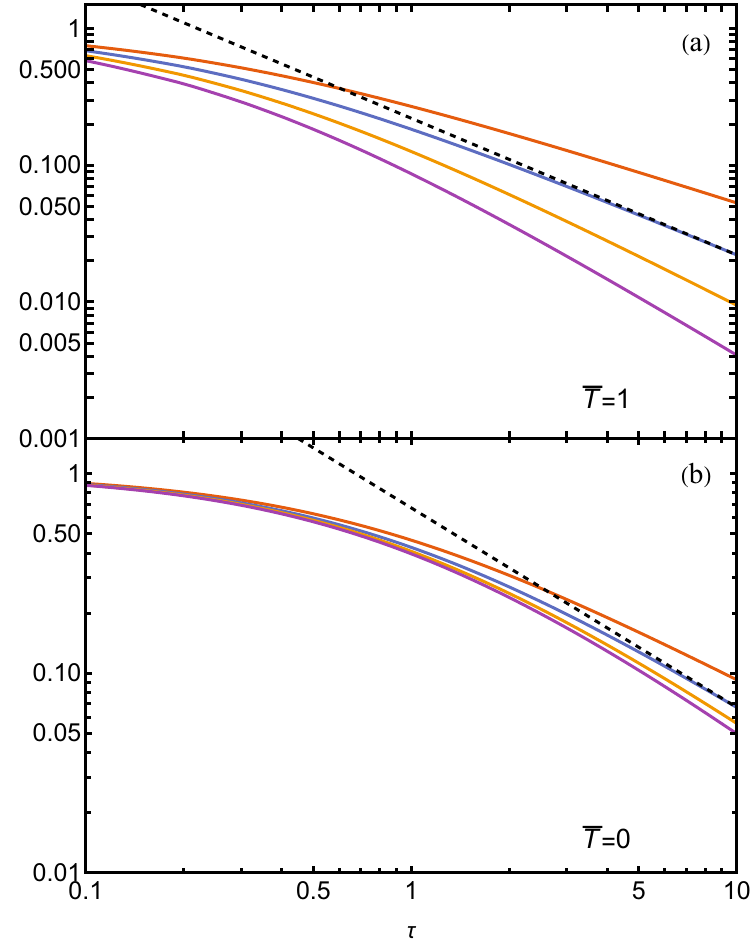}
    \caption{Log-log plot of $F_0(\tau)/F_0(0)$ to $F_3(\tau)/F_0(0)$ from top to bottom for (a) $\bar{T}=1$ and (b) $\bar{T}=0$ showing the power-law tails of different $F_n$. The black dashed lines are $1/\tau$ eye guides demonstrating the power law of $F_1$ always has $x_1=1$ in a long time.}
    \label{fig1}
\end{figure}

Generally, we conjecture all ${F}_j$ decays in the power-law manner, i.e. ${F}_j\xrightarrow[]{\tau\rightarrow\infty} A_j \tau^{-x_j}$. The conjecture is correct for high-initial-temperature exact solutions Eq.~(\ref{homo_noneq_high_temperature_Fj}). We also have numerical results in different temperatures verifying the conjecture shown in Fig.~\ref{fig1}.  Substituting this long-time ansatz into Eq.~(\ref{F_eqns}),
\begin{align}
    &x_1=1~\mathrm{and}~x_{j+1}-x_{j}=1-x_0,\label{long_time_exponential}\\
    &x_jA_j=A_j A_1+A_{j+1}A_0.\label{long_time_coefficient}
\end{align}
From Eq.~(\ref{long_time_exponential}), one sees that
\begin{equation}
    x_j=1+(j-1)(1-x_{0})
    \label{long_time_xj}
\end{equation}
is an arithmetic sequence with a common difference depending on $x_{0}$. Particularly, no matter the value of $x_{0}$, $x_{1}$ is always exactly $1$, which, again, has been verified by numerical results in Fig.~\ref{fig1}. This inspires us to construct approximation starting from ${F}_1$. A suitable functional form for ${F}_1$ is the Pad\'e approximant $[m/n]_{{F}_1}(\tau)$, which is defined as
\begin{equation}
\begin{split}
    &[m/n]_{{F}_1}(\tau)=\dfrac{\sum_{j=0}^{m}a_j \tau^j}{1+\sum_{k=0}^{n}b_k \tau^k},\\
    &\mathrm{satisfying}~\frac{d^{j}}{d\tau^{j}}[m/n]_{{F}_l}(\tau)=\frac{d^{j}}{d\tau^{j}}{F}_l(\tau)\\
    &\qquad\qquad(j=0,1,\cdots,n+m),
    \label{pade_def}
\end{split}
\end{equation}
since if $n=m+1$, $\lim_{\tau\rightarrow\infty}[m/n]_{{F}_1}(\tau)\propto1/\tau$. We expect in the limit $n=m+1\rightarrow\infty$, Eq.~(\ref{pade_def}) can reproduce ${F}_1$ in an exact form. However, due to the difficulty of counting $C[j;\{s_1,s_2,\cdots,s_{n+1}\}]$, practically we cannot construct Pad\'e approximant at a very high order.

The simplest construction is $[0/1]_{{F}_1}$, which only requires the information of ${F}_1(0)$ and $\frac{d}{d\tau}{F}_1(0)$. The latter can be directly read from Eq.~(\ref{F_eqns}), which is
\begin{equation}
    \frac{d}{d\tau}{F}_1(0)=-[{F}_1(0)]^2-{F}_2(0).
\end{equation}
Because ${1}/{(1-x)}=1+x+\mathcal{O}(x^2)$, one can directly write the $[0/1]_{{F}_1}$ to be
\begin{equation}
    {F}_1\simeq[0/1]_{{F}_1}=\cfrac{{F}_1(0)}{1+\left[{F}_1(0)+\cfrac{{F}_2(0)}{{F}_1(0)}\right]\tau}.
    \label{pade_F1}
\end{equation}

\begin{figure}[t]
    \centering
    \includegraphics[width=0.4\textwidth]{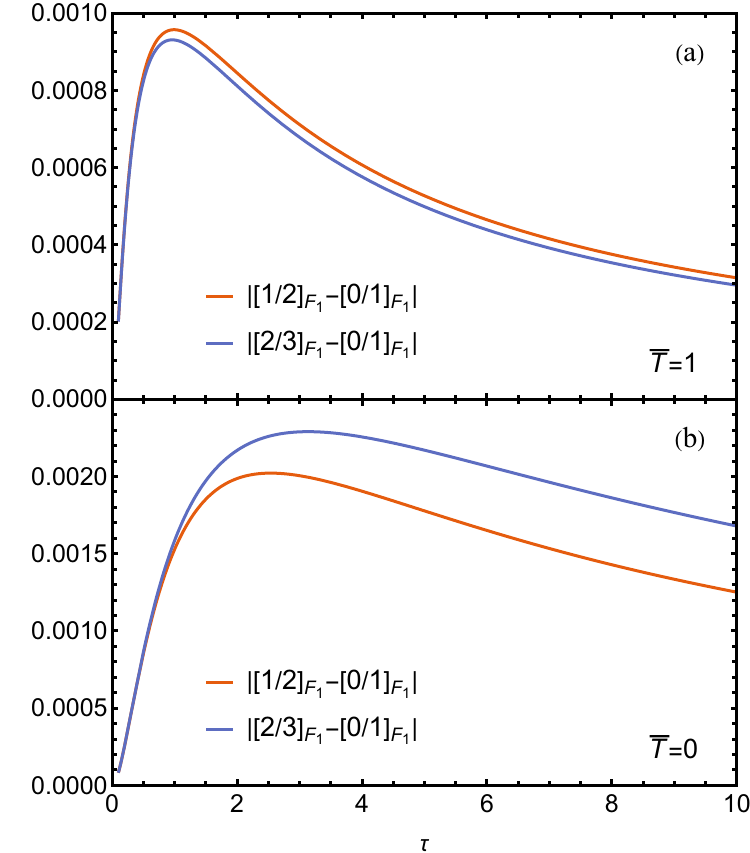}
    \caption{Two examples demonstrating how accurate lowest order Pad\'e Approximant is. (a) $\bar{T}=1$. The expressions used for three approximants are $[0/1]_{{F}_1}=\frac{1.6967}{1+4.3677\tau}$, $[1/2]_{{F}_1}=\frac{1.6967+1.0153\tau}{1+4.9661\tau+2.6396\tau^2}$ and $[2/3]_{{F}_1}=\frac{1.6967-165.80\tau-107.40\tau^2}{1-93.352\tau-490.09\tau^2-279.03\tau^3}$. (b) $\bar{T}=0$. The expressions used for three approximants are $[0/1]_{{F}_1}=\frac{0.6}{1+1.3143\tau}$, $[1/2]_{{F}_1}=\frac{0.6+0.19325\tau}{1+1.6364\tau+0.44144\tau^2}$ and $[2/3]_{{F}_1}=\frac{0.6+0.54310\tau+0.067246\tau^2}{1+2.2195\tau+1.3199\tau^2+0.15788\tau^3}$.}
    \label{fig2}
\end{figure}

\begin{figure}[t!]
    \centering
    \includegraphics[width=0.4\textwidth]{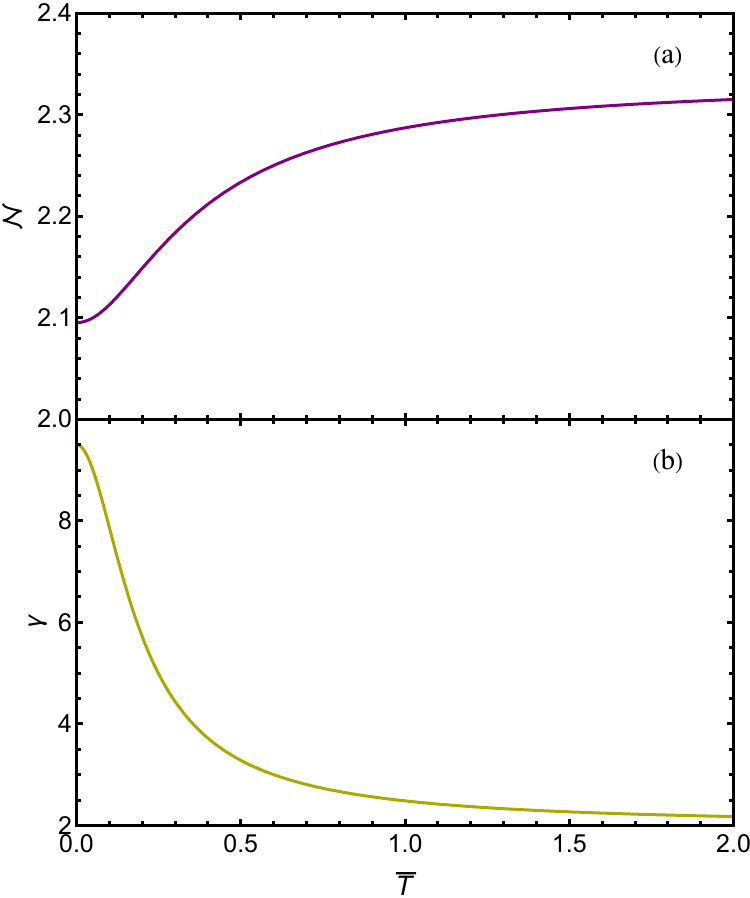}
    \caption{Initial temperature dependence of (a) $\mathcal{N}$ in Eq.~(\ref{homo_scN}) and (b) $\gamma$ in Eq.~(\ref{gamma}).}
    \label{fig3}
\end{figure}

One may suspect that Eq.~(\ref{pade_F1}) is not accurate enough, especially for a short time, since it is the lowest order approximation. However, as shown in Fig.~\ref{fig2}, higher-order Pad\'e approximants do not improve much. Therefore, we can believe Eq.~(\ref{pade_F1}) is a very reliable approximation of $F_1(\tau)$, and in the below discussion we will regard ${F}_1=[0/1]_{{F}_1}$. For $j=0$, Equation~(\ref{F_eqns}) reduces to
\begin{equation}
    \frac{d {F}_0(\tau)}{d \tau}=-2{F}_0(\tau) {F}_1(\tau),
    \label{F0_eqn}
\end{equation}
thus we immediately obtain $F_0(\tau)$ when we know $F_1(\tau)$. Substituting Eq.~(\ref{pade_F1}) into Eq.~(\ref{F0_eqn}), 
\begin{equation}
    \frac{N(\tau)}{N(0)}=F_0(\tau)=\left({1+\left[F_1(0)+\cfrac{F_2(0)}{F_1(0)}\right]\tau}\right)^{\frac{-2F_1^2(0)}{F_1^2(0)+F_2(0)}}.
    \label{pade_F0}
\end{equation}
Again, by comparing Eqs.~(\ref{pade_F0}) and (\ref{N_body_decay_ansatz_sol}), we can extract $\mathcal{N}$ for this general situation:
\begin{equation}
    \mathcal{N}=\frac{3}{2} + \frac{F_2(0)}{2F_1(0)^2}.
    \label{homo_scN}
\end{equation}
In the high-initial-temperature limit (i.e., Eq.~(\ref{Fj0highT})), as we expect, Eq.~(\ref{homo_scN}) has asymptotic value $\mathcal{N}=7/3$, which is consistent with the exact solution Eq.~(\ref{homo_scN_highT}) solved above. Furthermore, one can also check that Eq.~(\ref{pade_F0}) recovers Eq.~(\ref{homo_noneq_high_temperature_NoN0}) exactly. We can also extract the asymptotic value for the zero-initial-temperature value as well.  The asymptote of the polylogarithm function at the zero-temperature limit $z\rightarrow\infty$ is given by~\cite{wood1992computation}
\begin{equation}
    -\mathrm{Li}_s(-z)\xrightarrow[]{z\rightarrow+\infty}\frac{\ln(z)^s}{\Gamma(s+1)}.
\end{equation}
Correspondingly,
\begin{align}
    -\mathrm{Li}_{s_1}[\mathrm{Li}^{-1}_{s_2}(-x)]&\xrightarrow[]{x\rightarrow+\infty}\dfrac{[\Gamma(1+s_2)x]^{s_1/s_2}}{\Gamma(1+s_1)},\\
    {F}_j(0)&\xrightarrow[]{\Bar{T}\rightarrow0}\frac{3}{3+2j}.\label{Fj0zeroT}
\end{align}
Hence,
\begin{equation}
    \mathcal{N}\xrightarrow[]{\Bar{T}\rightarrow0}\frac{44}{21},
    \label{homo_scN_zeroT}
\end{equation}
which is still slightly larger than $2$. As shown in Fig.~\ref{fig3}(a), the two limits, Eqs.~(\ref{homo_scN_highT}) and (\ref{homo_scN_zeroT}) are smoothly connected by a monotonically decreasing $\mathcal{N}$ with lowering temperatures. Consequently, the two-body dissipation dynamics in homogeneous single-component Fermi gases over all the temperatures cannot be described by conventional two-body decay where $\mathcal{N}=2$.

Obtaining the full description of the momentum distribution dynamics is almost impossible for this arbitrary temperature scenario. The reason is that the inverse Mellin transform is generally too hard to perform. Below, we only show the infinite-time limit of momentum distribution dynamics. Practically, this is not a very relevant regime to experimental observation since, after a long time, the total number of particles left in the system will be too few to measure. However, this is still interesting from a pure theoretical aspect because long-term behavior tells whether the system can deviate from equilibrium. 

According to Pad\'e approximant Eq.~(\ref{pade_F1}) and its consequence, Eq.~(\ref{pade_F0}), we can explicitly express
\begin{equation}
    x_0=\frac{2F_1^2(0)}{F_1^2(0)+F_2(0)}.
\end{equation}
$A_0$ and $A_1$, the coefficient of decay power law in the long time limit
\begin{equation}
    A_0=\left[F_1(0)+\cfrac{F_2(0)}{F_1(0)}\right]^{\frac{-2F_1^2(0)}{F_1^2(0)+F_2(0)}},~A_1=\frac{F_1(0)^2}{F_1(0)^2+F_2(0)}.
\end{equation}
Equation~(\ref{long_time_coefficient}) provides a recursion relation of $A_j$, thus we can express a general $A_j$ with $A_0$ and $A_1$:
\begin{equation}
    A_j=(x_0-A_1)\left(\frac{A_0}{1-x_0}\right)^{1-j}\left(\frac{A_1-1}{x_0-1}\right)_{j-1}.
    \label{long_time_Aj}
\end{equation}

With Eqs.~(\ref{long_time_xj}) and (\ref{long_time_Aj}), we obtain the long time asymptote of $F_j(\tau)$, or the Mellin transform of $\left.N(\Bar{k},\tau)\right|_{\tau\rightarrow\infty}$,
\begin{equation}
\begin{split}
    &\left.\mathcal{M}[N(\Bar{k},\tau)](s)\right|_{\tau\rightarrow\infty}=(x_0-A_1)\left(\frac{A_0}{1-x_0}\right)^{\frac{3}{2}-\frac{s}{2}}\\
    &\qquad\qquad\times\left(\frac{A_1-1}{x_0-1}\right)_{\frac{s}{2}-\frac{3}{2}}\tau^{-\frac{1}{2}[(1-x_0)s+3x_0-1]}.
    \label{long_time_MNktau}
\end{split}
\end{equation}
as we note that
\begin{equation}
    \mathcal{M}[\exp(-c_1 k^2)k^{c_2}](s)=\frac{1}{2}c_1^{-\frac{s}{2}-\frac{c_2}{2}}\Gamma\left(\frac{s}{2}+\frac{c_2}{s}\right),
\end{equation}
the inverse Mellin transform of Eq.~(\ref{long_time_MNktau}) can be obtained as
\begin{equation}
\begin{split}
    &\left.N(\Bar{k},\tau)\right|_{\tau\rightarrow\infty}=\frac{2A_0^2(x_0-A_1)}{(x_0-1)^2\Gamma\left(\frac{A_1-1}{x_0-1}\right)}\tau^{1-2x_0}\\
    &\qquad\qquad\times\left(\sqrt{\frac{A_0}{1-x_0}\tau^{\frac{1}{2}-\frac{x_0}{2}}}\right)^{\frac{2(1+A_1-2x_0)}{x_0-1}}\\
    &\qquad\qquad\times\exp\left(-\frac{A_0\tau^{1-x_0}}{1-x_0}\Bar{k}^2\right)\Bar{k}^{\gamma},
\end{split}
    \label{Nktau_tauinfty}
\end{equation}
where 
\begin{equation}
    \gamma=\frac{2(A_1-1)}{x_0-1}-3.
    \label{gamma}
\end{equation}
First, it can be checked that Eq.~(\ref{homo_noneq_Nktau_dynamics}) is consistent with Eq.~(\ref{Nktau_tauinfty}) in the same limit, i.e., $\gamma=2$. However, $\gamma$ will no longer be $2$ for a lower initial temperature, making the profile clearly of a non-equilibrium shape. For example, using Eq.~(\ref{Fj0zeroT}), we find in the zero-initial-temperature limit, $\gamma=19/2$. Figure~\ref{fig3}(b) shows the general calculation result of $\gamma$. Together with the conclusion we made previously in Sec.~(\ref{high_T_exact}), the complete picture of the system's dissipation dynamics is: with a high enough initial temperature, the system can be kept under equilibrium without elastic collision, while for lower initial temperatures, generally the two-body loss drives system away from being thermalized.

\subsection{Effective Temperatures and Thermal Ansatz}
\label{Teff_and_thermal_ansatz}

\subsubsection{Dynamics of Effective Temperatures}
\label{dynamics_of_Teff}
\begin{figure}[t!]
    \centering
    \includegraphics[width=0.4\textwidth]{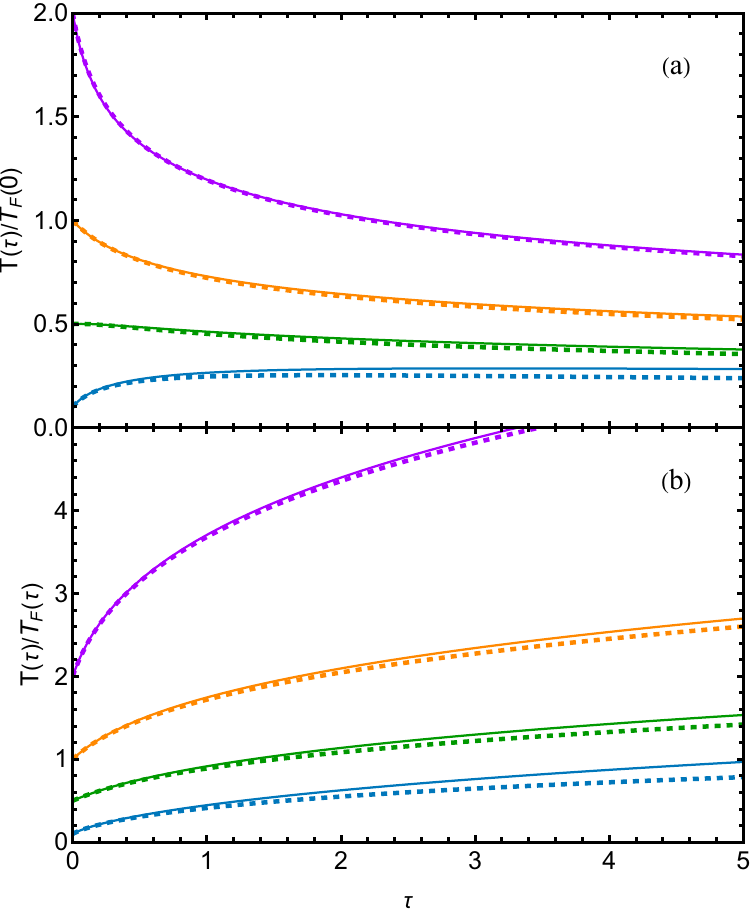}
    \caption{Dynamics of effective temperatures for systems with initial temperatures $\Bar{T}(0)=2, 1, 0.5$, and $0.1$ from above to below. Solid lines denote effective temperatures extracted from general non-equilibrium solutions of IQBE, i.e., Eqs.~(\ref{pade_F0}) and (\ref{pade_F1}). The dotted lines are the results of solving IQBE by assuming that the thermal ansatz always applies. (a) The dynamics of physical effective temperature $\Bar{T}(\tau)=T(\tau)/T_F(0)$. (b) The dynamics of reduced effective temperature $T(\tau)/T_F(\tau)$. 
    }
    \label{fig4}
\end{figure}

Two-body dissipation is a crucial experimental consideration due to its heating effect on systems, which poses a significant obstacle to cooling processes~\cite{ni2010dipolar,zhu2013evaporative}. This phenomenon, known as ``anti-evaporation," was originally studied in the context of harmonically trapped systems. The conventional explanation for this heating mechanism is as follows: the center of the harmonic trap, being much denser than the periphery, experiences a higher rate of particle loss. Since particles in this region possess lower potential energy compared to those in other areas, their loss leads to an increase in the average energy per particle, consequently raising the system's temperature.

Our investigation aims to determine whether a similar phenomenon occurs in homogeneous systems. However, since dissipative dynamics generally drive systems out of equilibrium, we must first establish appropriate effective temperatures. We define these by hypothetically thermalizing the systems for measurement purposes. Thus, a system's effective temperature $T(t)$ is defined as the temperature of a thermalized system with identical particle number and total energy. Mathematically, we describe this corresponding thermalized system using a thermal ansatz:
\begin{equation}
    N^\mathrm{th}(\Bar{k},\tau)\equiv F_0(\tau) N^*(\Bar{k},\tau),
    \label{thermal_ansatz1}
\end{equation}
where $N^*(\Bar{k},\tau)$ is the normalized thermal momentum distribution:
\begin{equation}
\begin{split}
    d\Bar{k} N^{*}(\Bar{k},\tau) &\equiv d\left(\frac{\Bar{k}}{\Bar{k}_F(\tau)}\right)\frac{3(\Bar{k}/\Bar{k}_F(\tau))^2}{\exp\left[\frac{(\Bar{k}/\Bar{k}_F(\tau))^2}{\Bar{T}(\tau)/\Bar{T}_F(\tau)}\right]z(\tau)^{-1}+1},\\
    z(\tau) &\equiv -\mathrm{Li}^{-1}_{3/2}\left(-\frac{4\Bar{T}_F(\tau)^{3/2}}{3\sqrt{\pi}\Bar{T}(\tau)^{3/2}}\right).
\end{split}
\label{thermal_ansatz2}
\end{equation}
This expression is constructed by replacing all $k_F$ and $T_F$ at $\tau=0$ with dynamic variables. The effective Fermi momentum $k_F(\tau)$ and effective Fermi temperature $T_F(\tau)$ are defined as:
\begin{equation}
    F_0(\tau)=\Bar{k}_F(\tau)^3=\Bar{T}_F(\tau)^{3/2}.
    \label{relation_F0_TF}
\end{equation}
Equation~(\ref{relation_F0_TF}) ensures that the thermal system maintains the same particle number as the non-equilibrium profile. The total energy for this thermal ansatz is given by:
\begin{equation}
\begin{split}
    E^\mathrm{th}(\tau) &= \int d\Bar{k} \Bar{k}^2 N^\mathrm{th}(\Bar{k},\tau)\\
    &= -\frac{9\sqrt\pi\Bar{T}(\tau)^{5/2}}{8}\mathrm{Li}_{5/2}\left[\mathrm{Li}^{-1}_{3/2}\left(-\frac{4F_0(\tau)}{3\sqrt{\pi}\Bar{T}(\tau)^{3/2}}\right)\right].
    \label{explicit_thermal_E}
\end{split}
\end{equation}
We then implicitly determine $\Bar{T}(\tau)$ by solving:
\begin{equation}
    F_1(\tau)=E^\mathrm{th}(\tau).
\end{equation}
Figure~\ref{fig4} illustrates the dynamics of effective temperatures for four different initial temperatures. To assess whether the system is ``cooled" or ``heated," we employ two distinct criteria: one examines changes in the physical effective temperature $\Bar{T}(\tau)=T(\tau)/T_F(0)$, while the other considers changes in the reduced effective temperature $\Bar{T}(\tau)/\Bar{T}_F(\tau)=T(\tau)/T_F(\tau)$. 
In Fig.~\ref{fig4}(a), based on the first criterion, we observe that the physical effective temperature decreases at high initial temperatures. This can be attributed to the momentum dependence in the inelastic collision integral Eq.~(\ref{Iinel}): particles with higher momenta are more likely to undergo inelastic collisions. Consequently, the chemical reaction in the system exhibits momentum selectivity, preferentially removing higher-energy particles and thus cooling the system. Conversely, at sufficiently low temperatures, inelastic collisions have the opposite effect, increasing the physical effective temperature as any perturbation to the Fermi Sea generates excitations.
Examining the reduced effective temperature in Fig.~\ref{fig4}(b) reveals that, regardless of the initial temperature, inelastic collisions consistently drive the system away from quantum degeneracy.

\subsubsection{IQBE Solution by Assuming Thermal Ansatz}
\label{IQBE_sol_thermal_ansatz}

\begin{figure*}[t!]
    \centering
    \includegraphics[width=0.7\textwidth]{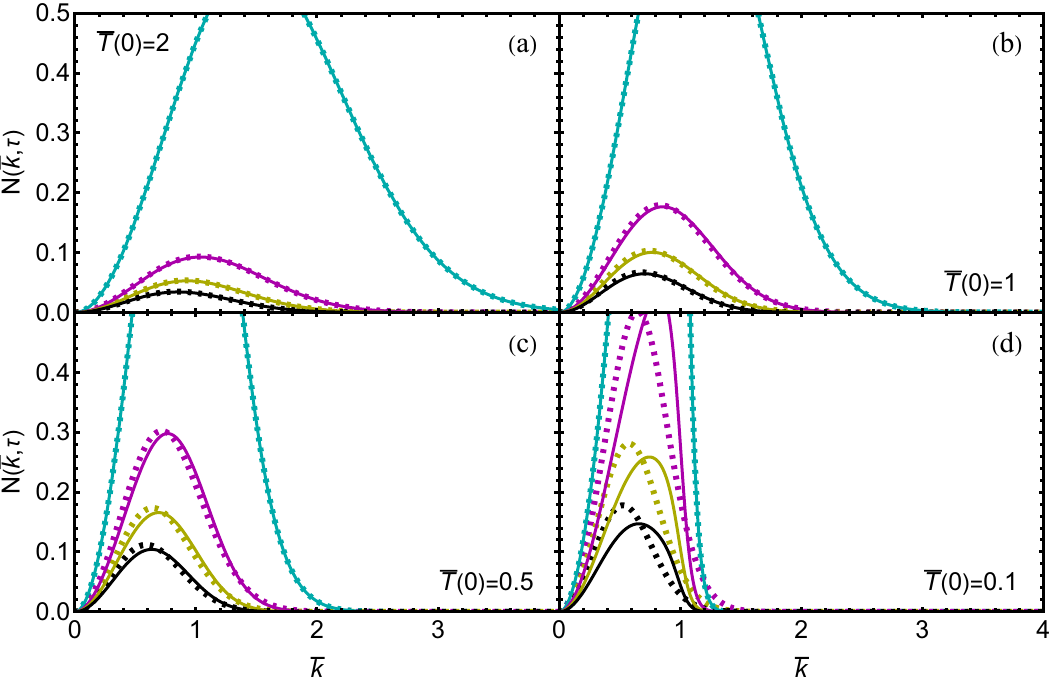}
    \caption{Comparison between thermalized momentum distribution (dotted lines) solved from the thermal ansatz and non-equilibrium momentum distribution (solid lines) solved from Eq.~(\ref{homo_boltzmann_eq_dimensionless}) at different time shots $\tau=0,2,5$ and $10$ denoted by cyan, magenta, yellow and black colors, respectively.}
    \label{fig5}
\end{figure*}

The thermal ansatz serves a dual purpose: it not only allows us to examine changes in effective temperatures but also enables us to assess the deviation of dissipation dynamics from the equilibrium limit. In this context, the equilibrium limit refers to dynamics where the system is assumed to thermalize instantaneously, such that $N(\Bar{k},\tau)=N^\mathrm{th}(\Bar{k},\tau)$ at every moment. Physically, this limit is equivalent to solving the IQBE with the elastic collision integral reintroduced under the constraint that elastic collisions occur on a much shorter timescale than inelastic collisions.

As discussed in Sec.~\ref{relax_tau}, realistic scenarios often present the opposite situation. Consequently, our analysis in Sec.~\ref{mellin_analysis} focuses on the IQBE while disregarding the elastic collision integral. However, if one were to retain the elastic collision integral, it is reasonable to expect that the exact solution would fall between the solution presented in Sec.~\ref{mellin_analysis} and that obtained from the instant thermalization assumption. Thus, the following analysis also provides an upper bound for potential interaction effects within the system.

In the thermal ansatz Eqs.~(\ref{thermal_ansatz1}) and (\ref{thermal_ansatz2}), we need to determine two unknown quantities: $F_0(\tau)$ and $\Bar{T}(\tau)$. For $F_0$, the equation of motion is straightforward:
\begin{equation}
\begin{split}
\frac{dF_0(\tau)}{d\tau} &= \int d\Bar{k} \frac{d}{d\tau} N^\mathrm{th}(\Bar{k},\tau)\\
&= \int d\Bar{k} d\Bar{q}(\Bar{k}^2+\Bar{q}^2)N^\mathrm{th}(\Bar{q},\tau)N^\mathrm{th}(\Bar{k},\tau) = -F_0^2 I_0,
\end{split}
\label{homo_thermal_F0evolve}
\end{equation}
where we utilize Eq.~(\ref{homo_boltzmann_eq_dimensionless}) in the second line. $I_0$ is explicitly expressed as:
\begin{equation}
I_0(\tau) = -\frac{9\sqrt\pi\Bar{T}(\tau)^{5/2}}{4F_0(\tau)}\mathrm{Li}_{5/2}\left[\mathrm{Li}^{-1}_{3/2}\left(-\frac{4F_0(\tau)}{3\sqrt{\pi}\Bar{T}(\tau)^{3/2}}\right)\right].
\end{equation}
The equation of motion for $\Bar{T}(\tau)$ is derived from the temporal change of energy per particle $\mathcal{E}^\mathrm{th}=E^\mathrm{th}/F_0$:
\begin{align}
\frac{d \mathcal{E}^\mathrm{th}}{d\tau} &= \frac{\partial \mathcal{E}^\mathrm{th}}{\partial F_0}\frac{d F_0}{d \tau}+\frac{\partial \mathcal{E}^\mathrm{th}}{\partial \Bar{T}}\frac{d \Bar{T}}{d \tau} = -F_0^2I_0\frac{\partial \mathcal{E}^\mathrm{th}}{\partial F_0}+\frac{\partial \mathcal{E}^\mathrm{th}}{\partial \Bar{T}}\frac{d \Bar{T}}{d \tau},\label{dEdtaudirect}\\
\frac{d \mathcal{E}^\mathrm{th}}{d\tau} &= \int d\Bar{k} \Bar{k}^2 \frac{dN^*(\Bar{k},\tau)}{d\tau} = -F_0 I_2+\mathcal{E}^\mathrm{th} F_0^2 I_0,\label{dEdtauintapproach}
\end{align}
where:
\begin{equation}
\begin{split}
I_2(\tau) &= \int d\Bar{k} d\Bar{q}\Bar{k}^2(\Bar{k}^2+\Bar{q}^2)N^*(\Bar{q},\tau)N^*(\Bar{k},\tau)\\
&= \frac{81\pi\Bar{T}(\tau)^5\mathrm{Li}^2_{5/2}\left[\mathrm{Li}^{-1}_{3/2}\left(-\frac{4F_0(\tau)}{3\sqrt{\pi}\Bar{T}(\tau)^{3/2}}\right)\right]}{64F_0(\tau^2)}\\
&\quad -\frac{45\sqrt{\pi}\Bar{T}(\tau)^{7/2}\mathrm{Li}_{7/2}\left[\mathrm{Li}^{-1}_{3/2}\left(-\frac{4F_0(\tau)}{3\sqrt{\pi}\Bar{T}(\tau)^{3/2}}\right)\right]}{16F_0(\tau)}.
\end{split}
\end{equation}
From Eqs.~(\ref{dEdtaudirect}) and (\ref{dEdtauintapproach}), we derive:
\begin{equation}
\frac{d\Bar{T}(\tau)}{d\tau} = \frac{({d\mathcal{E}^\mathrm{th}}/{dF_0})F_0^2I_0+\mathcal{E}^\mathrm{th}F_0I_0-F_0I_2}{({d\mathcal{E}^\mathrm{th}}/{d\Bar{T}})}.
\label{homo_thermal_Tevolve}
\end{equation}
The explicit forms of ${d\mathcal{E}^\mathrm{th}}/{dF_0}$ and ${d\mathcal{E}^\mathrm{th}}/{d\Bar{T}}$ are obtained from Eq.~(\ref{explicit_thermal_E}):
\begin{equation}
\begin{split}
\frac{d\mathcal{E}^\mathrm{th}}{dF_0} &= -\frac{2}{\sqrt{\pi}\sqrt{\Bar{T}}\mathrm{Li}_{1/2}\left[\mathrm{Li}^{-1}_{3/2}\left(-\frac{4F_0(\tau)}{3\sqrt{\pi}\Bar{T}(\tau)^{3/2}}\right)\right]}\\
&\quad +\frac{9\sqrt{\pi}\Bar{T}^{5/2}\mathrm{Li}_{5/2}\left[\mathrm{Li}^{-1}_{3/2}\left(-\frac{4F_0(\tau)}{3\sqrt{\pi}\Bar{T}(\tau)^{3/2}}\right)\right]}{8F_0^2},
\end{split}
\end{equation}
\begin{equation}
\begin{split}
\frac{d\mathcal{E}^\mathrm{th}}{d\Bar{T}} &= \frac{3F_0}{\sqrt{\pi}{\Bar{T}}^{3/2}\mathrm{Li}_{1/2}\left[\mathrm{Li}^{-1}_{3/2}\left(-\frac{4F_0(\tau)}{3\sqrt{\pi}\Bar{T}(\tau)^{3/2}}\right)\right]}\\
&\quad -\frac{45\sqrt{\pi}\Bar{T}^{3/2}\mathrm{Li}_{5/2}\left[\mathrm{Li}^{-1}_{3/2}\left(-\frac{4F_0(\tau)}{3\sqrt{\pi}\Bar{T}(\tau)^{3/2}}\right)\right]}{16F_0}.
\end{split}
\end{equation}
The coupled equations, Eqs.~(\ref{homo_thermal_F0evolve}) and (\ref{homo_thermal_Tevolve}), generally require numerical solutions. However, in the high-initial-temperature limit, following Eq.(\ref{polylog_highT}), they simplify to:
\begin{equation}
\left\{
\begin{matrix}
{dF_0(\tau)}/{d\tau} = -3\Bar{T}(\tau)F_0(\tau)^2 \\
{d\Bar{T}(\tau)}/{d\tau} = -F_0(\tau)\Bar{T}(\tau)^2
\end{matrix}
\right.
.
\label{homo_thermal_model_highT}
\end{equation}
Solving this set of equations yields:
\begin{equation}
F_0(\tau) = \frac{1}{(1+4\Bar{T}(0)\tau)^{3/4}},\quad \Bar{T}(\tau) = \frac{\Bar{T}(0)}{(1+4\Bar{T}(0)\tau)^{1/4}}.
\label{homo_thermal_model_highT_sol_F0_and_T}
\end{equation}
Notably, $F_0(\tau)$ in Eq.(\ref{homo_thermal_model_highT_sol_F0_and_T}) is identical to Eq.~(\ref{homo_noneq_high_temperature_NoN0}). Moreover, substituting Eq.~(\ref{homo_thermal_model_highT_sol_F0_and_T}) into Eqs.~(\ref{thermal_ansatz1}) and (\ref{thermal_ansatz2}) recovers Eq.~(\ref{homo_noneq_Nktau_dynamics}), corroborating our conclusion in Sec.~\ref{high_T_exact} that elastic collisions play no role in systems with high initial temperatures.
We can also analytically study ${T}(\tau)/{T}_F(\tau)$ using Eq.~(\ref{relation_F0_TF}):
\begin{equation}
\frac{{T}(\tau)}{{T}_F(\tau)} = \frac{T(0)}{T_F(0)}\left[1+\frac{4T(0)\tau}{T_F(0)}\right]^{\frac{1}{4}},
\end{equation}
which increases monotonically, consistent with our findings in Sec.~\ref{dynamics_of_Teff}.

For general situations, we numerically solve Eqs.~(\ref{homo_thermal_F0evolve}) and (\ref{homo_thermal_Tevolve}). Fig.~\ref{fig4} compares $\Bar{T}(\tau)$ and $T(\tau)/T_F(\tau)$ with the general non-equilibrium solution of the IQBE. Surprisingly, except for systems initially in deep quantum degeneracy ($T(0)/T_F(0)=0.1$), all cases show remarkable similarity. This demonstrates that for initial temperatures not excessively low, systems without elastic collisions remain close to equilibrium for reasonable time periods. Fig.~\ref{fig5} provides a more intuitive comparison of $N(\Bar{k},\tau)$ between non-equilibrium solutions calculated using Eq.~(\ref{homo_boltzmann_eq_dimensionless}) and results from the thermal ansatz.

\section{Dynamics of Harmonically Trapped Systems}

This section examines systems confined in a harmonic trap, where the dynamics are governed by Eq.~(\ref{boltzmann_eqn_form}). Unlike Eq.~(\ref{general_homo_IQBE}), which can be reduced to a one-dimensional equation [Eq.~(\ref{homo_boltzmann_eq_dimensionless})] due to the spherical symmetry of $n_\mathbf{k}$, Eq.~(\ref{boltzmann_eqn_form}) exhibits an irreducible six-dimensional spatial complexity. This complexity renders direct numerical solutions impractical.

As with homogeneous systems, we employ dimensionless quantities to simplify our notation. To facilitate the treatment of phase space as a unified whole, we introduce two distinct length scales for dimensionless representations: the inverse of the Fermi momentum $1/k_F^\mathrm{trap}$ and the Thomas-Fermi Radius $R_F^\mathrm{trap}$. These are defined as follows:
For one-dimensional systems (used in Sec.~\ref{FFA}):
\begin{equation}
    \frac{1}{k_F^\mathrm{trap}} = \frac{1}{\sqrt{2N(0)}}\sqrt{\frac{\hbar}{M \omega}}, \quad R_F^\mathrm{trap} = \sqrt{2N(0)}\sqrt{\frac{\hbar}{M \omega}}
    \label{1DkFRF}
\end{equation}
For three-dimensional systems:
\begin{equation}
    \frac{1}{k_F^\mathrm{trap}} = [48 N(0)]^{-\frac{1}{6}}\sqrt{\frac{\hbar}{M \omega}}, \quad R_F^\mathrm{trap} = [48 N(0)]^{\frac{1}{6}}\sqrt{\frac{\hbar}{M \omega}}
    \label{3DkFRF}
\end{equation}
In 1D systems, $\omega$ represents the angular frequency of the trap. For 3D cases, it denotes the geometric mean of angular frequencies in all three directions:
\begin{equation}
    \omega = (\omega_x \omega_y \omega_z)^{1/3}
\end{equation}
Our notation for dimensionless quantities follows these conventions:
\begin{itemize}
    \item Quantities in units of $1/k_F^\mathrm{trap}$ or its derivatives [such as $k_F^\mathrm{trap}$ (wavevector), $E_F^\mathrm{trap} = \frac{\hbar^2(k_F^\mathrm{trap})^2}{2M}$ (energy), $T_F^\mathrm{trap} = E_F^\mathrm{trap}/k_B$ (temperature)] are denoted with a ``bar" (e.g., $\bar{k}$).
    \item Quantities in units of $R_F^\mathrm{trap}$ or its derivatives are denoted with a ``tilde" (e.g., $\tilde{r}$).
\end{itemize}
For anisotropic 3D systems ($\omega_x \neq \omega_y \neq \omega_z$), we introduce a modified position vector $\mathbf{x}$, related to the physical position $\mathbf{r}$ by:
\begin{equation}
    \mathbf{x} = \left(\frac{\omega_x}{\omega}r_x, \frac{\omega_y}{\omega}r_y, \frac{\omega_z}{\omega}r_z\right)
\end{equation}
This formulation provides a consistent framework for analyzing trapped systems across different dimensionalities and trap geometries.

\subsection{Fast Flowing Approximation}
\label{FFA}

To address the computational challenges posed by the six-dimensional nature of the IQBE, we introduce the Fast-Flowing Approximation (FFA). This approach assumes that the distribution function $f(\mathbf{k},\mathbf{r},t)$ maintains hyperspherical symmetry throughout the entire phase space, such that $f(\mathbf{k},\mathbf{r},t)\equiv f(R,t)$, where:
\begin{equation}
    R = \sqrt{\bar{k}^2+\tilde{x}^2} = \sqrt{\sum_{i=x,y,z}(\bar{k}_i^2+\omega_i^2\tilde{r}_i^2/\omega^2)}.
\end{equation}
We assume that at $t=0$, the systems are prepared in thermalized states, which inherently possess hyperspherical symmetry:
\begin{equation}
    f(\mathbf{k},\mathbf{r},0) = f(R,0) = \frac{48}{\exp\left(\frac{R^2}{\Bar{T}}\right)(z^{\mathrm{trap}})^{-1}+1},
\end{equation}
where the fugacity in trapped systems is given by:
\begin{equation}
    z^{\mathrm{trap}} = -\mathrm{Li}_{3}^{-1}\left(-\frac{1}{6\Bar{T}^3}\right).
\end{equation}

The validity of this approximation for systems that have evolved over time can be justified by considering the disparate timescales involved in experimental setups. Harmonic trap dynamics typically operate on a much faster timescale than two-body relaxation processes. For instance, with an average trap frequency of $2\pi \times 100~\mathrm{Hz}$, the associated timescale is approximately $1~\mathrm{ms}$. In contrast, the typical relaxation process unfolds over several seconds. This significant separation of timescales implies that the molecular cloud's dynamics are substantially faster than the relaxation reactions.
Consequently, we can neglect inelastic collisions within short time intervals, allowing the system to reach a quasi-steady state. It can be readily demonstrated that the steady-state phase-space distributions, obtained by solving:
\begin{equation}
    \left[\frac{\hbar \mathbf{k}}{M}\nabla_\mathbf{r}-\frac{\nabla_{\mathbf{r}}U_{\mathrm{ext}}\cdot\nabla_\mathbf{k}}{\hbar}\right]f = 0,
\end{equation}
indeed exhibit the hyperspherical symmetry $f(\mathbf{k},\mathbf{r},t) = f(R,t)$.
This Fast-Flowing Approximation provides a tractable approach to modeling the complex dynamics of trapped systems while capturing the essential physics of the problem.

\subsubsection{Validating FFA with 1D analog}

To validate the Fast-Flowing Approximation (FFA), we initially applied it to a simplified one-dimensional version of the problem. The dimensionless form of the inelastic Boltzmann equation in this context is:
\begin{equation}
    \frac{df}{d(\omega t)} = \left[\Tilde{x}\partial_{\Bar{k}} - \Bar{k}\partial_{\Tilde{x}} - G\int\frac{d\Bar{q}}{2\pi}(\Bar{k}^2+\Bar{q}^2)f(\Bar{q},\Tilde{x})\right]f(\Bar{k},\Tilde{x}),
    \label{inelastic_boltzmann_1D}
\end{equation}
where $G$ represents the dimensionless reaction strength, with higher values indicating more rapid relaxation, it is important to note that Eq.~(\ref{inelastic_boltzmann_1D}) serves as an analog to the 3D problem we are focusing on and may not accurately describe a physical 1D or quasi-1D system.
Within the FFA framework, this equation simplifies to:
\begin{equation}
\begin{split}
    \frac{df(R)}{d\tau} = &-\frac{1}{4\pi^2}\int_0^{2\pi} d\theta \int_0^\infty d\Bar{q}(R^2\cos^2\theta+\Bar{q}^2)\\
    &\times f(\sqrt{R^2\sin^2\theta+\Bar{q}^2})f(R),
    \label{fast_flowing_approx_1D}
\end{split}
\end{equation}
where $\tau = G\omega t$ and $R = \sqrt{\Bar{k}^2 + \Tilde{x}^2}$.

\begin{figure}[t!]
    \centering
    \includegraphics[width=0.4\textwidth]{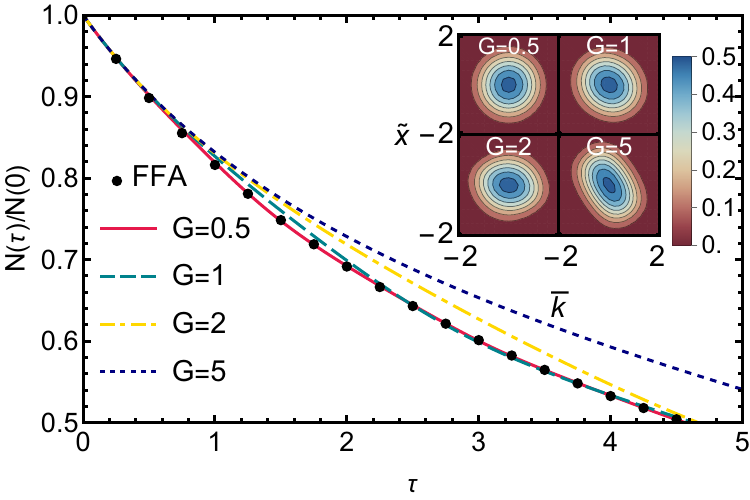}
    \caption{Comparison between brute-force numerical solutions of Eq.~(\ref{inelastic_boltzmann_1D}) with different $G$ and the result from Eq.~(\ref{fast_flowing_approx_1D}). The initial temperature is set to be $\Bar{T}=1$ in 1D, i.e. $f(\Bar{k},\Tilde{x},0)=[\exp(\Bar{k}^2+\Tilde{x}^2)/(e-1)+1]^{-1}$. Inset: Phase space density $f(\Bar{k},\Tilde{x},\tau)$ at $\tau=5$. }
    \label{fig6}
\end{figure}

We performed numerical solutions of Eq.~(\ref{inelastic_boltzmann_1D}) for various $G$ values and compared them with the predictions of Eq.~(\ref{fast_flowing_approx_1D}), as shown in Figure~\ref{fig2}. The results demonstrate improved agreement for smaller $G$ values, supporting the assumption that FFA is valid when the reaction rate is sufficiently slow. 
The inset of Fig.~\ref{fig6} provides an intuitive visualization of the phase space density at $\tau=5$ for different $G$ values. This reinforces the assumption that $f$ maintains spherical symmetry throughout the process, further validating the FFA approach.
These findings provide a solid foundation for applying the Fast-Flowing Approximation to more complex three-dimensional systems where direct numerical solutions are computationally prohibitive.

\subsubsection{3D IQBE under FFA}

We now extend our analysis to realistic 3D cases. First, we express Eq.~(\ref{boltzmann_eqn_form}) in dimensionless form:
\begin{equation}
\begin{split}
    &\frac{df(\Bar{\mathbf{k}},\Tilde{\mathbf{x}},\Bar{t})}{d\Bar{t}} = \left[-\frac{\hbar \omega}{E_F^\mathrm{trap}}\Bar{\mathbf{k}}\cdot\nabla_\mathbf{\Tilde{x}}+\frac{\hbar \omega}{E_F^\mathrm{trap}}\Tilde{\mathbf{x}}\cdot\nabla_\mathbf{\Bar{k}}\right]f(\Bar{\mathbf{k}},\Tilde{\mathbf{x}},\Bar{t})\\
    &+24\pi\mathrm{Im}(\Bar{v}_p)\int\frac{d^3\Bar{q}}{(2\pi)^3}(\Bar{q}^2+\Bar{k}^2)f(\Bar{\mathbf{k}},\Tilde{\mathbf{x}},\Bar{t})f(\Bar{\mathbf{q}},\Tilde{\mathbf{x}},\Bar{t}).
\end{split}
    \label{trap_boltzmann_eq_dimensionless}
\end{equation}
After non-dimensionalization, the normalization of $f(\Bar{\mathbf{k}},\Tilde{\mathbf{x}},\Bar{t})$ becomes:
\begin{equation}
    \int d^3\Tilde{x}\frac{d^3\Bar{k}}{(2\pi)^3}f(\Bar{\mathbf{k}},\Tilde{\mathbf{x}},\Bar{t})=\frac{N(t)}{48N(0)}.
\end{equation}
This additional factor arises from $k_F^\mathrm{trap}R_F^\mathrm{trap}=\left(48N(0)\prod_{i=x,y,z}{\omega_i}/{\omega}\right)^{1/3}$. For numerical convenience, we redefine the dimensionless phase space density as $\mathcal{F}(\Bar{\mathbf{k}},\Tilde{\mathbf{x}},\Bar{t})=48f(\Bar{\mathbf{k}},\Tilde{\mathbf{x}},\Bar{t})$.
Equation~(\ref{trap_boltzmann_eq_dimensionless}) can then be rewritten as:
\begin{equation}
\begin{split}
    &\frac{d\mathcal{F}(\Bar{\mathbf{k}},\Tilde{\mathbf{x}},\tau^\mathrm{trap})}{d\tau^\mathrm{trap}} =\\ &\bigg[\frac{2\hbar \omega}{\pi \mathrm{Im}(\Bar{v}_p)E_F^\mathrm{trap}}\Bar{\mathbf{k}}\cdot\nabla_\mathbf{\Tilde{x}}
    -\frac{2\hbar \omega}{\pi \mathrm{Im}(\Bar{v}_p)E_F^\mathrm{trap}}\Tilde{\mathbf{x}}\cdot\nabla_\mathbf{\Bar{k}}\bigg]\mathcal{F}(\Bar{\mathbf{k}},\Tilde{\mathbf{x}},\Bar{t})\\
    &-\int\frac{d^3\Bar{q}}{(2\pi)^3}(\Bar{q}^2+\Bar{k}^2)
    \mathcal{F}(\Bar{\mathbf{k}},\Tilde{\mathbf{x}},\tau^\mathrm{trap})\mathcal{F}(\Bar{\mathbf{q}},\Tilde{\mathbf{x}},\tau^\mathrm{trap}),
\end{split}
    \label{scF_trap_boltzman_eq}
\end{equation}
where $\tau^\mathrm{trap} = \frac{\pi}{2}\mathrm{Im}(\Bar{v}_p)\Bar{t} = -\frac{\hbar [k_F^\text{trap}(0)]^5}{\pi M}\mathrm{Im}(v_p)t$.

The Fast-Flowing Approximation (FFA) allows us to express $\mathcal{F}(\Bar{\mathbf{k}},\Tilde{\mathbf{x}},\tau^\mathrm{trap})$ solely in terms of the hyperradius $R=\sqrt{\Bar{k}^2+\Tilde{x}^2}$, i.e., $\mathcal{F}(\Bar{\mathbf{k}},\Tilde{\mathbf{x}},\tau^\mathrm{trap})=\mathcal{F}(R,\tau^\mathrm{trap})$. Consequently:
\begin{equation}
    \left[-\frac{2\hbar \omega \Bar{\mathbf{k}}\cdot\nabla_\mathbf{\Tilde{x}}}{\pi \mathrm{Im}(\Bar{v}_p)E_F^\mathrm{trap}}+\frac{2\hbar \omega \Tilde{\mathbf{x}}\cdot\nabla_\mathbf{\Bar{k}}}{\pi \mathrm{Im}(\Bar{v}_p)E_F^\mathrm{trap}}\right]\mathcal{F}(R,\Bar{t})=0.
\end{equation}
To convert other terms to hyper-spherical coordinates, we employ the necessary integral measures in Eq.~(\ref{scF_trap_boltzman_eq}):
\begin{equation}
\begin{split}
     &\int d^3\Tilde{x} d^3\Bar{k}\frac{d\mathcal{F}(\Bar{\mathbf{k}},\Tilde{\mathbf{x}},\tau^\mathrm{trap})}{d\tau^\mathrm{trap}} = \int d \Omega^{(5)} \frac{d\mathcal{F}(R,\tau^\mathrm{trap})}{d\tau^\mathrm{trap}}\\
     &= -\int d \Omega^{(5)}\int\frac{d^3\Bar{q}}{(2\pi)^3}(\Bar{q}^2+R^2h)\\
     &\quad \times\mathcal{F}(R,\tau^\mathrm{trap})\mathcal{F}(\sqrt{\Bar{q}^2+R^2(1-h)},\tau^\mathrm{trap})
\end{split}
\label{3D_trap_FFA_with_full_measure}
\end{equation}
where $\Omega^{(5)}$ is the solid angle in 5D with the measure:
\begin{equation}
    d\Omega^{(5)}=\sin^4(\theta_1)\sin^3(\theta_2)\sin^2(\theta_3)\sin(\theta_4)d\theta_1 d\theta_2 d\theta_3 d\theta_4 d\phi
\end{equation}
and $h$ is a function of $\Omega^{(5)}$:
\begin{equation}
    h[\Omega^{(5)}]=\cos^2\theta_1 + \sin^2\theta_1\cos^2\theta_2 + \sin^2\theta_1\sin^2\theta_2\cos^2\theta_3.
\end{equation}
The ranges of polar angles $\theta_1$ to $\theta_5$ are $0$ to $\pi$, and the range of azimuth angle $\phi$ is $0$ to $2\pi$. As $\int d\Omega^{(5)}=\pi^3$, Eq.~(\ref{3D_trap_FFA_with_full_measure}) simplifies to:
\begin{equation}
\begin{split}
    &\frac{d\mathcal{F}(R,\tau^\mathrm{trap})}{d\tau^\mathrm{trap}}= -\frac{2}{\pi^4}\int d\theta_1d\theta_2d\theta_3\sin^4\theta_1d\sin^3\theta_2d\sin^2\theta_3\\
    &\quad \times\int d\Bar{q} \Bar{q}^2(\Bar{q}^2+R^2h)\mathcal{F}(R,\tau^\mathrm{trap})\mathcal{F}(\sqrt{\Bar{q}^2+R^2(1-h)},\tau^\mathrm{trap}).
\end{split}
\label{3D_trap_FFA}
\end{equation}
Equation~(\ref{3D_trap_FFA}) represents the FFA of a 3D harmonically trapped system suitable for numerical analysis. To interpret results, we can use the following properties of $\mathcal{F}(R,\tau^\mathrm{trap})$ to obtain the normalized total number of particles and average energy per particle:
\begin{align}
    \frac{N(\tau^\mathrm{trap})}{N(0)} &= \frac{1}{8}\int_0^\infty dR R^5 \mathcal{F}(R,\tau^\mathrm{trap}),\\
    \frac{E(\tau^\mathrm{trap})}{N(\tau^\mathrm{trap})} &= \frac{1}{3}\frac{\int_0^\infty dR R^7 \mathcal{F}(R,\tau^\mathrm{trap})}{\int_0^\infty dR R^5 \mathcal{F}(R,\tau^\mathrm{trap})}.
\end{align}

\begin{figure}[t!]
    \centering
    \includegraphics[width=0.4\textwidth]{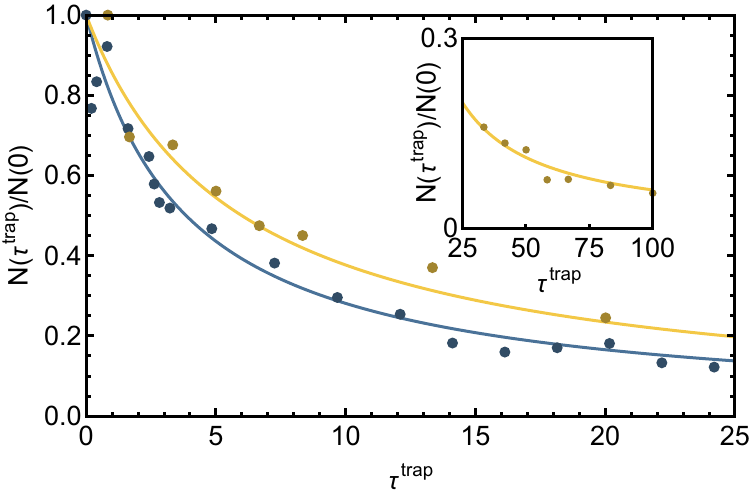}
    \caption{Reproducing experimental data in Ref.~\cite{demarco2019degenerate} by numerically solving Eq.~(\ref{3D_trap_FFA}). Yellow and blue colors represent systems with initial temperatures $\Bar{T}=1.26$ and $0.48$, respectively. Solid lines are our numerical results, and circle symbols are raw experimental data reported in Ref.~\cite{demarco2019replication}. }
    \label{fig7}
\end{figure}

Figure~\ref{fig7} demonstrates the application of this approach to reproduce experimental data from Ref.~\cite{demarco2019degenerate} by numerically solving Eq.~(\ref{3D_trap_FFA}). The results for systems with initial temperatures $\Bar{T}=1.26$ and $0.48$ are shown, with solid lines representing our numerical results and circle symbols indicating raw experimental data from Ref.~\cite{demarco2019replication}.

\begin{figure}[t!]
    \centering
    \includegraphics[width=0.4\textwidth]{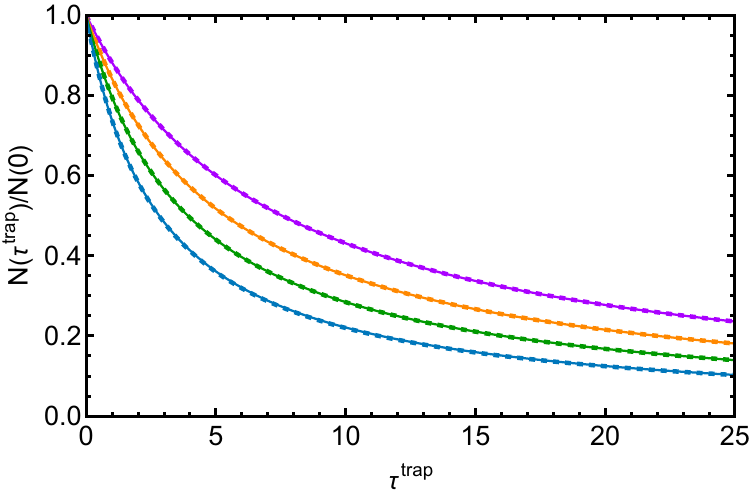}
    \caption{Particles dynamics obtained from the thermal ansatz Eqs.~(\ref{trap_thermal_F0evolve}) and (\ref{trap_thermal_Tevolve}) [dashed lines] and directly solving Eq.~(\ref{3D_trap_FFA}) [solid lines]. From top to bottom, $\Bar{T}(0)$ are set to be $2,1,0.5$ and $0.1$. }
    \label{fig8}
\end{figure}

\subsubsection{Comparison with Experimental Data}

To demonstrate the predictive power of Eq.~(\ref{3D_trap_FFA}), we reproduce the experimental measurements reported in Ref.~\cite{demarco2019degenerate}. Using the raw data provided in Ref.~\cite{demarco2019replication}, we analyze two decay measurement sets with initial reduced temperatures $\tilde{T}=1.26$ and $0.48$. The imaginary part of the scattering volume is set to $\mathrm{Im}(v_p)=-(118a_0)^3$, consistent with previous studies \cite{idziaszek2010universal,he2020exploring}.

Figure~\ref{fig7} presents a comparison between the dynamics predicted by Eq.~(\ref{3D_trap_FFA}) and the raw experimental data using the dimensionless unit system introduced in this section. Our theoretical calculations show excellent agreement with the experimental results, even over extended timescales. Specifically, we achieve good correspondence up to $\tau^\mathrm{trap}=100$ for the $\tilde{T}=1.26$ sample and $\tau^\mathrm{trap}=25$ for the $\tilde{T}=0.48$ sample. These dimensionless times translate to a physical duration of approximately 6 seconds, demonstrating the robustness of our approach in capturing long-term system behavior.

\subsection{Thermal Ansatz}

Equation~(\ref{3D_trap_FFA}) is significantly more complex than its homogeneous counterpart, for which we lack an analytical solution method. However, drawing on our experience with homogeneous systems, where the thermal ansatz effectively describes dynamics at high initial temperatures, we conjecture that a similar approach may provide a good approximation for trapped systems with high initial temperatures. 

\begin{figure*}[t!]
    \centering
    \includegraphics[width=0.7\textwidth]{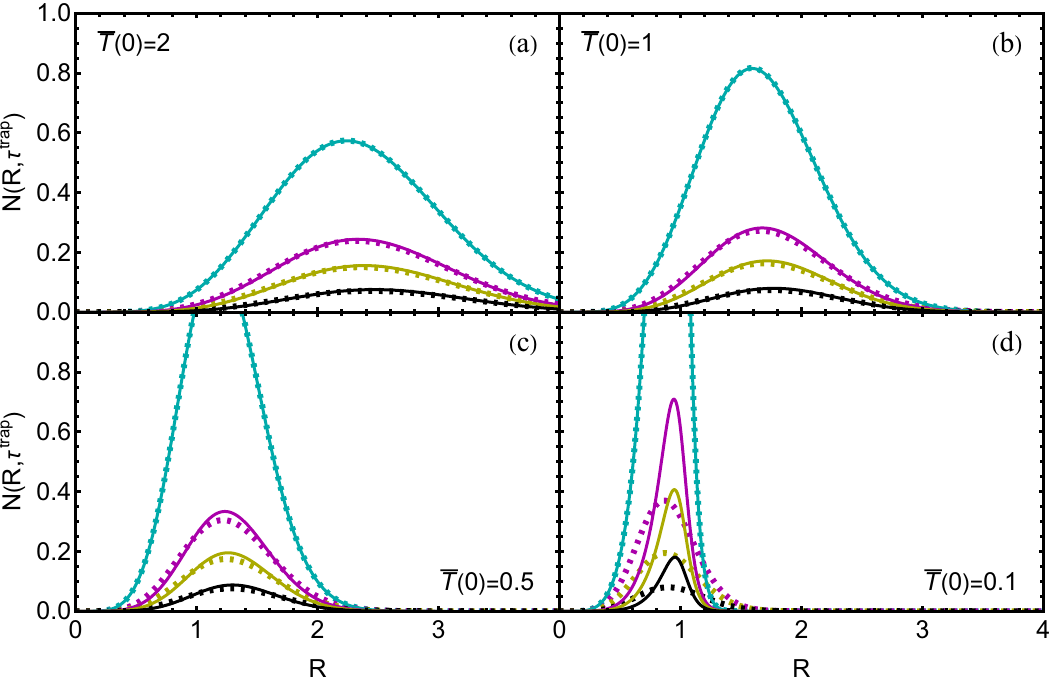}
    \caption{Comparison between radial phase-space distributions obtained from the thermal ansatz (dotted lines) and by directly solving Eq.~(\ref{3D_trap_FFA}) (solid lines) at different time shots $\tau^\mathrm{trap}=0,10,20$ and $50$ denoted by cyan, magenta, yellow and black colors, respectively.}
    \label{fig9}
\end{figure*}

Following the procedure in Sec.~\ref{dynamics_of_Teff}, we introduce the thermal ansatz:
\begin{equation}
    \mathcal{F}^\mathrm{th}(R,\tau^\mathrm{trap})\equiv F_0(\tau^\mathrm{trap}) \mathcal{F}^*(R,\tau^\mathrm{trap})
\end{equation}
into Eq.~(\ref{3D_trap_FFA}), where $F_0$ represents ${N(\tau^\mathrm{trap})}/{N(0)}$ and 
\begin{equation}
\begin{split}
    dR \mathcal{F}^*(R,\tau^\mathrm{trap}) &\equiv \frac{dR}{F_0(\tau^\mathrm{trap})}\frac{48}{\exp\left(\frac{R^2}{\Bar{T}(\tau^\mathrm{trap})}\right)(z^{\mathrm{trap}})^{-1}+1},\\
    z^{\mathrm{trap}}(\tau^\mathrm{trap}) &= -\mathrm{Li}_{3}^{-1}\left(-\frac{F_0(\tau^\mathrm{trap})}{6\Bar{T}^3(\tau^\mathrm{trap})}\right).
\end{split}
\end{equation} 
Following a derivation similar to that in Sec.~\ref{dynamics_of_Teff}, we obtain coupled equations for $F_0(\tau^\mathrm{trap})$ and $\Bar{T}(\tau^\mathrm{trap})$:
\begin{align}
    \frac{d F_0}{d\tau^\mathrm{trap}} &= -I_5 F_0^2, \label{trap_thermal_F0evolve}\\
    \frac{d\Bar{T}}{d\tau^\mathrm{trap}} &= \dfrac{({d\mathcal{E}^\mathrm{th}}/{dF_0})F_0^2I_5+\mathcal{E}^\mathrm{th}F_0I_5-F_0I_7}{({d\mathcal{E}^\mathrm{th}}/{d\Bar{T}})},\label{trap_thermal_Tevolve}
\end{align}
where
\begin{equation}
\begin{aligned}
    I_5=&-\frac{1}{4\pi^4}\int d\theta_1d\theta_2d\theta_3\sin^4\theta_1d\sin^3\theta_2d\sin^2\theta_3\\
    &\times\int dR\frac{288\sqrt{\pi}R^5\Bar{T}^{3/2}z^\mathrm{trap}}{F_0^2[\exp(R^2/\Bar{T})+z^\mathrm{trap}]}\\
    &\times\bigg\{2hR^2\mathrm{Li}_{\frac{3}{2}}\left[-\exp\left(\frac{(h-1)R^2}{\Bar{T}}z^\mathrm{trap}(\tau^\mathrm{trap})\right)\right]\\
    &+3\Bar{T}\mathrm{Li}_{\frac{3}{2}}\left[-\exp\left(\frac{(h-1)R^2}{\Bar{T}}z^\mathrm{trap}(\tau^\mathrm{trap})\right)\right]\bigg\}
\end{aligned}
\label{I5}
\end{equation}
and
\begin{equation}
\begin{aligned}
    I_7=&-\frac{1}{12\pi^4}\int d\theta_1d\theta_2d\theta_3\sin^4\theta_1d\sin^3\theta_2d\sin^2\theta_3\\
    &\times\int dR\frac{288\sqrt{\pi}R^7\Bar{T}^{3/2}z^\mathrm{trap}}{F_0^2[\exp(R^2/\Bar{T})+z^\mathrm{trap}]}\\
    &\times\bigg\{2hR^2\mathrm{Li}_{\frac{3}{2}}\left[-\exp\left(\frac{(h-1)R^2}{\Bar{T}}z^\mathrm{trap}\right)\right]\\
    &+3\Bar{T}\mathrm{Li}_{\frac{3}{2}}\left[-\exp\left(\frac{(h-1)R^2}{\Bar{T}}z^\mathrm{trap}\right)\right]\bigg\}.
\end{aligned}
\label{I7}
\end{equation}
The necessary expressions for $\mathcal{E}^\mathrm{th}$, ${d\mathcal{E}^\mathrm{th}}/{dF_0}$, and ${d\mathcal{E}^\mathrm{th}}/{d\Bar{T}}$ are also provided below:
\begin{align}
    &\mathcal{E}^\mathrm{th}=-\frac{6\Bar{T}^4\mathrm{Li}_4[-z^{\mathrm{trap}}(\tau^\mathrm{trap})]}{F_0},\\
    &\frac{d\mathcal{E}^\mathrm{th}}{dF_0}=\frac{6\Bar{T}^4\mathrm{Li}_4[-z^{\mathrm{trap}}(\tau^\mathrm{trap})]}{F_0^2}-\frac{1}{6\Bar{T}^2\mathrm{Li}_2[-z^{\mathrm{trap}}(\tau^\mathrm{trap})]},\\
    &\frac{d\mathcal{E}^\mathrm{th}}{d\Bar{T}}=\frac{F_0}{2\Bar{T}^3\mathrm{Li}_2[-z^{\mathrm{trap}}(\tau^\mathrm{trap})]}-\frac{24\Bar{T}^4\mathrm{Li}_4[-z^{\mathrm{trap}}(\tau^\mathrm{trap})]}{F_0},
\end{align}

Figure~\ref{fig8} presents numerical solutions of particle dynamics from Eqs.~(\ref{trap_thermal_F0evolve}) and (\ref{trap_thermal_Tevolve}). Comparing these results with those from Eq.~(\ref{3D_trap_FFA}), we find remarkably small differences even in the deep degeneracy regime. This close agreement might suggest that the thermal ansatz is an exact solution to Eq.~(\ref{3D_trap_FFA}). 

However, a careful examination of the radial phase-space distribution $N(R,\tau^\mathrm{trap})=R^5\mathcal{F}(R,\tau^\mathrm{trap})/8$, shown in Fig.~\ref{fig9}, reveals a more nuanced picture. For systems starting from high initial temperatures, the profiles remain approximately thermal throughout the evolution. In contrast, for low initial temperatures, despite the particle dynamics appearing similar to those given by the thermal ansatz, the system is, in fact, highly non-equilibrium.

\subsubsection{High-initial-temperature analytical solution}

For high initial temperatures, the integrals in Eqs.~(\ref{I5}) and (\ref{I7}) can be explicitly evaluated:
\begin{equation}
\begin{aligned}
    I_5 &\xrightarrow[]{\Bar{T}\rightarrow\infty} \frac{6}{\pi^{7/2}\sqrt{\Bar{T}}} \int d\theta_1d\theta_2d\theta_3
    \sin^4\theta_1 \sin^3\theta_2 \sin^2\theta_3 \frac{(2+h)}{(h-2)^4} \\
    &= \frac{0.190}{\sqrt{\Bar{T}}},
\end{aligned}
\end{equation}
and
\begin{equation}
\begin{aligned}
    I_7 &\xrightarrow[]{\Bar{T}\rightarrow\infty} \frac{2\sqrt{\Bar{T}}}{\pi^{7/2}} \int d\theta_1d\theta_2d\theta_3
    \sin^4\theta_1 \sin^3\theta_2 \sin^2\theta_3 \frac{(6+5h)}{(h-2)^5} \\
    &= 0.175\sqrt{\Bar{T}}.
\end{aligned}
\end{equation}
Consequently, Eqs.~(\ref{trap_thermal_F0evolve}) and (\ref{trap_thermal_Tevolve}) simplify to:
\begin{align}
    \frac{d F_0}{d\tau^\mathrm{trap}} &= -0.190481 F_0^2 \Bar{T}^{-1/2}, \label{trap_thermal_model_highT1}\\ 
    \frac{d\Bar{T}}{d\tau^\mathrm{trap}} &= 0.0158734 F_0 \Bar{T}^{1/2}.
    \label{trap_thermal_model_highT2}
\end{align}
Solving these equations yields:
\begin{align}
    F_0(\tau^\mathrm{trap}) &= \frac{1}{\left[1+{0.198418\tau^\mathrm{trap}}/{\Bar{T}(0)}^{1/2}\right]^{0.960}}, \label{trap_thermal_model_highT_F0}\\
    \Bar{T}(\tau^\mathrm{trap}) &= {\Bar{T}(0)}{\left[1+{0.198418\tau^\mathrm{trap}}/{\Bar{T}(0)}^{1/2}\right]^{0.0800}}. \label{trap_thermal_model_highT_T}
\end{align}

Notably, Eq.~(\ref{trap_thermal_model_highT_F0}) takes the form of Eq.~(\ref{N_body_decay_ansatz_sol}), describing an $\mathcal{N}$-body decay with:
\begin{equation}
    \mathcal{N} = 2.04167.
\end{equation}
This result contrasts with our findings in Sec.~\ref{high_T_exact}, where we demonstrated that for homogeneous systems, $\mathcal{N} = 7/3$. This difference underscores the significant impact of flowing dynamics in trapped systems on the two-body dissipation behavior.

\section{Summary}

In this companion paper, we present a comprehensive analysis of the inelastic quantum Boltzmann equation (IQBE) for single-component Fermi gases in both free space and harmonic traps. Our approach begins by deriving the IQBE from a non-Hermitian Hamiltonian and demonstrating that elastic collisions play a minimal role in the dynamics of typical experimental systems.
For homogeneous systems, we employ the Mellin transform technique to solve the IQBE, revealing that the dissipation follows an $\mathcal{N}$-body decay with a temperature-dependent $\mathcal{N}$. This finding challenges the conventional understanding of two-body loss dynamics.
To address harmonically trapped systems, we introduce a fast-flowing approximation (FFA) of the IQBE, enabling efficient numerical calculations. This method successfully reproduces experimental data without the need for fitting parameters, validating its efficacy and accuracy.
In both free space and trapped scenarios, we compare our calculations with solutions obtained using a thermal ansatz. Our results indicate that the system can achieve a quasi-thermalized state even in the absence of elastic collisions, except in deeply degenerate regimes. This observation provides new insights into the thermalization processes in these systems.

Our study offers a comprehensive understanding of two-body dissipative behavior in single-component Fermi gases. The results and methodologies presented here serve as valuable benchmarks for calibrating relevant Direct Simulation Monte Carlo (DSMC) simulations~\cite{wang2024simulations}. Furthermore, this work lays the groundwork for extending contact measurements using photo-excitation methods in $p$-wave BCS-BEC crossover studies.
By bridging theoretical predictions with experimental observations and providing novel analytical and numerical techniques, this research contributes significantly to the field of ultracold Fermi gas dynamics and opens new avenues for future investigations in related areas of quantum many-body physics.

\section{Acknowledgement}
We would like to acknowledge financial support from the National Natural Science Foundation of China under Grant No. 12204395, Hong Kong RGC Early Career Scheme (Grant No. 24308323) and Collaborative Research Fund (Grant No. C4050-23GF), the Space Application System of China Manned Space Program, and CUHK Direct Grant No. 4053676. We thank Doerte Blume and Kaiyuen Lee for their helpful discussions.


%